\documentclass[nofootinbib,10pt, twocolumn,prd,preprintnumbers,superscriptaddress,aps]{revtex4-2}
\usepackage{amsfonts}
\usepackage{graphicx}
\usepackage[colorlinks=true,linkcolor=red,citecolor=green,urlcolor=blue]{hyperref}
\usepackage{xcolor}
\usepackage{amsmath}
\usepackage{cases}
\usepackage{braket}
\usepackage[T1]{fontenc}
\usepackage{mathrsfs}
\usepackage{enumerate}
\usepackage{bm}
\usepackage{multirow}
\usepackage{comment}
\usepackage{makecell}
\usepackage{subfigure}

\usepackage[tiny, center, uppercase]{titlesec}
\usepackage{circuitikz}
\usepackage{multirow}

\usepackage{ulem}

\begin{document}


\title{The \texttt{frozen} vanilla model: Exploring dark sector interactions with delta effective theories}

\author{Mart\'in G. Richarte}
\email{martin@df.uba.ar (corresponding author)}
\affiliation{N\'ucleo Cosmo-ufes \& Departamento de F\'isica - Universidade Federal do Esp\'irito Santo, 29075-910 Vit\'oria, ES, Brazil}
\affiliation{Departamento de F\'isica, Facultad de Ciencias Exactas y Naturales,
Universidad de Buenos Aires, Ciudad Universitaria 1428, Pabell\'on I, Buenos Aires, Argentina}
\author{Luiz Filipe Guimar\~aes}
\affiliation{Departamento de F\'isica, Universidade Estadual de Londrina, Rod. Celso Garcia Cid, Km 380, 86057-970 Londrina, PR, Brazil}
\affiliation{N\'ucleo Cosmo-ufes \& Departamento de F\'isica - Universidade Federal do Esp\'irito Santo, 29075-910 Vit\'oria, ES, Brazil}
\author{Susana J. Landau}
\affiliation{CONICET - Universidad de Buenos Aires, Instituto de Física de Buenos Aires (IFIBA), Av. Intendente Cantilo S/N 1428 Ciudad Autónoma de Buenos Aires, Argentina}
\author{J\'ulio C. Fabris}
\affiliation{N\'ucleo Cosmo-ufes \& Departamento de F\'isica - Universidade Federal do Esp\'irito Santo, 29075-910 Vit\'oria, ES, Brazil}
\date{\today}

\begin{abstract}
 In this paper, we introduce a new interacting mechanism within the dark sector, encompassing both dark energy and dark matter, while grounding our analysis in the familiar framework of the $\mathrm{\Lambda CDM}$ model augmented by baryons and radiation components, including photons and neutrinos.  The interaction between dark energy and dark matter is confined to the perturbative level. One significant advantage of this proposal is that all geometric probes yield constraints consistent with the so-called vanilla model or the extended vanilla model, where dark energy has a constant equation of state, $w_{x}$. However, the introduction of this new interacting mechanism affects several theoretical signatures, involving contrast dark matter and dark energy densities. We perform an exploratory analysis of those effects in the CMB power spectra, matter-power spectra,  and redshift space distortions. For instance, it allows for a decrease/increment in the integrated Sachs-Wolfe (ISW) effect depending on the value taken by the interaction coupling. This effect could be observationally detected by looking for a cross-correlation between the ISW temperature fluctuations and the distribution of galaxies or quasars. At late times, the interaction in the dark sector becomes very effective, affecting the non-linear scale of structure formation.  We discuss how the estimators $f\sigma_{8}(z)$ and $S_{8}(z)$ are affected by different interacting couplings; indicating that $f\sigma_{8}(z)$ can show a relative change of up to $15\%$ compared to the concordance model at low redshifts. Finally, we show how the various terms in the dark energy pressure perturbation (both adiabatic and non-adiabatic) are relevant for different scales, demonstrating the absence of large-scale instabilities. 

\end{abstract}
\maketitle

\section{Introduction}
Within the domain of General Relativity (GR), the widely accepted cosmological model--often referred to as the 'concordance model' or $\mathrm{\Lambda CDM}$--represents the simplest model that provides a straightforward framework for interpreting observational data from various probes \cite{Weinberg:2008zzc, ratra}. At its core, the $\mathrm{\Lambda CDM}$-model combines a metric theory of gravity with the cosmological principle, depicting the universe as a homogeneous and isotropic spacetime described by the FLRW metric \cite{Weinberg:2008zzc}. It elegantly fits observations from cosmic microwave background radiation (CMB) \cite{planck2014}, large-scale structure surveys (LSS)\cite{ein}, and other sources. On the one hand, the homogeneous cosmological constant \cite{peeb1, wei1}, is usually interpreted as a vacuum energy density that drives the late-time accelerated expansion of the Universe \cite{sne1, sne2}.

In recent times, type Ia supernovae searches have increased the number of observed supernovae significantly. For instance, Pantheon$+$ compilation has substantially increased the sample size, including Cepheid distance measurements, while also improving systematic control \cite{panteo}. Whilst these new data appear to continue to favor the standard model over other choices, they show a discrepancy in the current Hubble parameter value ($H_0$)  in relation to the one inferred from the CMB data \cite{planck2014}.   On the other hand, the cold dark matter (CDM) component plays a pivotal role in describing the large-scale  cosmic structure, from galaxies to galaxy clusters, on scales beyond 100 $\text{Mpc}$, being responsible for their formation \cite{lid, prima, berto}.  Indeed, baryon acoustic oscillations generated in the photon-baryon fluid in the early universe \cite{ein}  were detected in the galaxy distribution using spectroscopic galaxies surveys \cite{sdss, sdss2, sdss3}.  The two-point correlation function computed from the catalogs of quasars remarkably pin points that these objects are also good tracers for the large-scale structure of the universe \cite{ein2, park}.

The remarkable success of the concordance model lies not only in its ability to fit observational data but also in its predictive prowess. The Planck satellite \cite{planck2014} revolutionized our understanding of cosmic microwave radiation through multifrequency observations \cite{planck2020a}, \cite{planck2020b}, producing stringent constraints on the six independent parameters of the $\mathrm{\Lambda CDM}$ model and derive additional bounds on the Hubble constant, $H_{0}$, the amplitude of density fluctuations $\sigma_{8}$,   the matter density, $\Omega_{m}$,  and $S_{8}=\sigma_{8}(\Omega_{m}/0.3)^{0.5}$.  In this way, the Planck Collaboration provided the state-of-the-art analysis of cosmic microwave radiation, including the detection of several galactic and extragalactic sources that contaminate the signal, such as polarized dust \cite{dust}. However, it is important to note that the latest results from the Kilo-Degree Survey (KiDS) have refined the estimation of the redshift distribution and expanded the survey area. As a result, the $S_8$ cosmological parameter now shows strong agreement with findings from the Planck satellite. The measurement yields $S_8 = 0.815^{+0.016}_{-0.021}$, reflecting a $0.73\sigma$ concordance with Planck \cite{newkids}.

 Although the concordance model has inherent simplicity, it is not immune to theoretical and observational challenges. There is a discrepancy between the observational value of the cosmological constant, denoted as  $\rho_{\text{obs}}=\Lambda \propto M^{2}_{p} H^{2}_{0}$,  and its prediction for the vacuum energy density of quantum field theory, expressed as $\rho_{\text{QFT}}=10^{122} \rho_{\text{obs}}$ \cite{wei1}, \cite{peeb2, sean, pad}; such a mismatch is known as the fine-tuning problem \cite{bur, ilia, sola} \footnote{A full debate on the cosmological constant problem is presented in \cite{jerome}.}. A somewhat weaker aspect is the so-called coincidence problem: Why is the ratio of dark matter to dark energy approximately equal, even though their dynamical evolution with redshift differs significantly? \cite{stein, stein2, no, migue}.  An interesting proposal to alleviate the coincidence problem was presented in \cite{di1, di2, di3, lu1} \footnote{A crucial point of the coincidence problem is to redirect our attention from the order of magnitude to how the onset of this behavior could affect structure formation. It is essential to recognize that non-linear structure formation must occur prior to the universe entering an accelerated expansion phase.}.
 
Significant discrepancies have emerged when comparing cosmological parameters obtained from CMB missions with those derived from late tracers. Comprehensive reviews addressing the tensions inherent to the concordance model can be found in the literature \cite{peri1, vale1, sun1}.  For example, the Planck mission has reported the current value of the Hubble constant as $H_{0}=67.4\pm 0.5 \text{km}\text{s}^{-1}\text{Mpc}^{-1}$ based on the concordance model \cite{planck2020a}. Meanwhile, direct local measurements, relying on SNeIa and cepheids, yield $H_{0}=73.04\pm 1.3\text{km}\text{s}^{-1}\text{Mpc}^{-1}$, as reported by the SHOES Collaboration \cite{shoes} \footnote{It should be noted that these so-called direct local measurements rely on the value of $q_{0}$ obtained from observations, in conjunction with the concordance model.}.  This discrepancy signals a tension $5\sigma$ between both collaborations. However, the source of such a mismatch remains unknown.  Further studies, based on observations from the James Webb Space Telescope in the near-infrared field, reveal that the situation worsens beyond several $\sigma$ thresholds. These findings cast doubt on the crowding effect in anchors with cepheids as a potential source of uncertainty \cite{riess2}.  Another crucial discrepancy pertains to the amplitude of the power spectrum, scaled by the square root of the matter density. The Planck mission reported $S_{8}= 0.834\pm0.016$, in contrast to the measurement of the DES-Y3 collaboration of $S_{8}= 0.766 ^{+ 0.020}_{-0.014}$ \cite{desy3}, \cite{desy3b}, which gives a lower value. A proposal to reconcile both values was reported in \cite{cal}, which envolves tempering with the amplitude of matter fluctuations on nonlinear scales. Indeed, the tensions arising from these divergent measurements stem from a fundamental distinction: the CMB data provides a snapshot frozen in time at the last scattering surface, whereas cosmic shear or weak lensing probes trace late-time cosmic evolution. This disparity necessitates extrapolation, thus assuming the cosmological constant paradigm when translating CMB to low redshift and therein lies the crux of the matter: conclusions drawn are inherently model-dependent (cf. \cite{vale2} and references therein).  An alternative explanation for these discrepancies might involve unidentified systematics or a distinct dark energy model, where the dark energy varies with the redshift \cite{des3}.  

Additional anomalies have been identified in the CMB data from early missions such as the Wilkinson Microwave Anisotropy Probe (WMAP) \cite{hansen}, which revealed a power asymmetry between the two hemispheres of the sky based on first-year data \cite{erik, hans2}. The Planck Collaboration later confirmed this hemispherical asymmetry, noting a dipolar modulation of significance $3\sigma$ for low multipoles ($\ell<60$) \cite{pla3}. Furthermore, the CMB shows an excess kurtosis for angular scales of approximately $10^{\circ}$, attributed to a cold spot in the southern Galactic hemisphere, which coincides with a region of positive dipolar modulation \cite{vielva}. A nearby spiral-rich group complex within this cold spot suggests that a local extragalactic foreground could explain the observed temperature depression \cite{lambas}.

An approach to extending the concordance model involves introducing an energy exchange mechanism aimed at mitigating cosmic coincidences \cite{ame1, wang, barrow, campo, hono, clem, yang, yang2, saulo, ada, mi, mi2, mi3, mi4, mi5, mi6, mi7, mi8, mi9, mi10, liu, ma3, ma4, rodrigo}. In certain cases, depending on the specific nature of the interaction, the Hubble tension can also be alleviated by an interaction proportional to the dark energy density \cite{olga, sunny, ema, pan, wy, mateo2, Yang:2018euj, sun2} or by a linear combination of dark matter and dark energy densities \cite{adaf}. An attractive scenario emerges with a minimal extension of the concordance model, in which the dark sector exchanges exclusively momentum at the perturbation level \cite{simpson, simpson2}. As a consequence of leaving the background unchanged, the elastic scattering between dark matter and dark energy helps to reconcile the interacting scenarios with the CMB data, and at the same time alleviate the $\sigma_{8}$ tension \cite{tram}. The elastic scattering model introduces a novel term into the dark matter Euler equation, involving the ratio of dark matter to dark energy, the scattering cross section, and the relative velocity between these dark components. A microscopic theory, rooted in a Lagrangian framework, can be expressed in terms of a modified quintessence model--known as model type 3--where a cubic kinetic term emerges in the dark matter framework \cite{skordis}. Cosmological constraints derived from various datasets suggest that this type 3 model remains highly competitive \cite{linton}. Furthermore, this model exhibits suppression in the growth of cosmic structure, a phenomenon further elucidated in \cite{fin}.  An analysis focusing on the role played by elastic interactions at small scales is presented in \cite{Asghari:2019qld}, highlighting the observational signatures left by peculiar velocities when considering the relative motion between baryons and dark matter. A slight reversal of this model is that the Hubble tension is not necessarily solved \cite{cardona}.

Inspired by Thomson scattering between baryons and photons during the early universe, a new interaction between dark energy and baryons was explored in \cite{teppa}.  This coupling does not alter the background cosmological model, but a relative velocity term emerges in the respective Euler equations for baryons and dark energy. A physical signature of the interaction is that it can potentially suppress cosmic structure growth, effectively preventing baryons from falling into the dark matter potential well \cite{teppa}.  While it is highly appealing, the prospects for directly detecting dark energy through cosmology appear unfeasible when considering realistic dark energy-baryon cross sections. In some cases, these cross sections are completely subdominant compared to Thomson scattering \cite{mota}. Although the effects linked to baryon and dark energy scattering appear to be quite subdominant within the framework of linear perturbation theory, N-body simulations have demonstrated that these effects can become significantly pronounced at the non-linear level \cite{mota2}. It should also be noted that specific interactions between dark matter and baryons in the early universe, based on phenomenological models or specific dark matter frameworks, were explored in the literature by several authors to some extent \cite{boehm1, boehm2, cora, poulin, yacine}.  Further consequences regarding such a coupling were obtained by analyzing the physics of the cosmic dawn \cite{jose, silk} and data from the EGDES collaboration \cite{ana}. Recently, bounds on dark matter-baryon interactions were reported by the Atacama Cosmology Telescope (ACT) through the combination of the Planck data with the ACT DR4 data \cite{liu}.  The impact of the dark matter-baryon interaction for the $S_{8}$ prior from DES-Y1, the BOSS DR12 BAO data, and the full Planck 2018 results was investigated in \cite{he}, leading to a $3\sigma$ significance of the non-vanishing interaction once the DES prior is included.  Taking into account a dark matter-neutrino interaction, it was argued that the Hubble and $\sigma_{8}$ tensions can be solved \cite{hivon}. However,  a constant scattering section for the dark matter-neutrino interaction shows a strong deviation from the concordance model when Lyman-$\alpha$ data are included \cite{mateo1}. 

 We are going to present a new mechanism within the dark sector, comprising dark matter and dark energy, which activates at the perturbation level while keeping the background unchanged. We focus on the concordance or extended $\mathrm{\Lambda CDM}$ model, where the dark energy equation of state remains constant at $w_{x}$. This interaction resembles a phase transition, linking a background model with zero energy exchange (the ‘vanilla’ model) to a perturbed cosmological model where the interaction is engaged. Expectedly, this mechanism will have significant effects primarily in the late universe, potentially observable through the CMB power spectrum, matter power spectrum, and clustering phenomena, such as redshift space distortion and number counts. Initially, the early universe aligns with the vanilla model in both background and perturbations. However, as time advances, an energy exchange initiates a transition to a distinct perturbative scenario within the concordance framework. This new model can be perceived as an effective theory that decouples from a fixed background. We can also draw parallels with elastic dark matter-dark energy models or momentum transfer models \cite{teppa, mota}, where the Euler equations are modified by Thomson-like scattering terms. However, this comparison is only partially accurate; while both scenarios operate at the perturbation level, the continuity equations for dark matter and dark energy change in our model, while the Euler equations remain unaffected as in the uncoupled case. A significant distinction also lies in the initial conditions of the radiation-dominated era: in our model, the dark matter potential velocity can be set to zero, which is not feasible in the momentum-transfer model.

The structure of our paper is organized as follows. In Sect. \ref{sec:bg}, we present the main features of background models (the vanilla model and the CPL model with a constant equation of state for dark energy), while in Sect. \ref{sec:cp}, we cover the basics of cosmological perturbations. In Sect. \ref{sec:nm}, we introduce a new interaction mechanism based on the perturbed effective theory for a fixed background, discussing the effective time scale when the interaction becomes efficient, along with the master equations for studying baryon and dark matter clustering. Sec. \ref{sec:ic} examines the initial conditions for dark-energy perturbations during the radiation era, focusing on super-horizon modes. In Sec. \ref{sec:ea}, we conduct a thorough numerical analysis of the interaction mechanism, scrutinizing both theoretical and observational signatures. We start by motivating parameter selection for numerical analysis (\ref{subsec:ps}), focusing on the equation of state parameter $w_x$ and coupling $\Sigma$, and categorizing models into four branches based on the signs of $(1 + w_x)$ and $\Sigma$. We examine the cosmological evolution of the interaction coupling (\ref{subsec:di}) across all branches and provide numerical results for branches I and III, employing a modified version of \texttt{CLASS} code \footnote{\url{https://github.com/lesgourg/class_public}}  \cite{ju1, ju2, ju3}. We explore the effects of the interaction coupling and dark energy equation of state on the CMB power spectrum \ref{subsec:ctt}, the late integrated Sachs-Wolfe effect \ref{subsec:isw}, the matter power spectrum from the \texttt{HALOFIT} method \ref{subsec:pdek}, and subsequent structure formation and clustering behaviors \ref{subsec:sforma}. Additionally, we investigate dark energy dynamics, including pressure perturbations, non-adiabatic contributions, and curvature perturbation evolution \ref{subsec:pzeta}. The analysis in \ref{subsec:i_iii} extends to branches II and IV in the next section, \ref{subsec:ii_iv}.  In \ref{sec_e} we summarize our principal findings, identifying, for each model, the best parameter ranges in terms of the relationship between observational signatures and current cosmological tensions. Finally, we summarize our conclusions in Sec. \ref{sec:sum}, encapsulating the key findings and implications.  In Appendix \ref{appendix:a}, we briefly describe the microscopic model introduced in the paper. Then, in Appendix \ref{appendix:b}, we discuss the relationship between the dark matter frame and the residual freedoms in synchronous gauge. Appendix \ref{appendix:c} illustrates the behavior of density contrasts and relative peculiar velocities. Appendix \ref{appendix:d} offers motivation for the use of parametric curves in the $f\sigma_{8}(z)$ and $S_{8}(z)$ planes to distinguish between different models.

\section{Background Evolution Equations}\label{sec:bg}

Our starting point is to consider a Universe described by the spatially flat FLRW metric, locally $\mathbb{R} \times \mathbb{R}^{3}$,  within the context of General Relativity (GR), where $G_{\mu\nu}=M^{-2}_{p}T_{\mu\nu}$ and $M^2_{p}=1/8 \pi G$ denote the square of the reduced Planck mass \cite{Weinberg:2008zzc}.  The Friedmann equation reads
\begin{equation}
H^2=\frac{1}{3M^2_p}\left(\bar{\rho}_r+\bar{\rho}_b+\bar{\rho}_{\text{dm}}+\bar{\rho}_{x} \right),
\end{equation}
where the labels $r, b, \text{dm}, x$ stand for radiation (photons+neutrinos), baryon,  dark matter, and dark energy,   respectively.  Introducing the dimensionless energy parameter  $\Omega_{i}=\rho_{i}/(3M^2_{p} H^{2})$, the Hubble function  can be recast as follows, 
\begin{equation}
H^2=H^2_0\left[\Omega_{r0}a^{-4}+\Omega_{m0}a^{-3}+\Omega_{x0}\right],\label{eq:FRE}
\end{equation}
where $\Omega_{r0}=\Omega_{\gamma0}+ \Omega_{\nu0}$ accounts for the neutrinos and photons contributions,  $\Omega_{m0}=\Omega_{b0}+\Omega_{\text{dm}0}$ denotes the amount of DM plus baryons, and  $\Omega_{x0}=1-\Omega_{r0}-\Omega_{b0}-\Omega_{\text{dm}0} $ represents the amount of DE, respectively.  In the following, the subscript `$0$' denotes their corresponding value at present,  where the scale factor is $a=1$. Given the fact that the different components do not interact with each other at the background level, all have a continuity equation of $\dot{\bar{\rho}}_{i}+3H(1+w_i)\bar{\rho}_{i}=0$, with $w_{b}=0$, $w_{r}=1/3$,  $w_{\text{dm}}\simeq 0$, and $w_{x}=-1$.   The matter content of each component that fills the universe will be modeled as a cosmic fluid with a barotropic equation of state, $p_{A}=w_{A}\rho_{A}$, or equivalently with a energy-momentum tensor (EMT)  given by $ T^{\mu}_{ \nu~A}= (\rho_{A}+p_{A})u^{\mu}_{A}u_{\nu ~A} + p_{A}\delta^{\mu}_{\nu}$.  Notice that the local energy density is the eigenvalue and $u^{\mu}_{A}$ the eigenvector that diagonalize the EMT, thus $ T^{\mu}_{ \nu~A} u^{\nu}_{ ~A}=-\rho_{A} u^{\mu}_{A}$.  A small deviation from the vanilla model could be considered as a dark energy with a constant equation of state.  Then, the density parameter behaves as $\rho_{x}=\rho_{x,~0}a^{-3(1+w_{x})}$.  If that is the case, the Hubble function reads $H=H_{0}\left[ \Omega_{r0}a^{-4}+\Omega_{m0}a^{-3}+\Omega_{x0} a^{-3(1+w_{x})}\right]^{1/2}$. 
\vspace{0.12cm}

\section{Cosmological Perturbation Equations}\label{sec:cp}
In this section, we revisit the fundamental elements of cosmological perturbation theory, specifically focusing on scenarios that incorporate an arbitrary interaction between dark components at the perturbative level. The reader may refer to \cite{Kodama:1984ziu} and \cite{Valiviita:2008iv}—and the references therein—for cosmological perturbation theory and to Ref. \cite{Hwang:2001qk} for gauge-ready formalism. We begin by writing the line element in the case of scalar perturbations \cite{Valiviita:2008iv, Majerotto:2009np, Valiviita:2009nu}, leaving aside the cases of vector and tensor perturbations,
\begin{eqnarray}\label{eq1}
ds^2&=a^2\Big( -(1+2\phi)d\tau^2+2\partial _i B d\tau dx^i+
  \nonumber\\
&\left[(1-2\psi)\delta_{ij}+2\partial_i\partial_j E\right]dx^i
dx^j \Big),
\end{eqnarray}
where  the conformal time defined as  $d\tau=dt/a(t)$. The functions $\phi$, $\psi$, $B$, and $E$ are the  contributions to the scalar sector under the $SO(3)$ group. Here,  $\psi$ is called the gauge-dependent curvature perturbation,  $\phi$ is the lapse function, and both $B$ and $E$ can be combined to obtain the scalar shear, $\sigma_{s}=a^{2}\dot{E}-aB$ \cite{Kodama:1984ziu}. For a comoving observer $A$,  the background four-velocity reads $\bar{u}^\mu=a^{-1}(1,0,0,0)$.  As usual, we express the spatial part of the velocity perturbation in terms of a velocity potential due to Helmholtz's theorem, namely, $\delta u^{i}=\partial^i v_A$.   Combining (\ref{eq1})  with the norm of the four-velocity vector  ($g_{\mu\nu}u_A^\mu u_A^\nu=-1$), we arrive at  the four-velocity,
\begin{eqnarray}
u^\mu_{A} &=a^{-1}(1-\phi, \partial^i v_A),\nonumber\\
u^A_{\mu} =g_{\mu\nu}u^\nu_{A}&=a(-1-\phi,\partial_i [v_A+B]).
\end{eqnarray}
Here we write the local volume expansion rate as $\theta = \vec{\nabla}\cdot \vec{v}$ and it becomes $\theta_A=-k^2(v_A+B)$ in the Fourier space.  The perturbed component of EMT  in the case of zero anisotropic tensor (in the dark sector) is  $\delta T^{0}_{0}=-\delta \rho=-\sum_{A}\delta\rho_{A}$,  $\delta T^{0}_{i}=-  (\bar{\rho}+ \bar{p})(B_{i}+v_{i})$, and $\delta T^{i}_{j}= \delta p \delta^{i}_{j}= \sum_{A}\delta p_{A} \delta^{i}_{j}$. Besides, the perturbed Einstein's field equation are
\begin{eqnarray}
\delta G^{0}_{0}&=&\nabla^{2}\psi-\mathcal{H}\nabla^{2}(B-E')-3\mathcal{H}(\psi' +\mathcal{H}\phi) \\ \nonumber
                         &=&4\pi  G a^{2} \delta T^{0}_{0},\\
\delta G^{0}_{i}&=& \nabla_{i}(\mathcal{H}\phi + \psi')=4\pi G a^{2} \delta T^{0}_{i},\\
\delta G^{i}_{j}&=& \Big[\psi'' + \mathcal{H} (\phi' +2\psi') + (\mathcal{H}^{2}+2\mathcal{H})\phi  \Big]\\ \nonumber
&=& 4\pi G a^{2} \delta T^{i}_{j},
\end{eqnarray}
By taking the trace of the $\delta G^{i}_{j}$ equation, we obtain
\begin{equation}
\psi'' + \mathcal{H} (\phi' +2\psi') + (\mathcal{H}^{2}+2\mathcal{H})\phi = 12\pi Ga^{2}\delta p,
\end{equation}
whereas, the divergence  of $\delta G^{0}_{i}$ implies the following relation for $\theta$:
\begin{equation}
\nabla^{2}(\mathcal{H}\phi + \psi')=-4\pi G a^{2} (p+\rho)\theta.
\end{equation}
We proceed to compute the perturbation of the local energy-momentum and separate them as follows.
\begin{widetext}
\begin{eqnarray}\nonumber
&a^{2}&\delta\big(\nabla_{\mu} T^{\mu 0}_{A}\big)=\delta \rho'_A+3\mathcal{H}(\delta\rho_A+\delta p_A)-3(\bar{\rho}_A+\bar{p}_A)\psi' 
+(\bar{\rho}_A+\bar{p}_A)\nabla^2(v_A+E') -2\left[\bar{\rho}'_A+3\mathcal{H}(\bar{\rho}_A+\bar{p}_A)\right]\phi, \\ 
&a^{2}&\delta \big( \nabla_{\mu}T^{\mu i}_{A}\big)=\partial^i \Big(\left[(\bar{\rho}_A+\bar{p}_A)(v_A+B)\right]'+4\mathcal{H}(\bar{\rho}_A+\bar{p}_A)(v_A+B)+
(\bar{\rho}_A+\bar{p}_A)\phi+\delta p_{A}
 -\left[\bar{\rho}'_A+3\mathcal{H}(\bar{\rho}_A+\bar{p}_A)\right]B\Big), ~ \label{peq} ~~~~
\end{eqnarray}
\end{widetext}
where the symbol $'$ stands for the derivative with respect to the conformal time and $\mathcal{H}=a'/a$ is the conformal Hubble parameter. At this point, we introduce the existence of an exchange of energy-momentum  exclusively at the  perturbative level \cite{Valiviita:2008iv} , 
\begin{equation}
\nabla_\mu T^{\mu\nu}_A=0,\quad \delta\nabla_\mu
T^{\mu\nu}_A=\delta Q_A^{\nu},
\end{equation}
and the general interaction term can be written  as covariant tensor of rank one, 
\begin{equation}
Q^{\mu}_A=Q_A u^{\mu}+F^{\mu}_A,
\end{equation}
where  $ F^{\mu}_A=a^{-1}(0,\partial^i f_A)$ represents the momentum transfer rate relative to $u^{\mu}_{A}$, therefore $u_{\mu A}F^{\mu}_A=0$\cite{Valiviita:2008iv}. The main difference from other proposals in the literature is that the exchange of energy takes place at the perturbative level $Q_A=\delta Q_A$, so $\bar{Q}_{A}=0$.   In another words, the components of the  interaction four-vector are:
\begin{eqnarray}\nonumber
Q^0_A=\delta Q_{A}u^0&=&a^{-1}(1-\phi)\delta Q_A ,\\ 
 \delta Q^0_A&=&a^{-1}\delta Q_A,
\end{eqnarray}
and
\begin{eqnarray}\nonumber
Q^i_A&=&\delta Q_Au^i+a^{-1} \partial^{i}f_A, \\
 \delta Q^i_A&=&a^{-1}\partial^i f_A.
\end{eqnarray}
As we might expect, the conservation of the perturbed energy-momentum  for the subsystem (dark matter + dark energy) implies that $\sum_{A} \delta Q_A=\sum_{A} f_A=0$ . The next step in the exposition of the cosmological perturbation formalism is  to inspect the modification of energy/momentum exchange in the energy and momentum balance equations,
\begin{eqnarray} 
\delta \rho'_{A} &+&3\mathcal{H}(\delta\rho_A+\delta p_A)-3(\bar{\rho}_A+\bar{p}_A)\psi'+ \\ \nonumber
(\bar{\rho}_{A} &+&\bar{p}_A)\nabla^2(v_A+E')= a\delta Q_A, ~~\\ \nonumber
\delta p_A&+&\left[(\bar{\rho}_A+\bar{p}_A)(v_A+B)\right]'+4\mathcal{H}(\bar{\rho}_A+\bar{p}_A)(v_A+B)\\ 
&+&(\bar{\rho}_A+\bar{p}_A)\phi= a f_{A}.~~~
\end{eqnarray}
It is important to stress that the above expressions slightly differ from those reported in \cite{Valiviita:2008iv}, as the terms multiplied by $\bar{Q}_{A}$ are zero in our case. 

As is well-known, the relation between $\delta \rho_A$ and $\delta p_A$ is not fixed. Even if there is a close relation between pressure and local energy of each component at the background level, there is no guarantee that this relation holds at the perturbative level.  To close the system of equations,  we need to specify a microscopic theory or phenomenological relation between these perturbative variables. One way to tackle this issue is by considering the adiabatic and intrinsic nonadiabatic perturbations. The sound speed of the $A$  fluid, say  $c^{2}_{sA}$,  is defined in the $A$ rest frame as \cite{Valiviita:2008iv}:
\begin{equation}
c^2_{sA}=\Big(\frac{\delta p_A}{\delta \rho_A}\Big)_{\text{rf}},
\end{equation}
where ${\text {rf}}$ indicates the rest frame. On the other hand,  the adiabatic  sound speed  for a fluid is given by,
\begin{equation}
c^2_{aA}=\frac{p'_A}{\rho'_A}=w_A+\frac{w'_A}{\rho'_A/\rho_A}.
\end{equation}
The rest frame of the   $A$  fluid is defined by the condition $\big(T^i_{0}\big)_{\text{rf}}=0= \big(T^0_{i}\big)_{\text{rf}}$. Some authors refer to this choice of gauge as the zero momentum gauge, provided $v_{A\text{rf}}=0$, and orthogonal because $B_{A\text{rf}}=0$.  To connect this particular gauge/frame to any other gauge, we perform an infinitesimal gauge transformation, $x^\mu\rightarrow x^\mu+(\delta \tau_A, \partial^i\delta x_A)$ \cite{Valiviita:2008iv}, which can be recast as 
\begin{eqnarray}
v_A+B&=&\Big(v_A+B\Big)_{\text{rf}}  +\delta \tau_A,\\ 
 \delta p_A&=&\Big(\delta p_A\Big)_{\text{rf}}  -p'_A\delta \tau_A, \\
\delta \rho_A&=&\Big(\delta \rho_A\Big)_{\text{rf}} -\rho'_A\delta \tau_A.
\end{eqnarray}
Using the gauge freedom available, we select  $\tau_A=v_A+B$ so that  the perturbation of the pressure becomes
\begin{eqnarray}\nonumber
\delta p_A&=&(\delta p_A)_{\text{rf}}  -p'_A\delta \tau_{A} \\ \nonumber
&=&c^2_{sA}\delta \rho_A+(c^2_{sA}-c^2_{aA})\rho'_A(v_A+B)\\ 
&=&c^2_{aA}\delta \rho_A+\delta p_{\text{nad} A},
\end{eqnarray}
where $\delta p_{\text{nad} A}=(c^2_{sA}-c^2_{aA})\left[\delta \rho_A+\rho'_A(v_A+B)\right]$ term represents the intrinsic non-adiabatic perturbation in the $A$-fluid \cite{Malik:2004tf}. Given  that the exchange of energy/momentum occurs at the perturbation level, the continuity equation remains unchanged [ $\bar{\rho}'_A=-3\mathcal{H}(\bar{\rho}_A+\bar{p}_A)$],  and using  the relation $\theta_A=-k^2(v_A+B)$ in Fourier space, the pressure perturbation acquires the following form \cite{Kodama:1984ziu, Malik:2004tf},

\begin{eqnarray}
\delta p_A&=&c^2_{sA}\delta \rho_A+(c^2_{sA}-c^2_{aA})\rho'_A(v_A+B) \\ \nonumber
&=&c^2_{sA}\delta \rho_A+(c^2_{sA}-c^2_{aA})\left[3\mathcal{H}(1+w_A)\bar{\rho_A}\right]\frac{\theta_A}{k^2}. \label{dpx}
\end{eqnarray}

The master equations that govern the dynamics of the velocity perturbation, denoted $\theta_{A}$, and the density contrast perturbation, $\delta_A=\delta \rho_A/\bar{\rho}_A$, in the context of an interaction vector that operates exclusively at the perturbative level, are articulated in the seminal works of \cite{Kodama:1984ziu, Malik:2004tf}:

\begin{eqnarray}\nonumber
\delta'_A&+&3\mathcal{H}(c^2_{sA}-w_A)\delta_{A}+(1+w_A)\theta_A+k^2(1+w_A)(B-E')\\ \nonumber ~~
&+&3\mathcal{H}\left[3\mathcal{H}(1+w_A)(c^2_{sA}-w_A)+w'_A\right]\frac{\theta_A}{k^2}-3(1+w_A)\psi' \\  &=& a\frac{\delta Q_A}{\bar{\rho}_A}, \label{meq}\\  \nonumber
\theta'_A&+&\mathcal{H}(1-3c^2_{sA})\theta_A-\frac{c^2_{sA}}{(1+w_A)}k^2\delta_A-k^2\phi \\ 
&=&-k^2f_A\frac{a}{(1+w_A)\bar{\rho}_A}. \label{meq2}
\end{eqnarray}
Once a specific form of $\delta Q_{A}$ and  $ f_{A}$ are chosen, the phenomenological interaction framework is established. To avoid violating the weak equivalence principle or constraints from the fifth force in the solar system, the usual approach is to set  $ f_{A}=0$ and only allows for  $\delta Q_{A} \neq 0$  \cite{Valiviita:2008iv}.  

\section{A new interacting mechanism}\label{sec:nm}
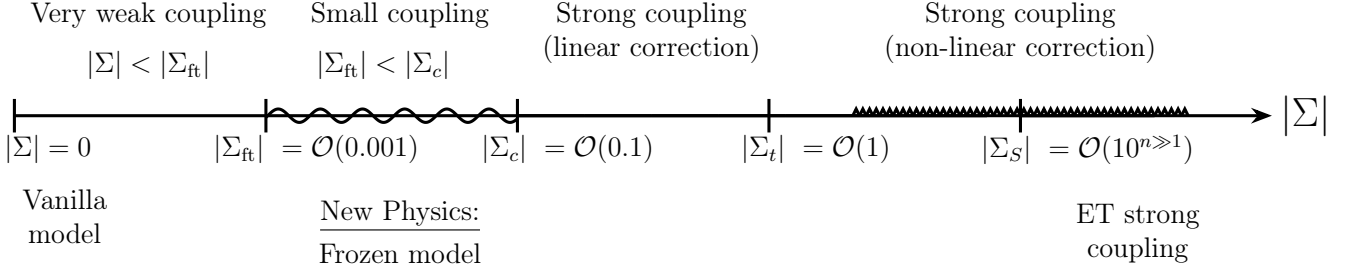
\begin{figure*}[!ht]
\centering
\resizebox{1\textwidth}{!}{ %
\begin{circuitikz}
\tikzstyle{every node}=[font=\LARGE]

\draw [line width=1.3pt, ->, >=Stealth] (1.25,9.75) -- (20,9.75);
\node [font=\LARGE] at (20.5,9.75) {$|\Sigma|$};
\draw [line width=1.3pt, short] (1.25,10) -- (1.25,9.5);
\node [font=\large] at (1.75,9.25) {$|\Sigma| = 0$};
\node [font=\large] at (2,8.5) {Vanilla};
\node [font=\large] at (2,8) {model};
\draw [line width=1.3pt, short] (5,10) -- (5,9.5);
\draw [line width=1.3pt, short] (8.75,10) -- (8.75,9.5);
\draw [line width=1.3pt, short] (12.5,10) -- (12.5,9.5);
\draw [line width=1.3pt, short] (16.25,10) -- (16.25,9.5);
\node [font=\large] at (5.75,9.25) {$    |\Sigma_{\rm{ft}}| ~ =\mathcal{O}(0.001)$};
\node [font=\large] at (9.5,9.25) {$  |\Sigma_c| ~ =\mathcal{O}(0.1)$};
\node [font=\large] at (13.2,9.25) {$  |\Sigma_{t}| ~ =\mathcal{O}(1$)};
\node [font=\large] at (17.25,9.25) {$    |\Sigma_S| ~ =\mathcal{O}( 10^{n \gg  1})$};
\node [font=\large] at (18,8.25) {ET strong};
\node [font=\large] at (18,7.75) {coupling};
\node [font=\large] at (16.25,11.25) {Strong coupling};
\node [font=\LARGE] at (10.75,12.5) {$\delta$-ET};
\node [font=\large] at (3.25,11.25) {Very weak coupling};
\node [font=\large] at (7,8.25) {\underline{New Physics:}};
\node [font=\large] at (7,7.7) {Frozen model};
\node [font=\large] at (6.75,10.5) {$  |\Sigma_{\rm{ft}}| < |\Sigma_c|$};
\draw[domain=5:8.75,samples=100,smooth, line width=1.3pt] plot (\x,{0.1*sin(10*\x r -0 r ) +9.75});
\foreach \x in {0,...,6}{
  \draw [ line width=1.3pt] (8.75+\x*1,9.75) -- ++(0.5,0) -- ++ (0.5, 0);
}
\foreach \x in {0,...,1}{
  \draw [ line width=1.3pt] (16.25+\x*1,9.75) -- ++(0.5,0) -- ++ (0.5, 0);
}
\foreach \x in {0,...,49}{
  \draw [ line width=1.3pt] (13.75+\x*0.1,9.75) -- ++(0.05,0.1) -- ++ (0.05, -0.1);
}
\node [font=\large] at (10.75,11.25) {Strong coupling};
\node [font=\large] at (3.25,10.5) {$|\Sigma| < |\Sigma_{\rm{ft}}|  $};
\node [font=\large] at (7,11.25) {Small coupling};
\node [font=\large] at (10.75,10.75) {(linear correction)};
\node [font=\large] at (16.25,10.75) {(non-linear correction)};
\end{circuitikz}
}
\caption{The different extensions of the vanilla model under the perturbative ET approach are shown upon the value of the coupling parameter $\Sigma$. Each region in  $\Sigma$ corresponds to distinct modifications in the conservation equations and the Einstein field equations for a given background.}
\label{fig:eft}
\end{figure*}

 We introduce a new interacting mechanism within a covariant framework. The aforementioned proposal is framed as an effective theory for cosmological perturbations, termed $\delta$-ET, situated within the context of an interacting dark sector [cf. Fig. \ref{fig:eft}]. In this exploration, we will adopt the effective theory approach in conjunction with Einstein's field equations, while also ensuring that the conservation of energy-momentum is upheld.   To grasp the rationale behind the $\delta$-ET framework, we must first examine the so-called \texttt{vanilla model} and the covariant perturbed field equations that dictate its dynamical evolution beyond the fixed background (associated with the concordance model): 
\begin{eqnarray}
\label{CA1}
&\delta G^{\texttt{vm}}_{\mu\nu}= \kappa^{2} \delta T^{\texttt{vm}}_{\mu\nu}, \\  
&\delta \Big(\nabla^{\mu}T^{\texttt{vm},\rm{sm}}_{\mu\nu}\Big)=0,
\end{eqnarray}
where $\kappa^{2}=8\pi G$.   However, the\texttt{ frozen model}, which arises as an extension of the \texttt{vanilla model}, is distinguished by a nearly identical set of perturbed field equations. This similarity suggests that the dynamics of the frozen model retain essential features of the original framework, allowing for a deeper exploration of their implications in cosmological perturbations. In that sense,  the covariant perturbed field equations within the $\delta$-ET framework for the\texttt{ frozen model} can be recast as,
\begin{eqnarray}
\label{CA2}
\delta G^{\texttt{fm}}_{\mu\nu}&= \kappa^{2}  \delta T^{\texttt{fm}}_{\mu\nu}, \\  
\delta\Big(\nabla^{\mu}T^{\texttt{fm}, \rm{sm}}_{\mu\nu}\Big)&=0,~~~~~~~~~\\
\delta\Big(\nabla^{\mu}T^{\texttt{fm}, \rm{x}}_{\mu\nu}\Big)&=\delta Q^{\texttt{fm},\rm{x}}_{\nu},
 \\  
\delta\Big(\nabla^{\mu}T^{\texttt{fm}, \rm{dm}}_{\mu\nu}\Big)&=\delta Q^{\texttt{fm}, \rm{dm}}_{\nu}.
\end{eqnarray}

Notice that the covariant interaction vector is nonzero solely for the dark sector (dark matter plus dark energy), thus $\delta Q^{\texttt{fm}, \rm{dm}}_{\nu}+\delta Q^{\texttt{fm},\rm{x}}_{\nu}=0$ \footnote{It should be clarified that the label \texttt{vm} refers to the vanilla model, whereas \texttt{fm} denotes the frozen model.}. Interestingly enough, this interaction manifests itself only at the perturbation level for these components and that is the reason behind the term $\delta$-ET, as mentioned before \footnote{See. Appendix \ref{appendix:a}.}. Additionally, we have the standard energy-momentum conservation equations for each component of the vanilla model considered separately. In contrast, both the \texttt{frozen vanilla model} and the \texttt{vanilla model} share the same background, typically represented by either the $\Lambda$CDM model or a CPL model with a constant equation of state. Thus, we can argue that their theoretical treatments differ significantly at the perturbative level \cite{Gubitosi:2012hu}. This distinction is essential for exploring the observational signatures and theoretical implications of our proposed framework across various physical observables.

In considering the physical rationale for choosing a specific covariant interaction vector among various options, we can assert that a key guiding principle is that this interaction should not violate any established symmetries in nature \cite{lu1, wang, barrow, clem}. If, however, it does break a symmetry, it must do so within the constraints imposed by current observational bounds. This criterion ensures that our theoretical framework remains consistent with empirical data while exploring new dynamics \cite{wang, clem, Valiviita:2008iv}.
For instance, we might concentrate on $\delta$-ET, which leads to modifications of the vanilla model that are relevant only in recent epochs; thereby leaving the early universe and Big Bang nucleosynthesis unaltered. During the late pivotal phases of the universe's evolution, the interplay between dark matter and dark energy becomes essential for accurately describing the large-scale structure of the universe and its observed accelerating expansion. In addition, these late phases can be tested in a plethora of surveys and future missions.

 We proceed by presenting the contrast and velocity evolution equation for the system (dark matter + dark energy) in the synchronous gauge,  obtained by choosing $\phi=B=0$, $\psi=\eta$, and $k^{2}E=-\frac{h}{2}-3\eta$ \cite{Hwang:2001qk}.  As mentioned earlier, these system of equations are coupled through the cosmic expansion of the universe and the local exchange of energy ($\delta Q_{A}\neq 0$). In the context of dark energy perturbations, it is important to recognize that a model with a constant $w_x$ cannot be treated as strictly barotropic. Nonadiabatic contributions must be considered to accurately capture the dynamics (see Appendix A in \cite{wayne}). Consequently, we adopt $c^{2}_{a~x}=w_{x}<0$ and $c^{2}_{sx}=1>0$ \cite{wayne}.  In the case of cold dark matter, the relevant physical parameters are $w_{\text{dm}}\simeq 0$ and $c^{2}_{a~\text{dm}}=c^{2}_{s~\text{dm}}=w_{\text{dm}}$. The dynamical equations for the cold dark matter are given by, 
\begin{eqnarray}\label{sync2a}
\delta'_{\text{dm}}&+& \theta_{\text{dm}} +\frac{h'}{2}=a\frac{\delta Q_{\text{dm}}}{\bar{\rho}_{\text{dm}}},\\
\theta'_{\text{dm}}&+&\mathcal{H}\theta_{\text{dm}}=0.  \label{sync2b}
\end{eqnarray}
 Besides,  the dynamical equations for DE  can be recast as follows, 
\begin{eqnarray}\label{sync1}
\delta'_{x}&+&(1+w_x) \big(\frac{h'}{2} +\theta_x \big)+  3\mathcal{H}(c^2_{sx}-w_{x})\delta_{x}\\ \nonumber
&+& 9\mathcal{H}^{2}(1+w_{x})(c^2_{sx}-w_{x})\frac{\theta_{x}}{k^2}=a\frac{\delta Q_{x}}{\bar{\rho}_{x}}, \label{sync1x}\\
\theta'_{x}&+&\mathcal{H}(1-3c^2_{sx})\theta_{x}-\frac{c^2_{sx}}{(1+w_{x})}k^2\delta_{x}=0. 
\end{eqnarray} 
To ensure the self-consistency of the perturbation framework, we incorporate the perturbation of the Hubble function in $\delta Q_{x}$, as introduced in \cite{gavela}.
In the synchronous gauge, the energy exchange perturbation can be expressed as follows:

\begin{eqnarray}\label{IFV}
\delta Q_{x}=-\Sigma (3H\bar{\rho}_{x})[\mathcal{K}+\delta_{x}]. 
\end{eqnarray}

Here, the term $3H\bar{\rho}_{x}$ serves as a convenience factor, although it is not strictly necessary; therefore, one could simply consider $\delta Q_{x}=-\Sigma[\mathcal{K}+\delta_{x}]$. The $\mathcal{K}$-term encapsulates the perturbation of the Hubble term \cite{gavela}: 

\begin{eqnarray}\label{eq:kappa}
\mathcal{K}=\Big(\frac{\frac{h'}{2}+\theta_{T}}{3\mathcal{H}}\Big),
\end{eqnarray}

where the velocity perturbation for the dark sector components is given by  $\theta_{T}=\sum_{A}(1+w_{A})\bar{\rho}_{A} \theta_{A}/(1+w_{T})\bar{\rho}_{T}$ being $\bar{\rho}_{T}=\bar{\rho}_{x}+\bar{\rho}_{\text{dm}}$. In case of a synchronous gauge with a cold dark matter frame ($\theta_{\text{dm}}=0$), the latter expression becomes $\theta_{T}=\theta_{x}/(1 +\mathcal{R}_{I})$ with $\mathcal{R}_{I}=\bar{\rho}_{\text{dm}}/ \bar{\rho}_{x}$~\footnote{It is worth noting that the interaction is not explicitly dependent on $k$, but it implicitly dependent via its terms $\delta_x$, $h'$ and $\theta_x$, which are all Fourier modes.}.  Notice that under a gauge transformation, $\delta Q_{A}$ remains unchanged, making it a physical gauge-invariant quantity \cite{Kodama:1984ziu}. In relation to Eq. (\ref{sync2a}) and Eq. (\ref{sync1}), local energy conservation implies $\delta Q_{\text{dm}}=+\Sigma (3H\bar{\rho}_{x})[\mathcal{K}+\delta_{x}]$.  At this point, we could identify the typical time scale associated with the interaction by noticing that $a \delta Q_{x}/\bar{\rho}_{x} = -3 \Sigma \mathcal{H} [\mathcal{K}+\delta_{x}]$ so the effective coupling becomes $\Sigma_{\rm{eff}}=3\Sigma \mathcal{H}$, so that $\Sigma$ is dimensionless. Then, the new time scale and the strength of the interaction are given by $t_{\Sigma}=\Sigma^{-1}_{\rm{eff}}$. In a similar fashion, using that $a \delta Q_{c}/\bar{\rho}_{\rm{dm}} = 3 \Sigma \mathcal{H} \mathcal{R}_{I} [\mathcal{K}+\delta_{x}]$ another equivalent time-scale can be expressed as $t_{\Sigma}=\Sigma^{-1}_{\rm{eff}} \mathcal{R}^{-1}_{I}$. Both quantities, $\Sigma^{-1}_{\rm{eff}}$ and $\Sigma^{-1}_{\rm{eff}} \mathcal{R}^{-1}_{I}$, can be used to measure the coupling in the dark sector, provided that these combinations appear in the master equations.   In other words, we could anticipate that interactions will become increasingly significant in the late universe, specially because the background will not change accordingly as is fixed to the vanilla model. The typical time scales for the interactions, compared to the conformal Hubble time, can be illustrated in Fig. \ref{fig:int}.  Although  our proposal it bears some resemblance to elastic dark-matter-dark energy models or momentum transfer models \cite{teppa, mota}, where the Euler equations are adjusted by scattering terms, this analogy has its limitations. In our framework, the continuity equations governing dark matter and dark energy evolve, whereas the Euler equations remain unchanged, akin to the uncoupled scenario. Another crucial difference emerges in the initial conditions during the radiation-dominated era: our model allows the dark-matter potential velocity to be set to zero, which is not an option in momentum-transfer models.

The master equation for neutrinos and photons remains unchanged in comparison to the standard case. For example, after the decoupling from the baryon-photon fluid \footnote{ It is essential to highlight additional terms present in the baryon master equation. Firstly, the master equation governing the baryon density contrast is expressed as $\dot{\delta}_{b}=-\theta_{b}-\dot{h}/2$. Secondly, the master equation for the peculiar velocity can be reformulated as $\dot{\theta}_{b}=-\mathcal{H}\theta_{b}+ c^{2}_{sb}k^{2}\delta_{b}+R a n_{e}\sigma_{T}(\theta_{\gamma}-\theta_{b})$, where $R=4\bar{\rho}_{\gamma}/3\bar{\rho}_{b}$ \cite{Ma:1994dv}. However, it is important to note that these terms do not play a role in describing the interacting dark sector.}, the   dynamical equations for baryons are  \cite{Ma:1994dv},
\begin{eqnarray}\label{sync3a}
\delta'_{b}&+&\frac{h'}{2} +\theta_{b} =0, \nonumber\\
\theta'_{\text{b}}&+&\mathcal{H}\theta_{\text{b}}=0.\label{sync3b}
\end{eqnarray}

We recover the usual master equations in the synchronous gauge of the $\mathrm{\Lambda CDM}$ model for $\delta Q_{\text{dm}}=0$  \cite{Ma:1994dv}.  Interestingly, the cold dark-matter velocity perturbation equation does not change compared to the uncoupled case, where the interaction vanishes. Combining the Euler equations for dark matter and baryons (\ref{sync2b}-\ref{sync3b}), we obtain $\Theta_{\text{dm-b}}+ \mathcal{H} \Theta_{\text{dm-b}}=0$, then the relative velocity is $\Theta_{\text{dm-b}}=\rm{cte}$. Further, the Euler equation for the dark matter component (\ref{sync2b}) indicates there is no transfer of momentum in the dark matter frame, so it follows geodesics.   Therefore, it is possible to set $\theta_{\text{dm}}=0$ in the dark matter frame (for a full discussion on the cold dark matter frame, see Appendix \ref{appendix:b}).

To obtain a second-order differential equation for dark matter density contrast, we start by calculating $\delta''_{\text{dm}}$ from Eq. (\ref{sync2a}) and replacing $\theta'_{\text{dm}} $  in terms of   $\delta'_{\text{dm}}$ , 
\begin{equation}\label{constradm}
\delta''_{\text{dm}}+ \mathcal{H} \delta'_{\text{dm}}= \mathcal{H} a\frac{\delta Q_{\text{dm}}}{\bar{\rho}_{\text{dm}}} + \Big(a\frac{\delta Q_{\text{dm}}}{\bar{\rho}_{\text{dm}}} \Big)' +4\pi Ga^{2}(\delta \rho+3\delta p),
\end{equation}
where we used the fact that in the synchronous gauge, after combining two of the Einstein field equations, we obtain $h'' +\mathcal{H}h=-8\pi Ga^{2}(\delta \rho+3\delta p)$.  Similarly, by working the master equations for baryons, we arrive at a second order differential equation for the baryons contrast density,
\begin{equation}\label{constrab}
\delta''_{\text{b}}+ \mathcal{H} \delta'_{\text{b}}= 4\pi Ga^{2}\Big( \delta_{b} \bar{\rho}_{\text{b}} + \delta_{\text{dm}}\bar{\rho}_{\text{dm}}\Big).
\end{equation}
We will adopt a framework where $\bar{Q}_{\text{dm}}= 0$ is at the background level, ensuring that the background evolution equation for the density parameter aligns with that of the vanilla model. However, the dynamics changes significantly when we account for perturbations in dark matter and dark energy, particularly under the condition that an exchange is allowed ($\delta Q_{\text{dm}}\neq 0$). This introduces a new mechanism in which the perturbed system can be viewed as exhibiting a phase transition relative to the background model, which leads us to designate this approach as the \texttt{frozen vanilla model}.

\section{Behavior in the radiation epoch: Initial conditions}
\label{sec:ic}

We begin by examining the initial conditions, particularly emphasizing the behavior of modes that are well outside the horizon, or in the superhorizon limit ($k\ll \mathcal{H}$), during the radiation era. In this regime, the scale factor evolves as $a(\tau)\simeq \sqrt{\Omega_{r0}}H_{0}\tau$, following the methodology established by Ma and Bertschinger \cite{Ma:1994dv}. 
 The only potential issue would be the presence of a growing mode due to the interaction, but this is not the case for the interaction we are considering (cf. \cite{Valiviita:2008iv}). As usual, the metric perturbation in the synchronous gauge will evolve according to the growing mode, $h=C_{\gamma}(k\tau)^{2}$, where $C_{\gamma}$ can depend on the mode $k$ at most and its specific value is fixed at inflation. In the case of the $\delta Q_{x}$ interaction, the velocity of dark matter can effectively be set to zero, as indicated by Eq. (\ref{sync2b}).  
 
 It is appropriate to approximate the density contrast using its particular (inhomogeneous) solution corresponding to the attractor solution, provided $\Sigma \mathcal{R}_{I}\ll 1$ during the radiation era, 
\begin{eqnarray}\label{sync2ap}
\delta'_{\text{dm}}+\frac{h'}{2}\simeq 0. 
\end{eqnarray}
Due to this behavior (\ref{sync2ap}), the contrast density of dark matter yields $\delta_{\text{dm}}=h/2$, provided that the integration constant vanishes. Within the same limit, the dark energy master equation can be written,
\begin{eqnarray}\label{syncz}
\delta'_{x}&=&-(1+w_{x}) \frac{h'}{2}-  3\mathcal{H}(c^2_{sx}-w_{x})\delta_{x} ~~\\ ~~\nonumber
&-& 9\mathcal{H}^{2}(1+w_{x})(c^2_{sx}-w_{x})\frac{\theta_{x}}{k^2} ~~\\ ~~\nonumber
&-&\Sigma\big(\frac{h'}{2} + 3\mathcal{H} \delta_{x}\big),\nonumber\\
\theta'_{x} &=&-\mathcal{H}(1-3c^2_{sx})\theta_{x}+\frac{c^2_{sx}}{(1+w_{x})}k^2\delta_{x}.
\end{eqnarray}
It becomes apparent that the inhomogeneous solution, driven by $h$, serves as the attractor for these equations, and it is expressed as follows:
\begin{eqnarray}\label{at1}
\delta_{x}&=&\frac{h(1+w_{x}) (1+w_{x}-\Sigma)}{(12 w^{2}_{x}-2w_{x}-3\Sigma -3\Sigma w_{x}-14  )},\\ 
\frac{\theta_{x}}{k} &=& \frac{ h (k\tau) (1+w_{x}-\Sigma)}{(12 w^{2}_{x}-2w_{x}-3\Sigma -3\Sigma w_{x}-14  )},\label{at2}
\end{eqnarray}
where we chose $c^2_{sx}=1$ for simplicity. A similar attractor was reported in \cite{gavela}. As we might expect, Eq. (\ref{at1}) and Eq. (\ref{at2}) tell us that the dark energy velocity perturbation is subdominant in relation to the contrast of the dark energy density, $\theta_{x}/k = \mathcal{O}(k/\mathcal{H})\delta_{x}$, in the superhorizon limit within the radiation era. For $\Sigma=0$ and $c^2_{sx}=1$, we recover the standard initial conditions for dark energy as reported in \cite{balle:2010}.

\section{Numerical Exploratory Analysis}\label{sec:ea}

\subsection{Parameter selection}\label{subsec:ps}
In this section, we will perform an exploratory analysis of the frozen model, juxtaposing it with the vanilla model and the Chevallier-Polarski-Linder (CPL) parameterization \cite{Chevallier:2000qy, Linder:2002et} with $w_0 \equiv w_x$ and $w_a = 0$, i.e. the so-called $w\mathrm{CDM}$ scenario. It is important to emphasize that for a fixed $w_{x}$, the introduction of this model merely adds an additional parameter, $\Sigma$, which characterizes the strength of the interaction (or coupling) between dark matter and dark energy perturbations~\footnote{From this point forward, we will streamline our terminology by omitting the term 'perturbations' and referring solely to a dark matter-dark energy interaction. However, it is crucial to bear in mind that the energy transfer occurs between the perturbations of dark matter and those of dark energy.}. Thus, this represents a minimal extension of the concordance model and can be interpreted as an effective theory at the first-order perturbation level \cite{Gubitosi:2012hu}.

\begin{figure}
    \centering
    \includegraphics[width=0.75\linewidth]{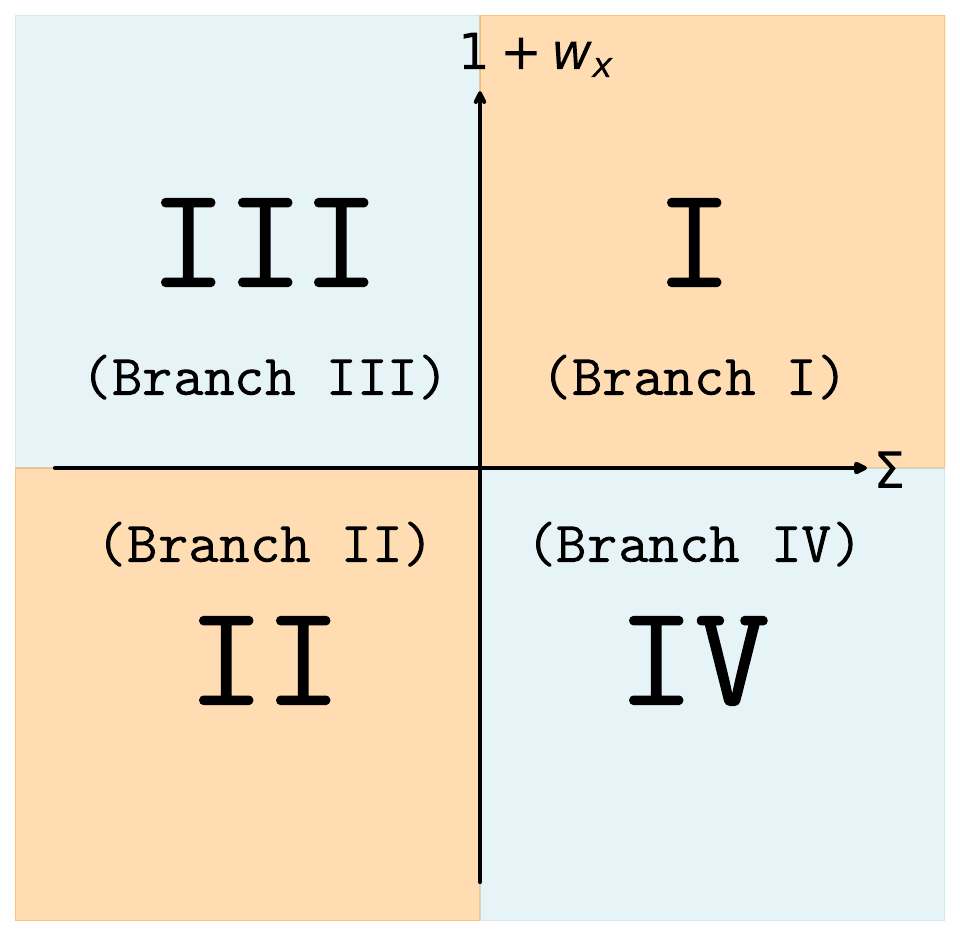}
    \caption{The classification of models based on their position in the parameter space $(1 + w_x) \times \Sigma$ reveals distinct behavior among different branches. Branches I and III are associated with non-phantom models, whereas branches II and IV correspond to phantom models. In light-blue, branches I and II represent models that $(1 + w_x)$ and $\Sigma$ have the opposite sign, while in light-orange, we present branches III and IV, in which $(1 + w_x)$ and $\Sigma$ have the same sign.}
    \label{fig:models}
\end{figure}

The effects of the \texttt{frozen (vanilla) model} are examined through the resolution of its Boltzmann equations, utilizing our modified \texttt{CLASS} \cite{ju1, ju2, ju3} code. In our analysis, we consider a diverse range of models [cf. Fig. \ref{fig:models} and Table \ref{tab:models}]: half of them are characterized by a nonphantom fluid with an equation of state, $w_x > -1$, while the other half exhibits phantom behavior with $w_x < -1$. Specifically, we set the equation of state parameter to $w_{x}=\{-0.9, -0.8\}$ for the non-phantom scenario and $w_{x}=\{-1.1, -1.06\}$ for the phantom case, \cite{olga, mateo2, adaf, Asghari:2019qld}.  The chosen values for the $w_{x}$ parameter in both phantom and non-phantom scenarios are in strong agreement with the statistical analysis performed using the combined Planck 2018+BAO+Pantheon+ observational data. This analysis encompasses a range of interacting dark energy models as detailed in \cite{wxs}. Such an alignment reinforces the viability of these models in light of current cosmological observations. Throughout all models, we maintain a fixed sound speed, $c_{sx}^2 = 1$. This approach enables us to comprehensively investigate the dynamical implications of the frozen model under these distinct regimes.

It is established that interacting dark matter-dark energy models could suffer from a well-known instability \cite{Valiviita:2008iv} for some interaction coupling, which in turn constrains the coupling parameter $\Sigma$ to possess the opposite sign of $(1+w_{x})$. In such scenarios, phantom models produce a positive coupling constant ($\Sigma>0$), while nonphantom models result in a negative coupling constant ($\Sigma<0$). However, in our analysis, we notice that such an instability is never present. Consequently, they are the focus of this work and, for each set of models, both in the phantom and non-phantom regimes, the parameter $\Sigma$ spans a range from negative to positive values, specifically $\{\pm10^{-3}, \pm10^{-2}, \pm10^{-1} \}$. For models with very low coupling, we anticipate that the frozen models will closely resemble the corresponding CPL model with constant $w_{x}$. In Fig. \ref{fig:models} and Table \ref{tab:models}, we show in blue the branches whose values $(1+w_{x})$ and $\Sigma$ have opposite signs, while the branches whose quantities present the same sign are displayed in orange. 

In addition, we adopt the best-fit parameters for $\mathrm{\Lambda CDM}$ derived from the Planck 2018\cite{planck2020b} data (68 \% TT,TE,EE+lowE+lensing): $h = 0.6736$, $\tau_{\rm{reio}} = 0.0544$, $\Omega_{\rm{b}} h^2  = 0.02237$, $\Omega_{\rm{cdm}}  h^2= 0.1200$, $A_{\rm{s}}  = 2.1 \times 10^{-9}$, $n_{\rm{s}}  = 0.9649$.  We incorporate a small tensor-to-scalar ratio of $r_{0.002} = 0.055$ (95 \%, TT+lowE+lensing+BK15+BAO) to assess the implications of the power spectrum associated with the B modes \cite{planck2020b} in the context of our model. 

As will be discussed later, we separate our analysis into large, intermediate, and small scales. In units of $h/\rm{Mpc}$, large scales are defined as $k < 10^{-3}$, intermediate scales as $10^{-3} < k < 10^{-1}$, and small scales as $k > 10^{-1}$. When plotting the transfer functions and the interactions, we select three modes corresponding to each of these regimes.

\subsection{Dynamical behavior of the interaction coupling}\label{subsec:di}

\begin{table}
    \centering
    \begin{tabular}{|c|c|c|} \hline
        Branch & $1 + w_x$ & $\Sigma$ \\\hline
        I & $>0$ & $>0$ \\\hline
        II & $<0$ & $<0$ \\\hline
        III & $>0$ & $<0$ \\\hline
        IV & $<0$ & $>0$ \\\hline
    \end{tabular}
    \caption{The classification of different models based on the signs of $(1 + w_x)$ and $\Sigma$ establishes clear distinctions between the branches. Branches I and III, which represent non-phantom models, can be derived by reversing the interaction coupling. In a similar fashion, branches II and IV, associated with phantom models, are obtained by considering the transformation $\Sigma\rightarrow -\Sigma$.}
    \label{tab:models}
\end{table}

\begin{figure*}[!htbp]
    \centering
    \begin{minipage}[b]{0.475\textwidth}
        \centering
        \includegraphics[width=\textwidth]{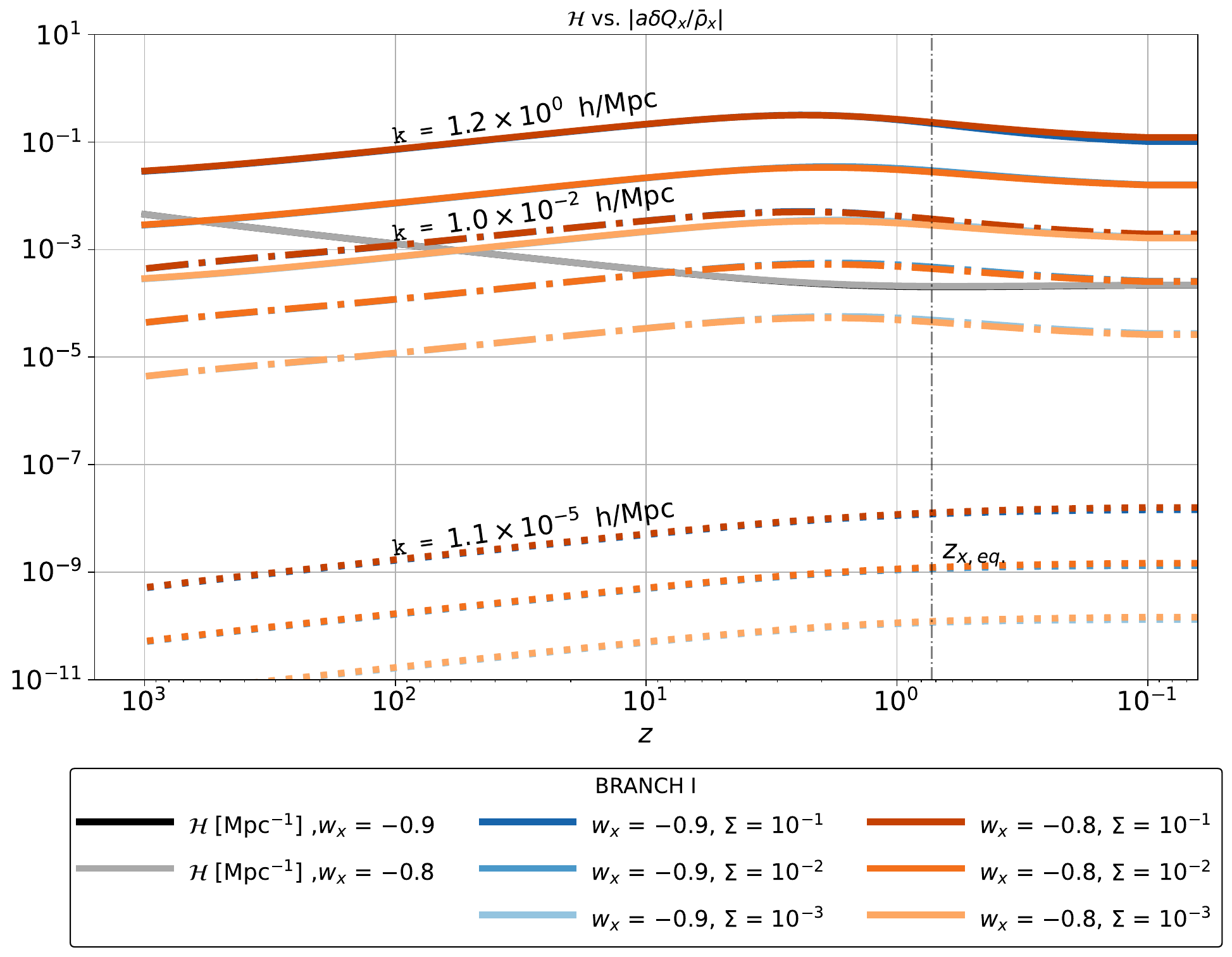}
    \end{minipage}
    \hfill
    \begin{minipage}[b]{0.475\textwidth}
        \centering
        \includegraphics[width=\textwidth]{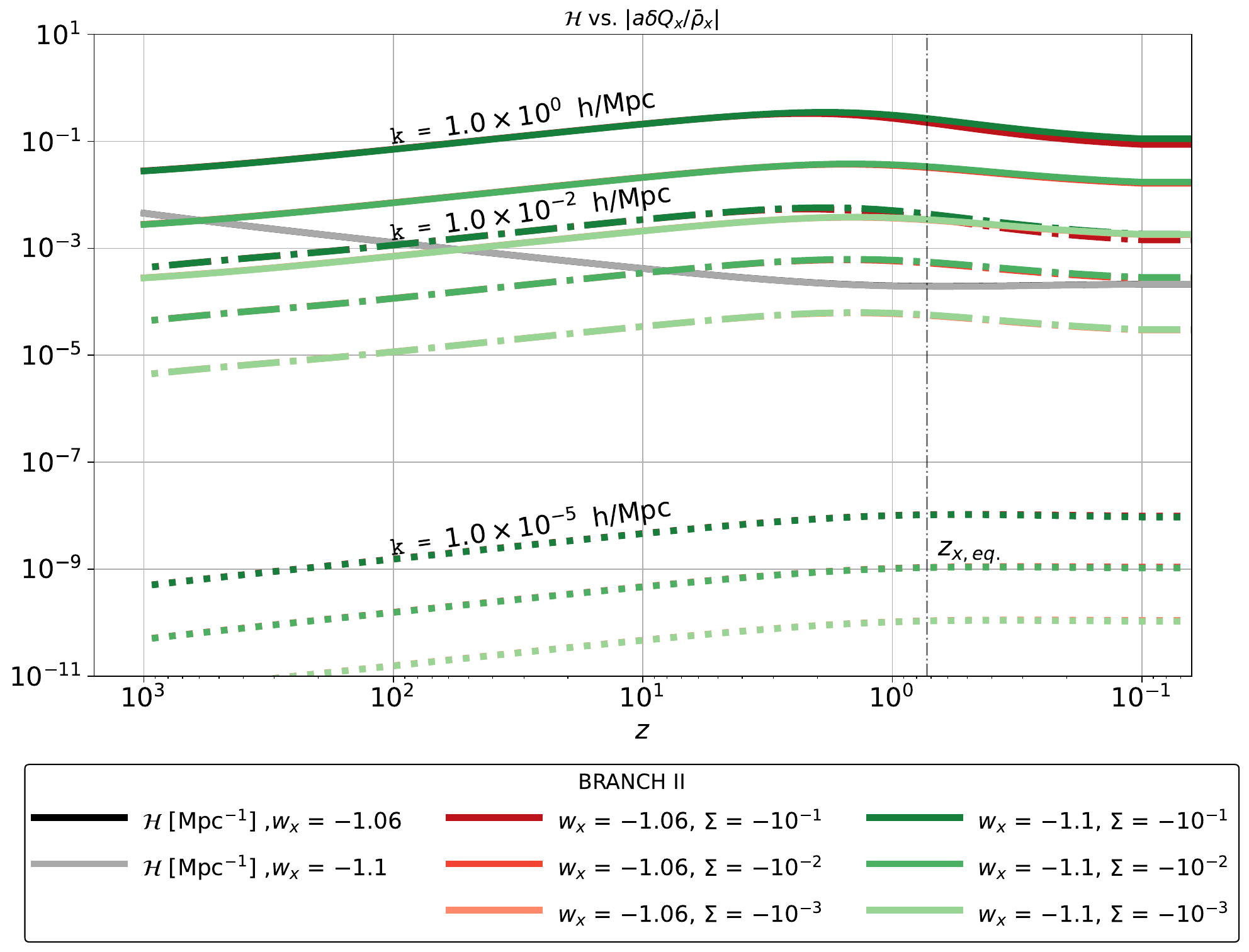}
    \end{minipage}

    \vspace{0.5cm}

    \begin{minipage}[b]{0.475\textwidth}
        \centering
        \includegraphics[width=\textwidth]{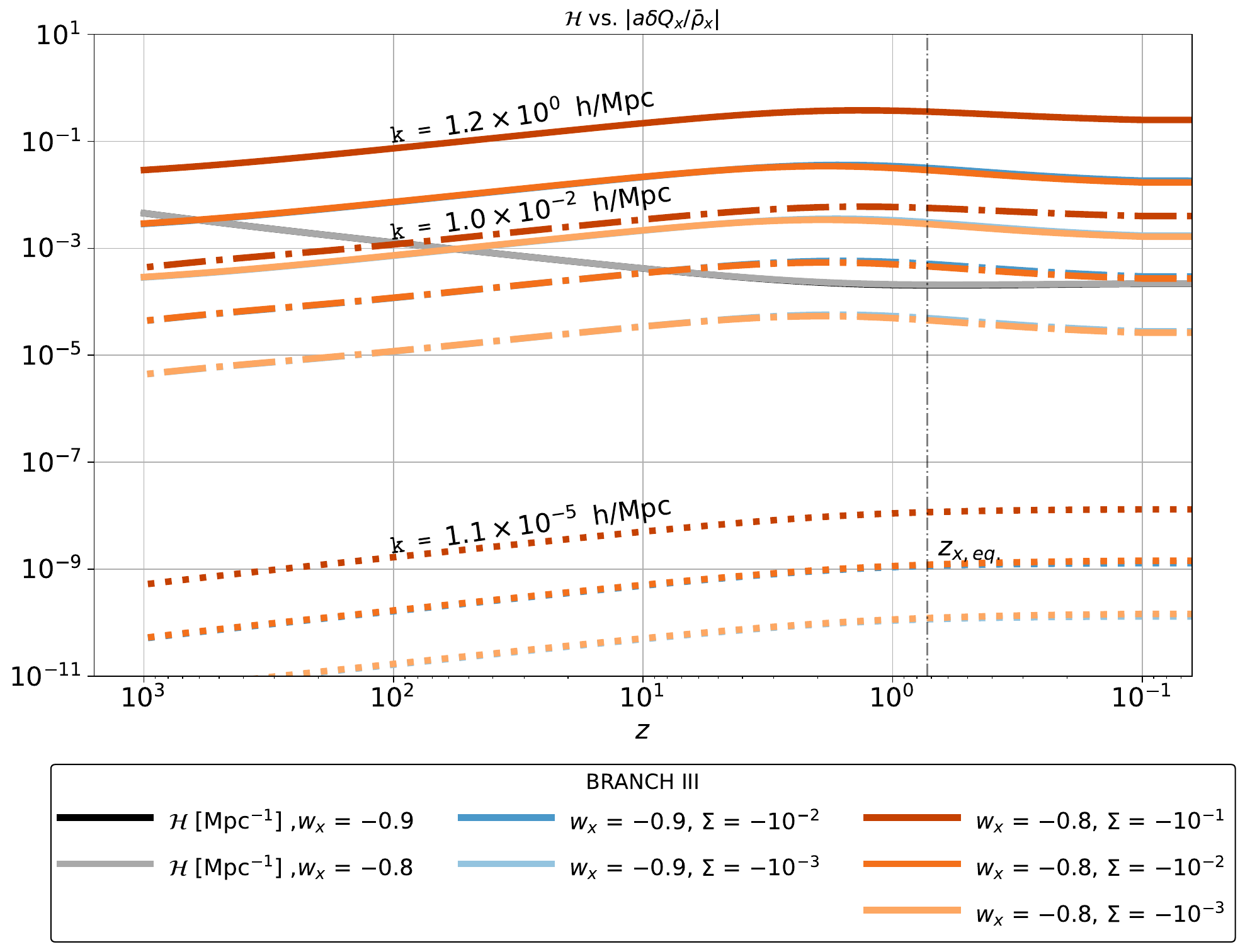}
    \end{minipage}
    \hfill
    \begin{minipage}[b]{0.475\textwidth}
        \centering
        \includegraphics[width=\textwidth]{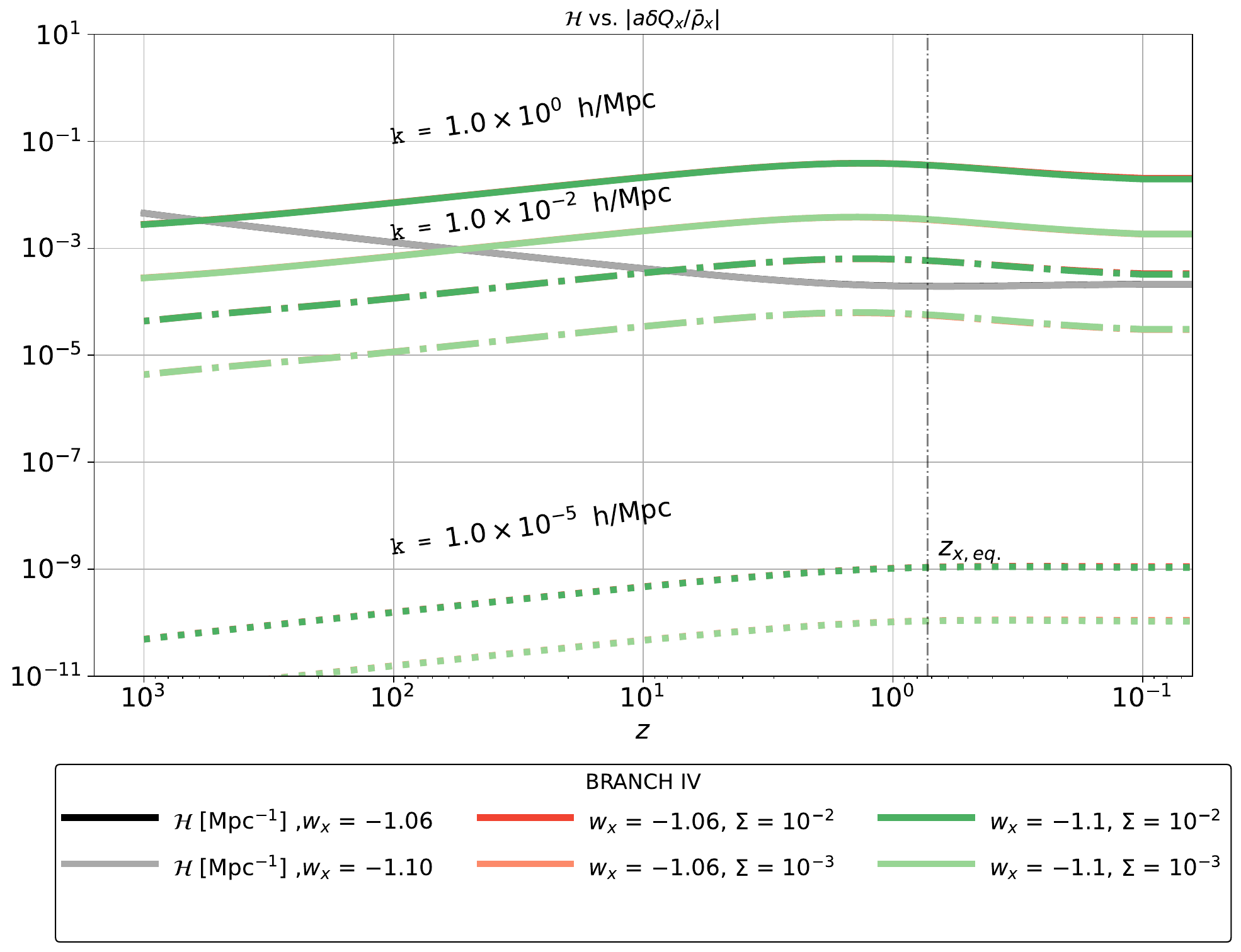}
    \end{minipage}
    \caption{We examine the cosmological evolution of the absolute value of the interaction strength within the dark energy sector of the frozen vanilla model, comparing it to the conformal Hubble function $\mathcal{H}$ across all models under investigation. The interaction exhibits a scale-dependent variation, showcasing larger values for small-scale modes. Notably, the interaction becomes significant for these small-scale modes much earlier, deep in the matter-dominated epoch, compared to intermediate scales, where it becomes efficient closer to dark energy domination. Large-scale modes, on the other hand, remain unaffected by the interaction. This behavior is consistent across all models analyzed. 
    Additionally, it is important to note that the interaction strength $\delta Q_x$ is negative when $\Sigma > 0$. We have numerically proven that 
    $h^{'} \gg \theta_x$, $h^{'} \gg \delta_x$ at early times, and $h^{'} > \delta_x$ at late times, which leads to $\delta Q_x$ to have the opposite sign of $\Sigma$.}
    \label{fig:int}
\end{figure*}

A crucial aspect of the  \texttt{frozen (vanilla)} model relates to the dynamical behavior of the interaction between dark matter and dark energy.  We need to evaluate the conditions under which this interaction is considered inefficient or efficient in relation to the conformal Hubble factor. The latter evaluation will offer important insights into the robustness of the minimal extension of the vanilla model from a physical standpoint.

Understanding these dynamics will be essential for elucidating the implications of our model within the broader cosmological context. To address this, we start by comparing the value of $a \delta Q_x / \bar{\rho}_x$ for the dark energy sector with the conformal Hubble function $\cal{H}$ [cf. Fig. \ref{fig:int}]. First, we observe that the interaction (\ref{IFV}) is scale-dependent since it involves $\delta_{x}$, $h'$, and $\theta_{T}$, which are inherently Fourier modes that depend on $k$-modes. This scale dependence implies that the nature of the interaction varies across different scales, influencing the behavior of the dark energy sector in a nuanced manner.  To be more precise, small-scale perturbations enter the horizon earlier, resulting in a more rapid growth compared to larger scales that reenter the horizon at a later time. Since the interaction is dependent on the gravitational potential and density contrast, this earlier entry leads to a more pronounced interaction for small-scale modes. As a result, we also find that in the intermediate regime, the strength of the interaction is greater than that observed for large scales. For small scales, the interaction becomes significant during the first stages of the matter-dominated era, as the density contrast amplifies sufficiently from the onset of matter domination onward. In contrast, intermediate scales, which enter the horizon at a later stage, only experience relevant interactions during the late-time dark matter domination, where the interaction is efficient in the sense that $|a\delta Q_{x}/\bar{\rho}_{x}|\gg \cal{H}$. Consequently, we expect deviations from the $\mathrm{\Lambda CDM}$ and CPL models to manifest primarily on small scales, particularly within the matter power spectrum.  We substantiate our previous findings by computing the interaction term for the dark matter component, which agrees with Fig. \ref{fig:int}.

We note that the interaction strength $\delta Q_x$ is negative when $\Sigma > 0$. This conclusion arises from our numerical findings, which demonstrate that $h^{'}$ is positive and serves as the dominant term in the $\mathcal{K}$ term in equation (36). Therefore, $\delta Q_x$ must have the opposite sign of $\Sigma$.

\subsection{BRANCH I/III} \label{subsec:i_iii}

\subsubsection{CMB Power Spectra}\label{subsec:ctt}

\begin{figure*}[!htbp]
    \centering
    \begin{minipage}[b]{0.475\textwidth}
        \centering
        \includegraphics[width=\textwidth]{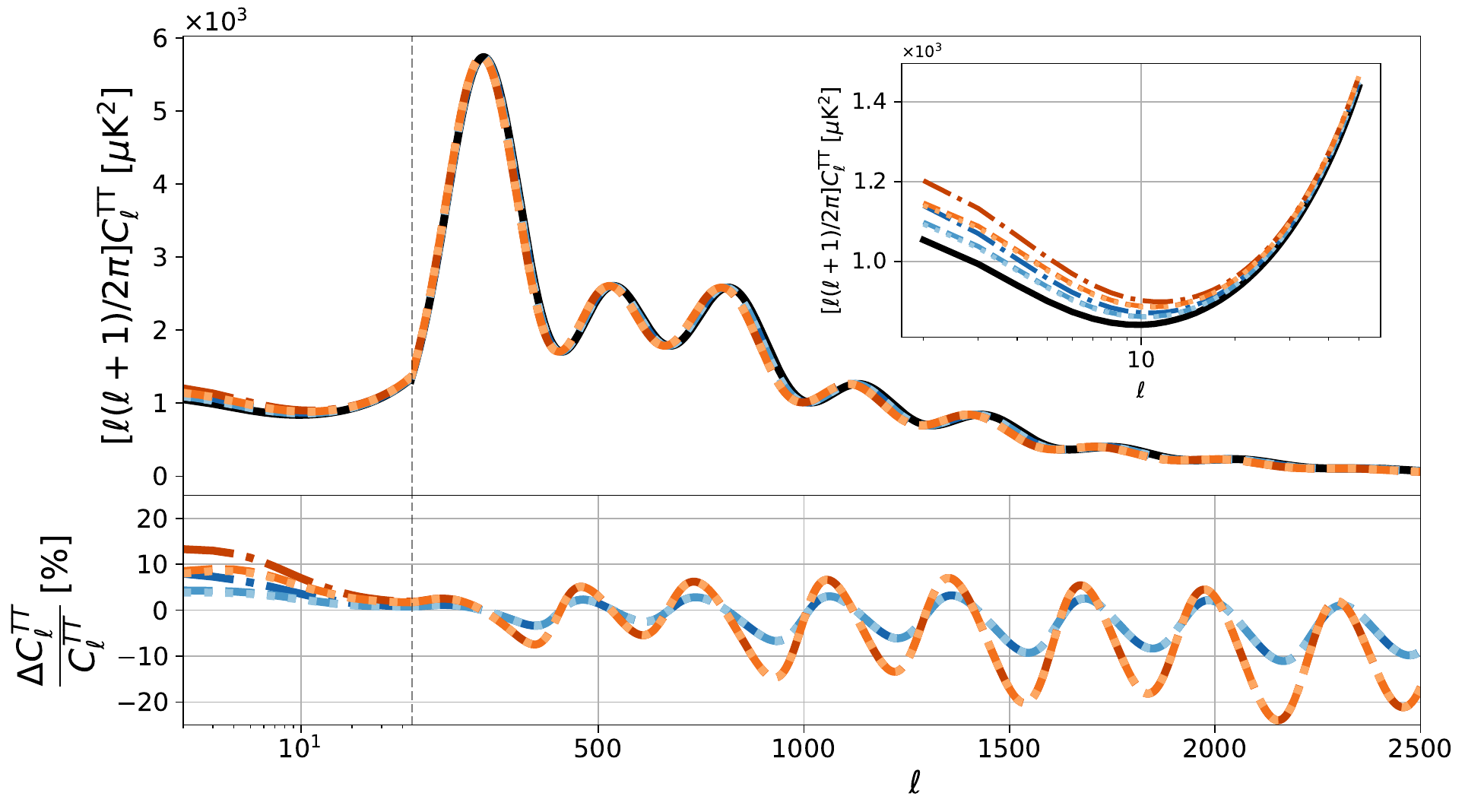}
    \end{minipage}
    \hfill
    \begin{minipage}[b]{0.475\textwidth}
        \centering
        \includegraphics[width=\textwidth]{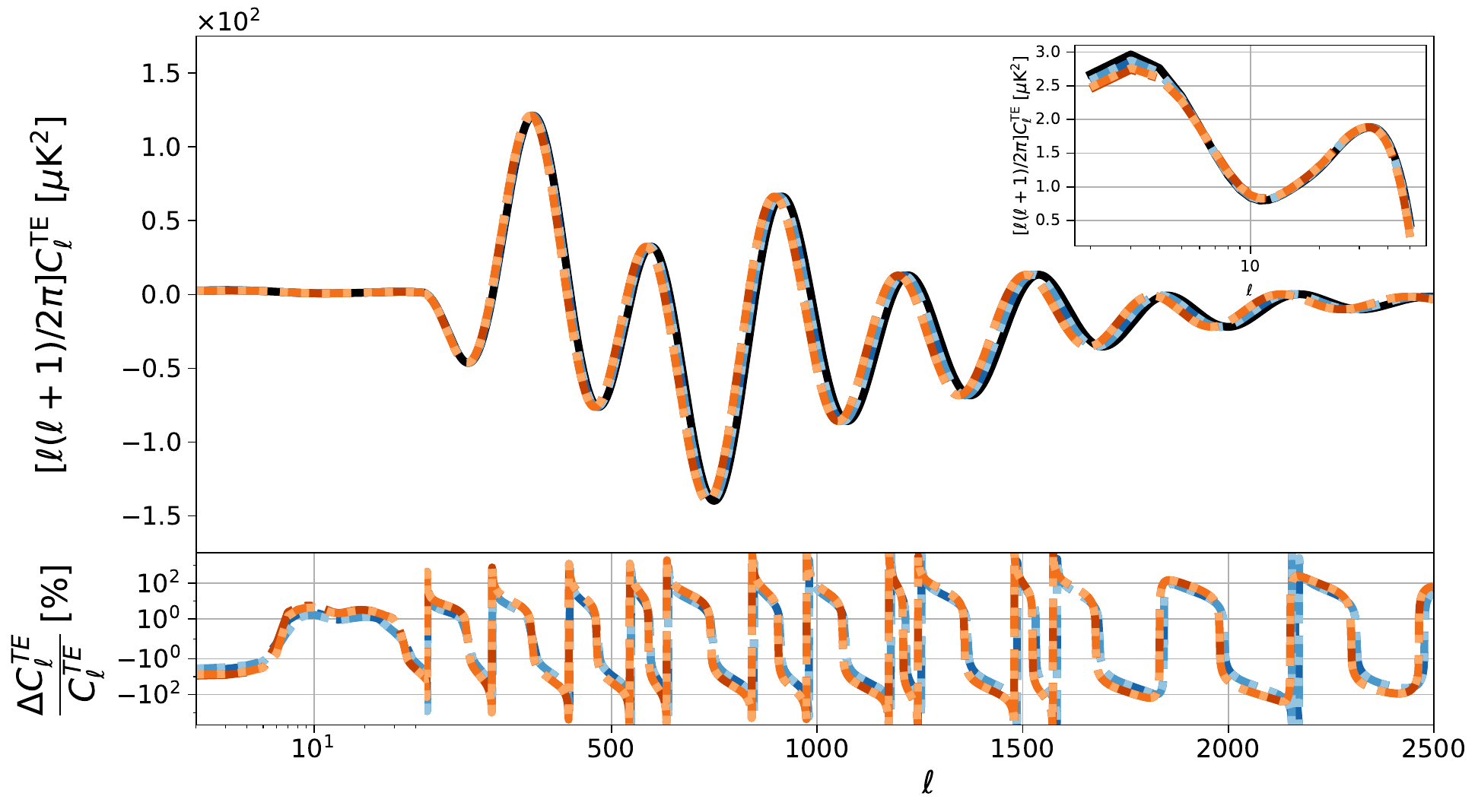}
    \end{minipage}

    \vspace{0.5cm}

    \begin{minipage}[b]{0.475\textwidth}
        \centering
        \includegraphics[width=\textwidth]{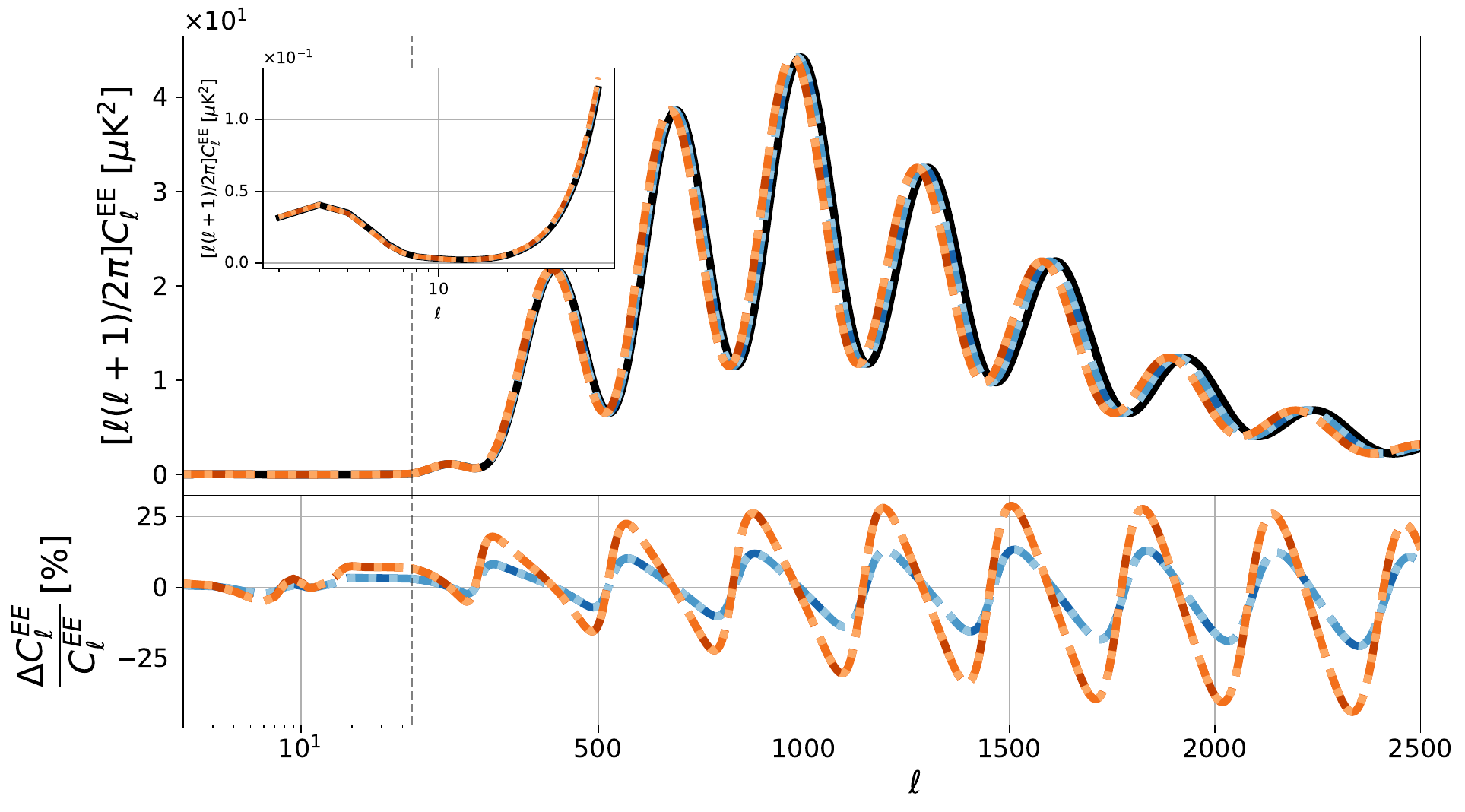}
    \end{minipage}
    \hfill
    \begin{minipage}[b]{0.475\textwidth}
        \centering
        \includegraphics[width=\textwidth]{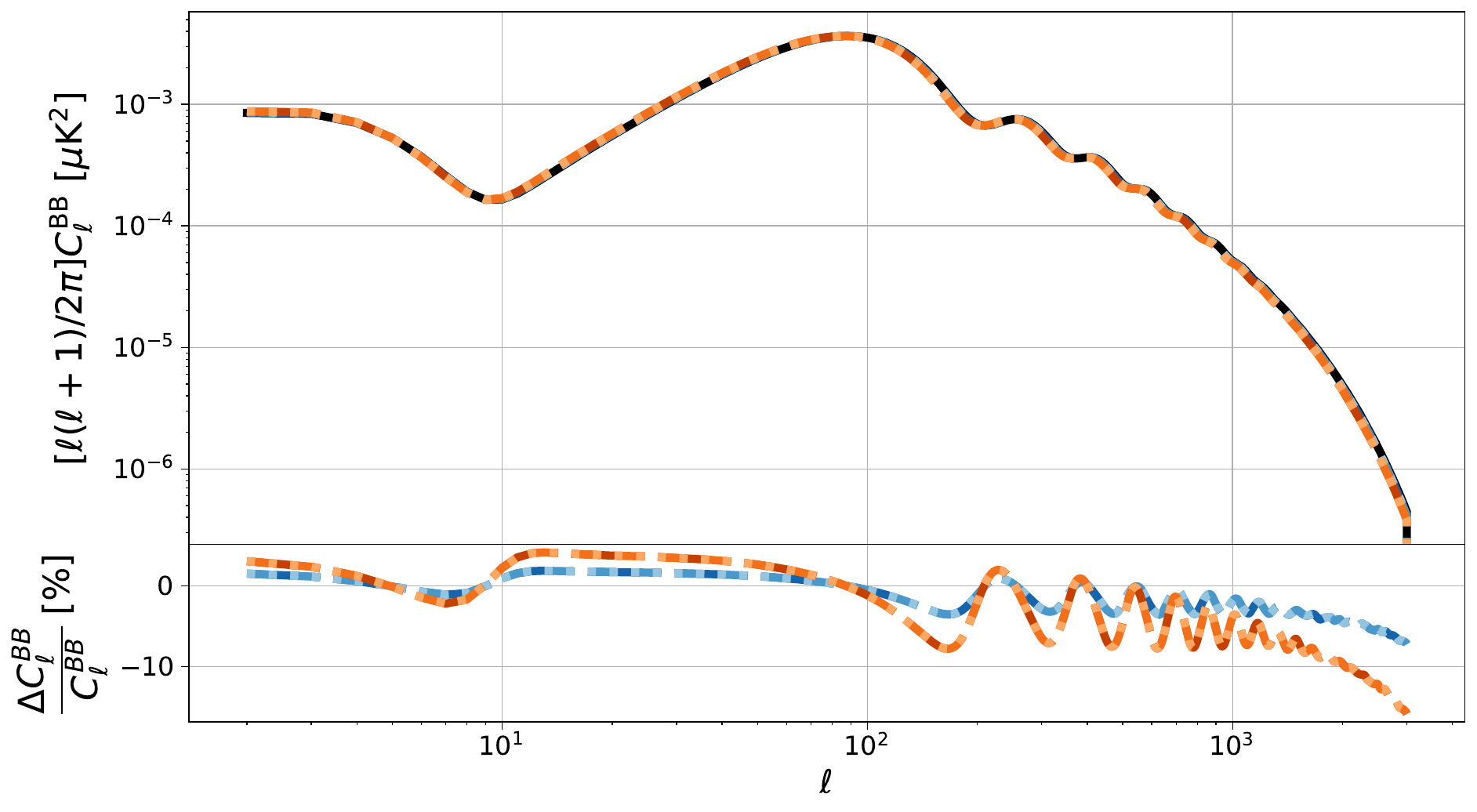}
    \end{minipage}

    \vspace{0.25cm}

    \begin{minipage}[b]{0.5\textwidth}
        \centering
        \includegraphics[width=\textwidth]{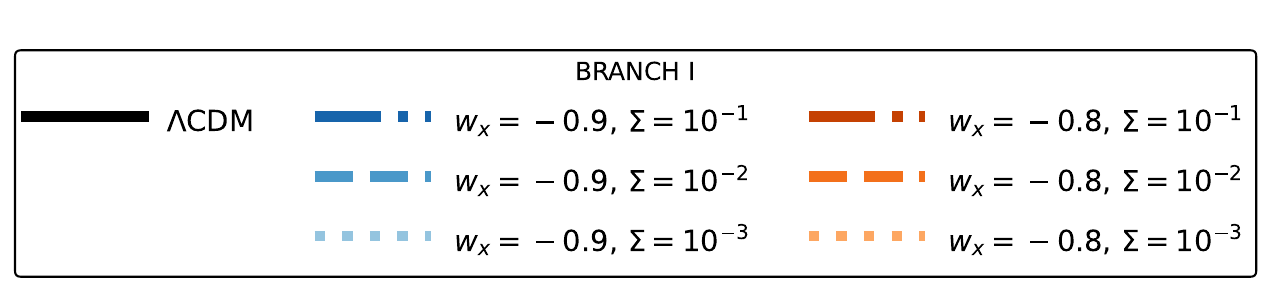}
    \end{minipage}
    
    \caption{
    The CMB angular power spectrum for the $TT, TE, EE$, and $BB$ two-point correlation functions for $\mathrm{\Lambda CDM}$ and the \texttt{Frozen-branch-I} models for a particular range of $\Sigma$ and $w_x$, as indicated. The difference from $\mathrm{\Lambda CDM}$ in each spectrum is illustrated in the lower panel of each plot. This percentage difference emphasizes the role played by interactions in the dark sector, and how such mechanism translates into new features. In the top-left corner plot, we highlight that the $C_{\ell}^{TT}$ presents a larger deviation from $\mathrm{\Lambda CDM}$ at large scales for models with positive coupling constant $\Sigma$.The main contribution of the coupling $\Sigma$ happens at low-$\ell$, as best shown in the $TT$ power spectrum, via the ISW effect, which is enhanced for positive $\Sigma$ in branch I.}
    \label{fig:cl_class_I}
\end{figure*}

We now shift our focus to exploring the observational signatures linked to the aforementioned interaction and how this interaction manifests itself in various physical observables. In this section, we focus on a set of non-phantom models, specifically with a fixed equation of state parameter $w_x =\{-0.9, -0.8 \}$. The results are presented in the context of the CMB power spectrum, where we examine the correlation/cross-correlation functions $TT$, $EE$, $TE$, and $BB$. The corresponding power spectra, shown as a function of the multipole, $\ell$, while varying the coupling $\Sigma$, are illustrated in Fig. \ref{fig:cl_class_I}. 

The coupling has no effect on the $TT$-correlation function at $\ell > 40$, neither for the other correlation functions $EE$, $TE$, and $BB$~\footnote{
A shift in the position of the peaks is evident, as illustrated in Fig. \ref{fig:cl_class_I}. This behavior is characteristic of a dark energy fluid adhering to the CPL prescription, and we have verified this phenomenon for $\Sigma = 0$  independently. In fact, one can observe this behavior by noting that the positions of the peaks remain unaffected by variations in $\Sigma$ as well.}. The only regime where the interaction is notable is for the $TT$ correlation function at low-$\ell$, which is investigated in the next section. 

For both $TT$ and $EE$, we note that the position of the peaks is shifted, which makes the difference to $\mathrm{\Lambda CDM}$ reach around $20 \%$. This value is a lot higher than the Planck sensitivity. However, such a difference between $\mathrm{\Lambda CDM}$ and the models depicted in Fig. \ref{fig:cl_class_I} arises solely from the difference in $w_x$, independently of the value taken by $\Sigma$ and the particular choice of $\Omega_{\mathrm{cdm}} = 0.1200$. In other words, this effect appears due to the fluid nature of dark energy, following the CPL parametrization with constant $w_x \neq -1$, and the chosen dark matter density. The greater the gap between $w_x$ and the phantom divide, the larger the difference between the CMB power spectra compared to $\mathrm{\Lambda CDM}$ results. For the $TE$ spectrum, the relative difference seems to indicate divergences, see the bottom panel of the top-right plot in \ref{fig:cl_class_I}. However, that is just a feature of the fact that the $TE$ spectrum changes sign at its acoustic oscillation scales. It is not a physical feature.

\subsubsection{Late ISW signature}\label{subsec:isw}

\begin{figure}[h]
    \centering
    \includegraphics[width=0.95\linewidth]{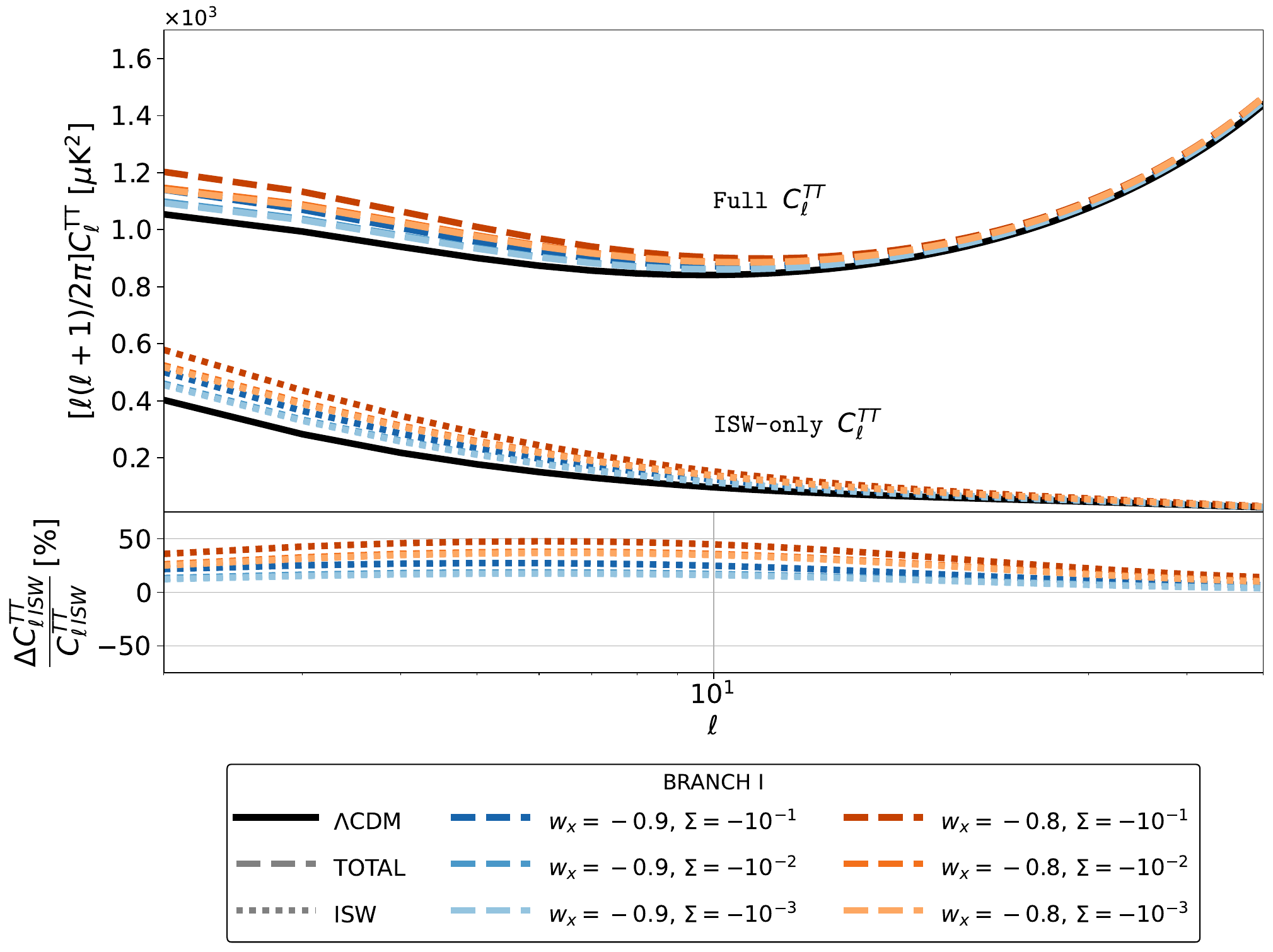}\\
    [1cm]
    \includegraphics[width=0.95\linewidth]{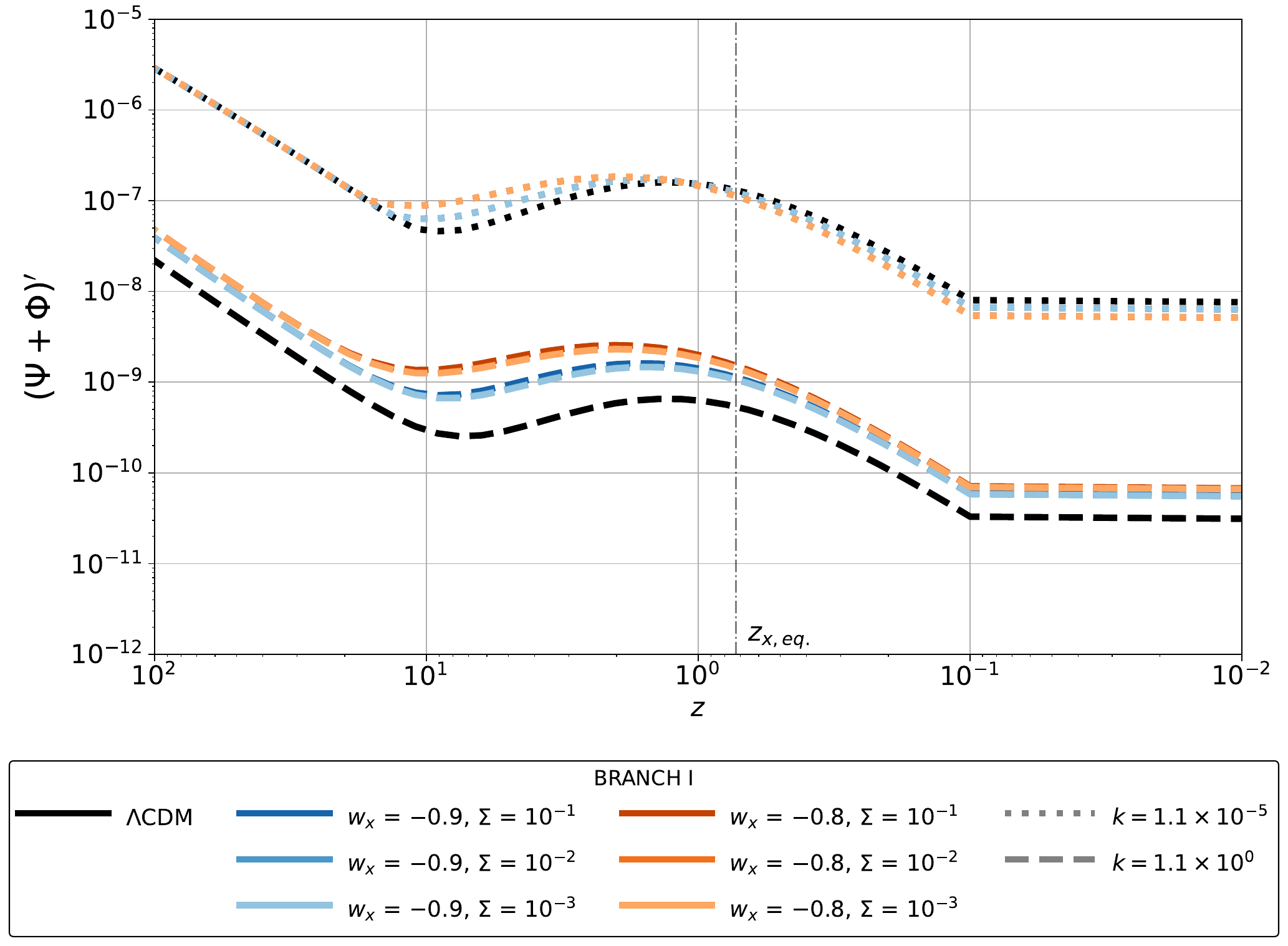}
    \caption{\textbf{Top panel:} In the top, the CMB TT power spectrum and the contribution from the late-time integrated Sachs-Wolfe effect for low $\mathit{l}$. The lower panel of the plot details the percentage difference between the ISW effect between $\mathrm{\Lambda CDM}$ and \texttt{Frozen-branch-I}. \textbf{Bottom panel:} The derivative of the gravitational potential $(\Psi + \Phi)$. For branch I, the stronger the coupling $\Sigma$, the larger the effect.}
    \label{fig:LISW_class_I}
\end{figure}

Another promising observational signature to consider in our exploration of effective field theories called \texttt{frozen} is the amplitude of secondary effects within the CMB temperature-temperature  power spectrum. Certain phenomena occur between the last scattering surface and the present day, which could be enhanced by the universe's expansion. In particular, the integrated Sachs-Wolfe effect (ISW) arises due to the differential redshift experienced by photons as they traverse time-evolving potential perturbations from the last scattering surface to the present day \cite{isw1}. This is a valuable indicator of new physics or appealing mechanisms, primarily due to its dependence on variations in the gravitational potential during recent epochs when the dark energy component began to dominate \cite{isw2}.  The shift in the CMB temperature is given by \cite{isw2}:
\begin{eqnarray}\label{eqisw1}
\frac{\Delta T_{\texttt{ISW}}}{T_{\rm{CMB}}}=\frac{2}{c^2}\int^{\tau_0}_{\tau_{lss}}d\tau \dot{\Psi}(\textbf{x}(\tau), \tau)
\end{eqnarray}
where the integral is along the photon path from LSS to today , $\Psi$ is the time-varying Newtonian potential, while  $\textbf{x}=\hat{\textbf{n}}\chi$ stands for the comoving coordinates [$\chi=c(\tau_{0}-\tau_{lss})$], and  $T_{CMB}$ is the mean CMB temperature.  In the framework of linear theory, the auto-correlation of the ISW effect represented in the $C_{\ell}$ power spectrum requires the knowlegde the total matter power spectrum at the present time $P(k)$ \cite{isw3, isw4}:
\begin{eqnarray}\label{eqisw2}
C^{ISW}_{\ell}=\frac{2}{\pi} \int{k^{2}dk P_{m}(k) |W^{ISW}_{\ell}(k)|^{2}},
\end{eqnarray}
where the window function is given by 
\begin{eqnarray}\label{eqisw2}
W_{\ell}=\frac{T_{CMB}}{a} 3H^{2}_{0}\Omega_{m}\int{d\tau \dot{\big(D/a\big)} k^{-2} j_{\ell}[\chi(\tau)k]}.
\end{eqnarray}
Here, $D(a)$ refers to the linear growth factor, $j_{\ell}$ is the Bessel function of the first kind. If the Limber approximation holds for a small angle ($\chi k_{L}=\ell+1/2$) and slow-varying terms inside the integral (\ref{eqisw2}) can be written as 

\begin{eqnarray}\label{eqisw3}
C^{ISW}_{\ell}=\frac{T^{2}_{CMB}}{a^2} [3H^{2}_{0}\Omega_{m}]^{2} \int{d\tau \frac{\chi^{2}}{c} [\dot{(D/a)}]^{2} P_{m}(k_{L}}).~~
\end{eqnarray}
Eq. (\ref{eqisw3}) is applicable in the regime of low redshift and low multipoles. It indicates that the ISW effect is influenced by the total matter content, the growth function, and consequently by the matter power spectrum. This part of the signal directly influences the gravitational potential experienced by photons, leading to a characteristic signature in the ISW correlations \cite{Weinberg:2008zzc}. In this context, we find that the ISW effect is enhanced when $\delta Q_{x}<0$. Within our model branch-I, the late ISW effect is sensitive to both the interaction coupling ($\Sigma$) and the equation of state, $w_{x}$. Moreover, we observe that  increasing $|\Sigma|$ can amplify  the late-time ISW effect by more than $10\%$ [see Fig. \ref{fig:LISW_class_I}].  In contrast, the effect is reversed for branch III provided $\Sigma$ has a different sign. Hence, the ISW effect is suppressed as the coupling strength increases, which leads to a lower power spectrum compared to the $\mathrm{\Lambda CDM}$ model.  Since the late ISW effect is particularly significant on large scales and is influenced by cosmic variance, a promising approach would be to investigate the cross-correlation function between the ISW temperature fluctuations and the distribution of galaxies or quasars. This analysis could provide a robust confirmation of the observational signatures proposed by our model, specifically through the correlation $C^{\rm{gal}\times ISW}_{\ell}$ \cite{sabino}.

Anomalies detected in the low multipole region of the CMB ($\ell<20$) may indeed have indirect connections to the ISW effect, particularly if they point to unexpected large-scale structures or dynamic processes that shape the gravitational potentials impacting the CMB at late times. This prompts us to consider how the resulting tension could be rigorously tested through cross-correlation analyses between lensing CMB data and large-scale structure tracers, such as galaxy surveys, under both the vanilla and frozen models \cite{planck2016isw}. Moreover, we can further evaluate our model by examining the ISW effect in the vicinity of the Eridanus supervoid—an extensive underdense region within the cosmic web. This analysis is relevant, as the ISW might explain a fraction of the temperature decrement observed in the cold spot \cite{planck2016isw}.

A range of interacting dark matter–dark energy models has investigated the cosmological signatures arising from interaction terms at both the background and perturbation levels \cite{gavela, Yang:2018euj, sunny, DiValentino:2019ffd, olga}. For instance, models characterized by a stable interaction proportional to dark matter \cite{Yang:2018euj}, a positive coupling constant, and a phantom-like dark energy equation of state leads to a reduction of the ISW effect, while a interaction that is proportional to the total dark sector energy density leads to enhancement of the effect. Our model is not included in any of the dark sector interaction works above.

\subsubsection{Matter power spectrum}\label{subsec:pdek}

We compute the non-linear power spectrum at redshift $z=0$ incorporating the corrections at small scales using  the \texttt{HALOFIT} procedure, as detailed in \cite{Takahashi:2012em}. To be more specific,  it offers a way to compute $P_{\rm{nl}}(k)$ based on the linear power spectrum $P_{\rm{l}}(k)$, which is calibrated against $N-$body simulation to capture nonlinear growth of structure across a wide range of scales and is computationally less intensive than full $N-$body simulations \cite{Takahashi:2012em}. The halo model describes a collection of discrete gravitationally bound structures known as halos,  primarily consisting of dark matter. In doing so, the halo mass function uses the Press-Schechter formalism with the NFW model for the profile $\rho_{\rm{dm}}$. But more importantly, the correlation halo-halo is used to compute the non-linear halo clustering \cite{pnkl1, pnkl2}. 
\begin{figure}[h] 
    \centering
    \includegraphics[width=0.95\linewidth]{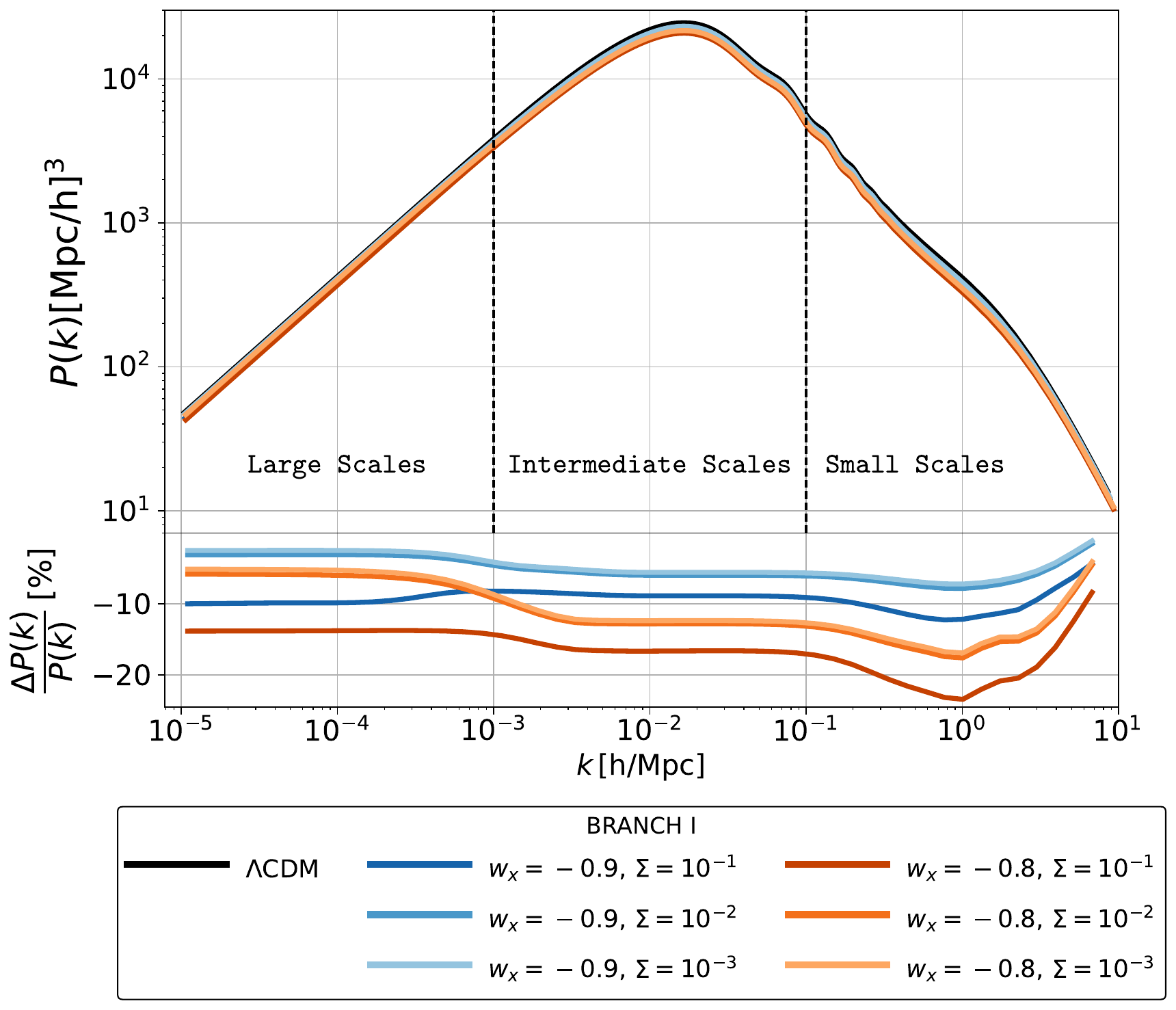}\\
    [1cm]
    \includegraphics[width=0.95\linewidth]{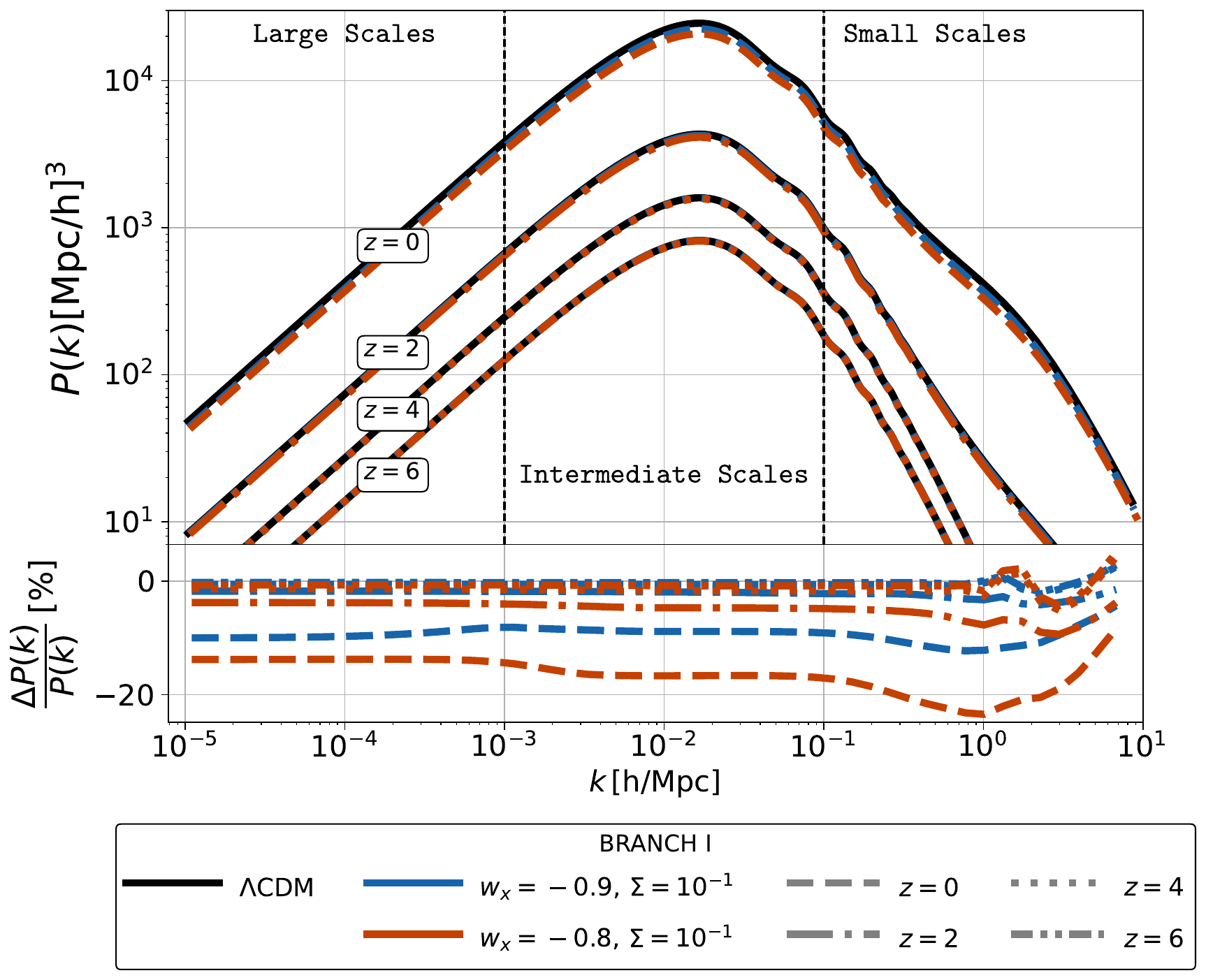}
    \caption{
    \textbf{Top panel}: The non-linear matter power spectrum at $z = 0$ for $\mathrm{\Lambda CDM}$ and the \texttt{Frozen-branch-I} models as a function of wavenumber $k$. In the lower panel, we present the $\%$ difference to $\mathrm{\Lambda CDM}$ in the non-linear power spectrum. Stronger coupling leads to stronger deviation from $\mathrm{\Lambda CDM}$. \textbf{Bottom panel}: Evolution of the non-linear power spectrum for four different redshift values (in different line styles), for the coupling $\Sigma = -0.1$ and $w_x = \{ -0.9, -0.8\}$. The difference to $\mathrm{\Lambda CDM}$ grows in time, reaching the maximum value at $z = 0$ for both $w_x$. That corroborates the result that the interaction is relevant only at late-times.
    }
    \label{fig:Pk_class_I}
\end{figure}
$N$-body simulations reveal that nonlinear effects significantly enhance the amplitude of the power spectrum compared to linear predictions in the range of scales $0.1<k({\rm{Mpc}}/h)<10$. This amplification arises from the transfer of power from larger scales through the interacting mechanism. To account for these non-linearities accurately is crucial for correctly detecting observational signatures on smaller scales. To address this, we use the \texttt{HALOFIT} routine within the \texttt{CLASS} package, enabling a more precise modeling of the power spectrum's behavior across various scales. The results are illustrated in Fig. \ref{fig:Pk_class_I}, showcasing the refined structure of the power spectrum. This highlights the significance of these corrections in accurately modeling the distribution of matter, particularly in the context of interactions within the dark sector.  In fact, the frozen branch I exhibits less power compared to the $\mathrm{\Lambda CDM}$ framework. This trend is particularly pronounced for the positive and stronger values of $\Sigma$, where the deviations from $\mathrm{\Lambda CDM}$ become even more significant (around 10 \%). As previously highlighted, this behavior arises as a natural consequence of the interactions occurring in the dark sector, altering the dynamics of structure formation and leading to a suppression of the matter power spectrum within these scenarios. The opposite effect occurs in branch III, in which stronger coupling leads to a larger matter power spectrum. However, this amplification is small, leading only to a $5\%$ increase in comparison to the vanilla model. The amplification of the matter power spectrum in branch III occurs for both dark energy equations of state analyzed.

In the bottom of Fig. \ref{fig:Pk_class_I} we show the evolution of the matter power spectrum for different values of the redshift $z$. The difference in relation to $\mathrm{\Lambda CDM}$ grows in time, as is shown in Fig. \ref{fig:int}: the power spectrum deviates only strongly from the results $\mathrm{\Lambda CDM}$ for late times, $z = 0$. For branch III the same follows for the enhancement of the power spectrum.

\subsubsection{Structure formation: $f \sigma_8$}\label{subsec:sforma}

The clustering properties of matter, which include both dark matter and baryons (luminous), can be effectively described using a variety of physical observables \cite{Weinberg:2008zzc}. One approach we have examined is the matter power spectrum, or the two-point correlation function of the random density field, which reveals its distinctive characteristics across small, intermediate, and large scales. Another method of measuring the mass power spectrum involves analyzing the peculiar velocities arising from significant fluctuations in the mass density field. It is important to note that the velocity field reflects the total mass distribution, which includes not just the luminous matter but also the contributions from dark matter \cite{ber1}, both quantities are linked through the continuity equation. When mapping the distribution of galaxies, it is essential to consider not only their recession velocity resulting from the expansion of the universe but also their peculiar velocities. This consideration leads to the conclusion that the apparent distribution of galaxies in redshift space differs from their actual distribution in real space \cite{ber1}.   
Distortions in galaxy distributions can be classified into two main types based on the nature of the peculiar motions involved. Virialized structures elongate along the line of sight in redshift space, rather than spherical (Fingers of God). In contrast,  peculiar velocities, resulting from in-fall toward significant cosmic structures, enhance the density contrast along the line of sight, which is called the Kaiser effect, which is associated with the linear growth of density fluctuations and can be quantified analytically. In fact, the density field in redshift space for galaxies can be expressed in terms of the total matter constraint density as follows \cite{kai1, kai2}:
\begin{eqnarray}\label{kaiser1}
\delta^{s}_{\texttt{g}}(k,\mu)=b\delta^{x}_{\texttt{m}}[1+\beta \mu^{2}],
\end{eqnarray}
where $\mu=\cos(\theta)$ denotes the angle between the line-of-sight in  real space $\textbf{x}$ and the vector $\textbf{k}$. Here, we assumed that galaxies are biased tracers of the underlying density field, namely $b=\delta_\texttt{g}/\delta_\texttt{m}$, so for $b$ constant the matter power spectrum in both frames are connected by $P_{\texttt{g}}(k, \mu)=b^{2}P_{\texttt{m}}$. The distortion factor (filtering) is $\beta=f_{\texttt{m}}/b$ \cite{kai1}, and the linear growth of the matter perturbation corresponds to $f_{\texttt{m}}(z,k)=(\delta'_{m}(z,k)/\delta_{m}(z,k))\mathcal{H}^{-1}$ \cite{kai3} \footnote{It is worth noting that, in principle, we could employ a gauge-invariant density contrast related to the density contrast in the synchronous gauge, expressed as, $\Delta^{\rm{GI}}_{m}=\delta^{\rm{sync}}_{m}+\frac{3\mathcal{H}\theta_{m}}{k^2}$.  However, under the subhorizon limit ($\mathcal{H}\ll k$), we find that $\Delta^{\rm{GI}}_{m}=\delta^{\rm{sync}}_{m}$.}.  The total density contrast is $\delta_{m}=f_{\rm{dm}}\delta_{\rm{dm}}+ f_{\rm{b}}\delta_{\rm{b}}$, where $f_{\rm{dm}}=\bar{\rho}_{\rm{dm}}/[\bar{\rho}_{\rm{dm}}+\bar{\rho}_{\rm{b}}]$ and $f_{\rm{b}}=\bar{\rho}_{\rm{b}}/[\bar{\rho}_{\rm{dm}}+\bar{\rho}_{\rm{b}}]$.  To obtain  $\delta_{\rm{m}}$ , we need to solve the system of coupled equations given by
(\ref{constradm})-(\ref{constrab}),  or equivalently,  we could solve a master equation for the total matter on sub-horizon scales ($k\gg \mathcal{H}$) in order to extract numerically $\delta'_{m}$ and $\delta_{m}$:
\begin{eqnarray}\label{coupledm}
\delta''_{m}+\mathcal{H}\delta'_{m}-[4\pi G a^{2} (\bar{\rho}_{\rm{dm}}+\bar{\rho}_{\rm{b}})]\delta_{m}=f_{\rm{dm}}\mathcal{S},
\end{eqnarray}
where the source term includes the different contributions from the interaction:
\begin{eqnarray}\label{coupledm2}
\mathcal{S}=
\mathcal{H} a\frac{\delta Q_{\text{dm}}}{\bar{\rho}_{\text{dm}}} + \Big(a\frac{\delta Q_{\text{dm}}}{\bar{\rho}_{\text{dm}}} \Big)'.
\end{eqnarray}
In the limit where the interacting mechanism becomes inefficient, $\mathcal{S} \rightarrow 0$, we recover the standard result of the vanilla model or the CPL model with a constant $w_{x}$,  within the synchronous gauge, employing conformal time as the time foliation.

The RSD data from surveys employs the smoothed  amplitude of the mass/density fluctuations  in a sphere of comoving
radius  $R=8h^{-1}\rm{Mpc}$\cite{kai3} defined as:
\begin{eqnarray}\label{kaiser3}    
\sigma_{8}(z)=\Bigg[\frac{1}{2\pi^2}\int^{\infty}_{0}k^{2}dk W^{2}_{8}P_{m}(k,z)\Bigg]^{1/2},
\end{eqnarray}
where $W_{8}(k)$ is the Fourier transform
of the spherical top-hat window function with radius $R=8 h^{-1}\rm{Mpc}$. The physical reason for adopting $f\sigma_{8}(z)$ over other possibilities is that the latter combination does not depend on the galaxy bias at the linear level \cite{kai3}. 

\begin{figure*}[!htbp]
    \begin{minipage}[b]{0.475\textwidth}
        \centering
        \includegraphics[width=\textwidth]{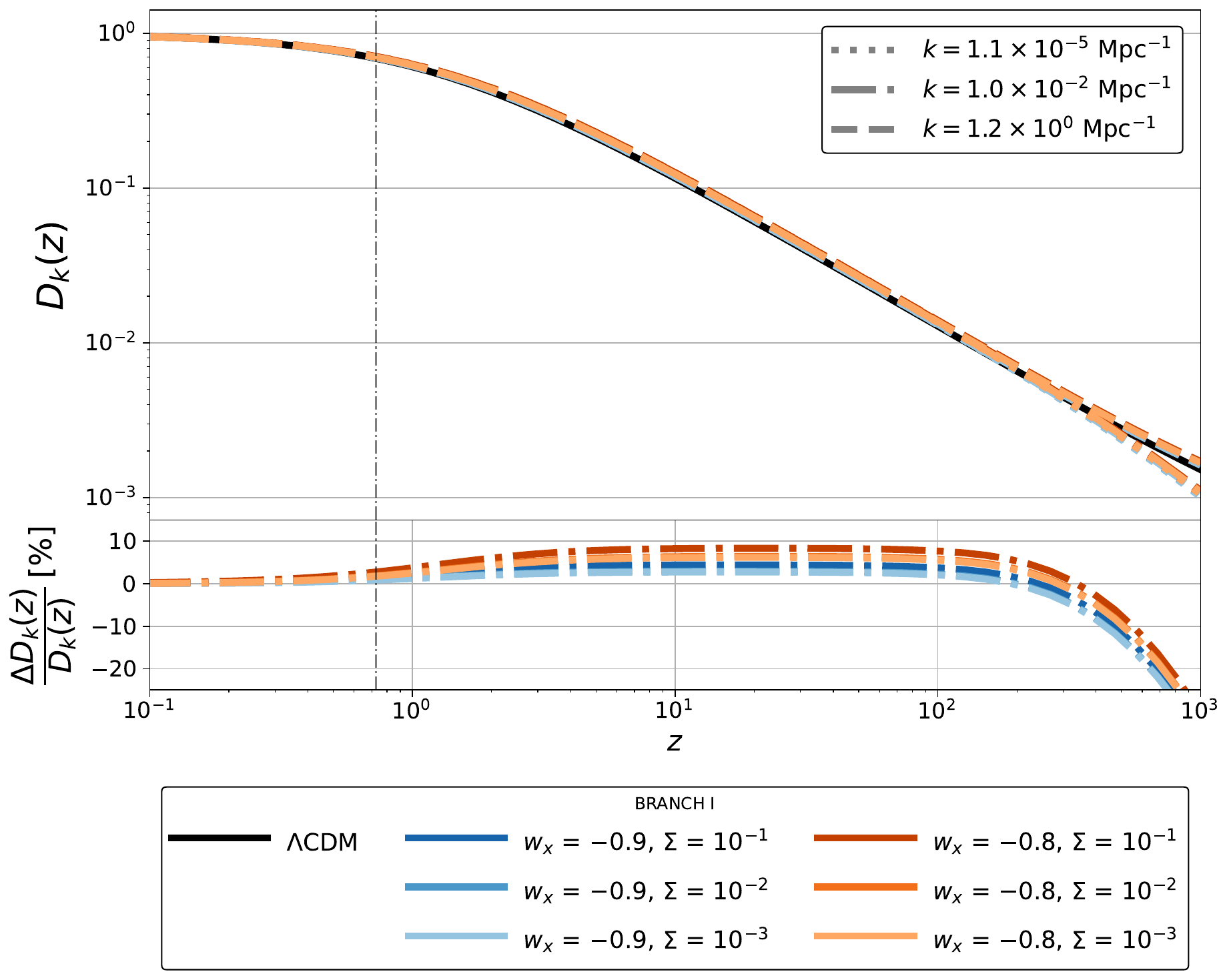}
    \end{minipage}
    \hfill
    \begin{minipage}[b]{0.475\textwidth}
        \centering
        \includegraphics[width=\textwidth]{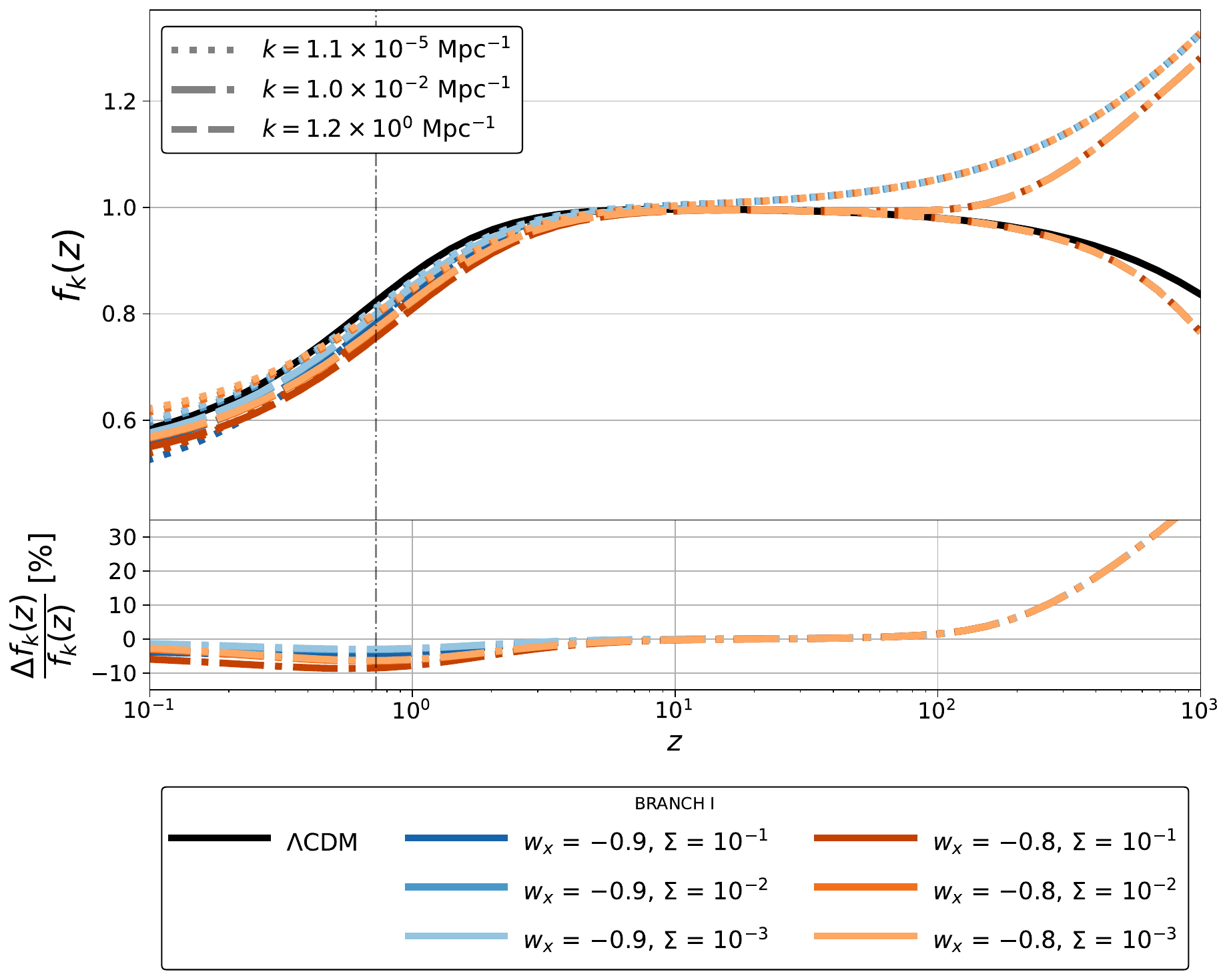}
    \end{minipage}

    \vspace{0.5cm}

    \begin{minipage}[b]{0.475\textwidth}
        \centering
        \includegraphics[width=\textwidth]{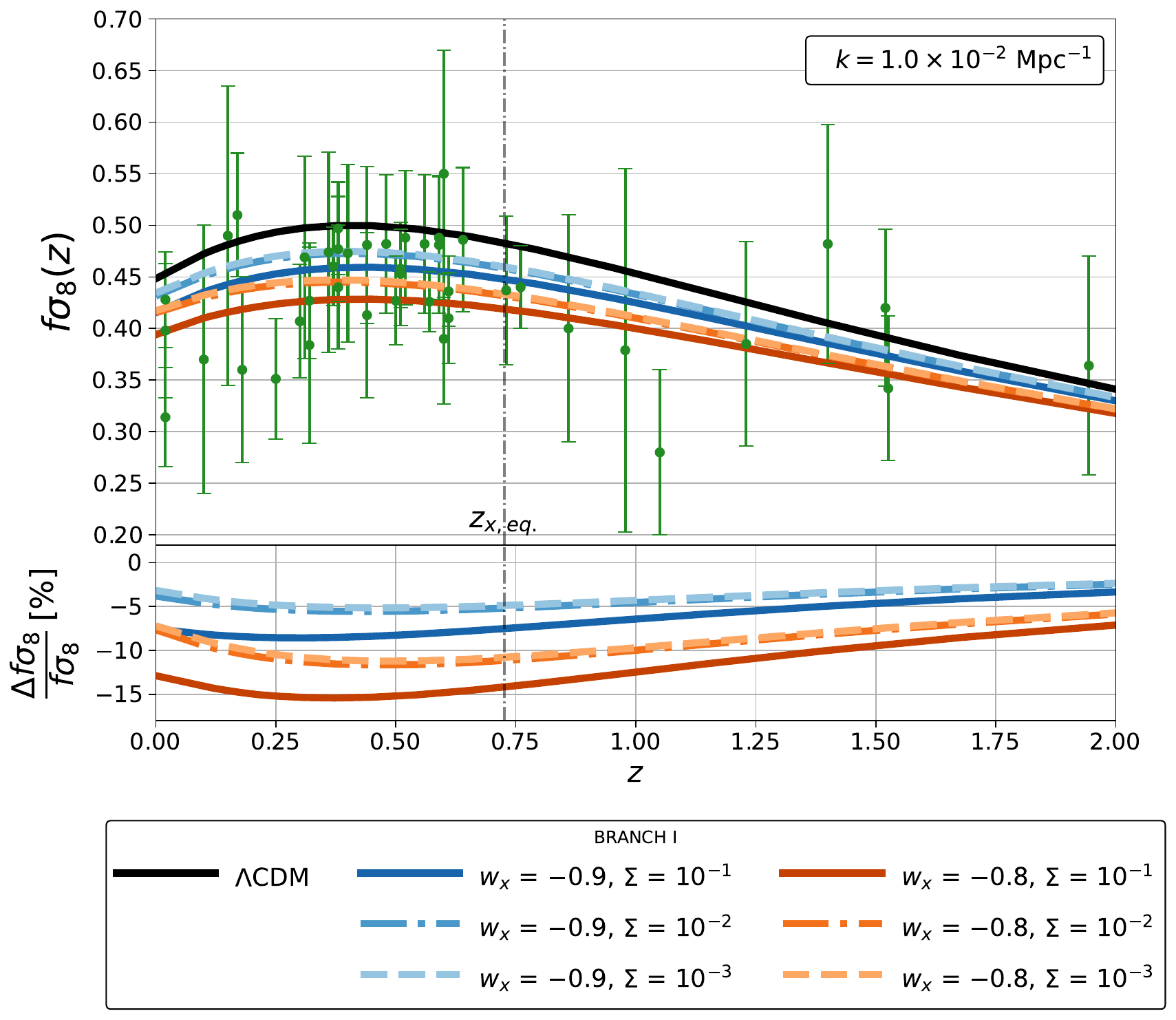}
    \end{minipage}
    \hfill
    \begin{minipage}[b]{0.475\textwidth}
        \centering
        \includegraphics[width=\textwidth]{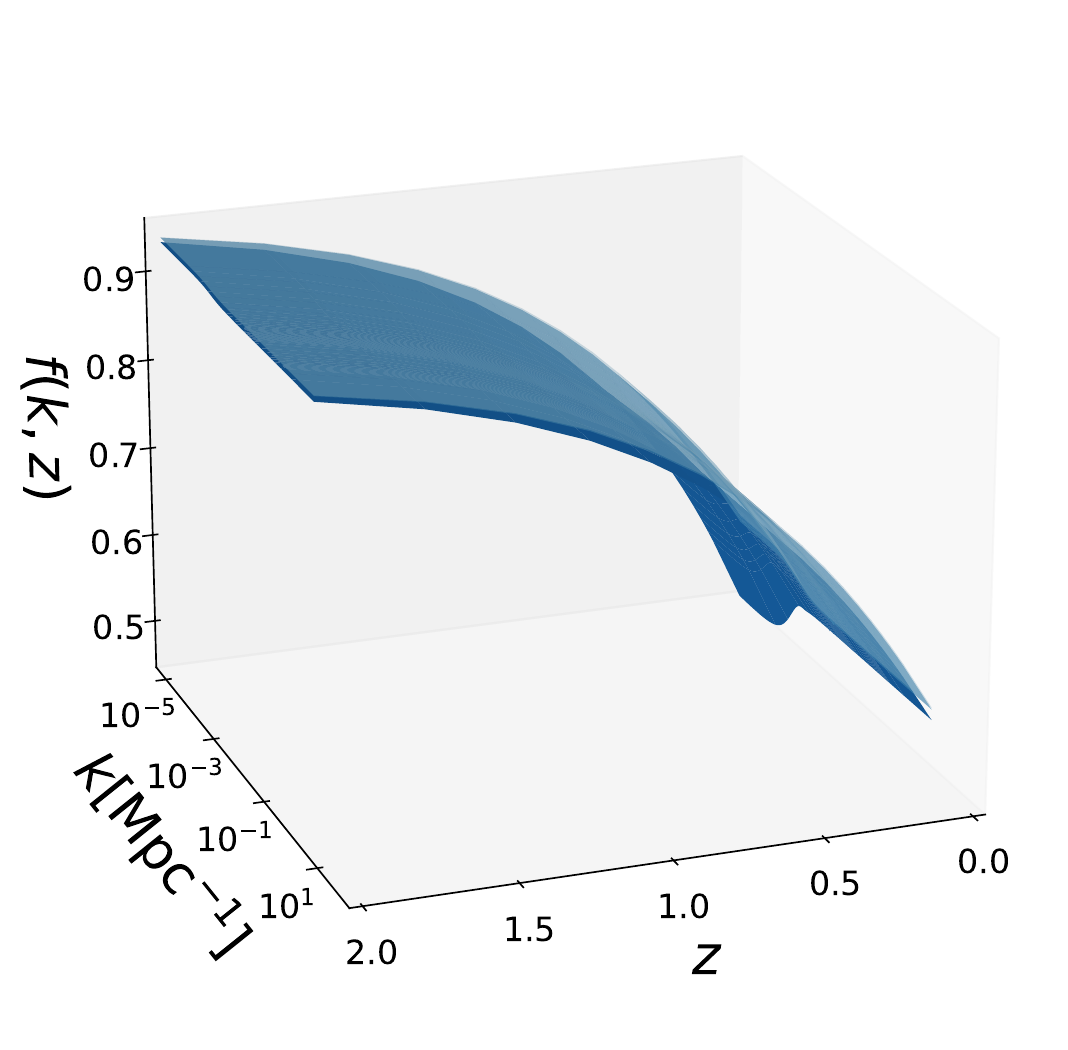}
    \end{minipage}

    \caption{\textbf{Top panel}: \textit{Left}: The growth factor $D_{k}(z)$ for models in branch I, for three different $k$ values, following the same prescription: large ($k \sim 10^{-5} \mathrm{Mpc}^{-1}$), intermediate ($k \sim 10^{-2} \mathrm{Mpc}^{-1}$) and small ($k \sim  \mathrm{Mpc}^{-1}$) scales. We also present the percentage difference to $\mathrm{\Lambda CDM}$ for intermediate scales, which is only relevant for very large $z$. We note an almost scale independence of $D_{k}(z)$, and therefore the difference to the reference model is consistent across scales. \textit{Right}: The growth factor $f_{k}(z)$ for models in branch I, for the chosen large, intermediate, and small scales. There is a large scale dependence for $f_{k}(z)$ for large $z$. However, for $z < 10$ -- which is the relevant range for observations -- the difference is below $10\%$ across the models. We also note an almost scale independence of $f_{k}(z)$ for this low redshift range. \textbf{Bottom panel}: \textit{Left}: The cosmological evolution of the $f \sigma_8$ observable in terms of redshift for $k = 1.0 \times 10^{-2} \mathrm{MPc}^{-1}$ for models in \texttt{branch I}. We highlight the percentage difference to $\mathrm{\Lambda CDM}$, which can grow larger than 10 $\%$ at late times for models with strong coupling $|\Sigma|$. Here, the redshift $z_{x,eq}$ signifies the point at which the density of dark matter is equal to the density of dark energy. In our model, $f \sigma_8$ varies slightly with wavenumber in the chosen region of $k$ and $z$, as explained next. \textit{Right}: The 3D plot for $f_{k}(z)$, for $w_x = -0.9$, highlighting the fact that $f_{k}(z)$ does not vary much with $k$ for the relevant redshift region, $z < 2$. Therefore, there is an almost scale-independence of $f_{k}(z)$, and, consequently, of $f \sigma_8(z)$ as well.}
    \label{fig:fs8}
\end{figure*}

From the matter density contrast given in (\ref{coupledm}), we can build the linear growth function $D(z)$:
\begin{eqnarray}
    D(z) \equiv \delta_m(z)/\delta_m(z = 0),
\end{eqnarray}
where, at present time, $D(z = 0) = 1$. The linear growth function $D$ serves to quantify the linear growth of cosmic structures. In the case of the $\mathrm{\Lambda CDM}$ model, $D$ is not scale dependent. However, for the \texttt{frozen} model, $D$ depends on the $k$ modes, as clearly indicated in the source term of Eq. \eqref{coupledm2}. Consequently, we will denote this scale-dependent growth function by $D_{k}(z)$ throughout the paper. 
Similarly, the growth rate, which is also scale-dependent, will be referred to as
$f_{k}(z)$, defined as follows:
\begin{eqnarray}
    f_{k}(z) \equiv \frac{\mathrm{d} \log D_k(a)}{\mathrm{d} \log a}.
\end{eqnarray}
By numerically solving equation (\ref{coupledm}) using \texttt{CLASS}, we can effectively reconstruct the linear growth function $D_{k}(z)$, the growth rate $f_{k}(z)$, and, consequently, the $f\sigma_8(z)$ observable as a function of redshift and wavenumber $k$ for various values of $w_{x}$  and different interaction couplings, $|\Sigma|=\{ 0.1,0.01, 0.001\}$. We also select three scales -- large, intermediate, and small -- to represent the scale dependence of these functions. The results of this analysis for branch I are illustrated in Fig. (\ref{fig:fs8}), highlighting the dependencies and variations introduced by these parameters in the context of cosmic evolution. 
Several observations merit discussion. The results for $D_{k}(z)$ and $f_{k}(z)$ at small redshift, see, in particular, the bottom-right plot in Fig. \ref{fig:fs8}, indicate small-scale dependence of the models. Therefore, for $f \sigma_8$ we plot only the intermediate scale results, which will closely resemble all scales. The clustering properties of the perturbed ET model, called frozen, closely resemble those of the concordance model, as expected. However, we find that the $f\sigma_8(z)$ curves display lower amplitudes for branch I. The ratio $\Delta f \sigma_{8}/ f \sigma_{8}$ can reach more than $10\%$ for the interaction couplings of $|\Sigma|=\{0.01, 0.001\}$ if $w_x = -0.8$. 
~\footnote{The behavior of $f \sigma_8$ closely resembles that of the matter power spectrum $P_k$, as expected from \eqref{kaiser3}: a larger (smaller) matter power spectrum leads to stronger (weaker) clustering.} Notably, the greater discrepancies appear at low redshift, particularly after surpassing the critical value, $z_{x,eq}$. In branch III, suppression of $f\sigma_8(z)$ at late times is attenuated for all $\Sigma$. However, in branch I, suppression is enhanced for large coupling $|\Sigma| \sim 0.1$, see Fig. \ref{fig:fs8}.

We will compare our findings on the estimation of  $f \sigma_8$  and $S_8$ in the context of other interacting dark energy models, particularly where the background is not constrained to the vanilla model or additional component are included.   According to the Planck collaboration, the joint analysis of Planck data, including $TT+ TE+EE + \rm{low}E + \rm{lensing} + BAO$ , yields a value of $\sigma_{8}=0.824\pm 0.010$ \cite{planck2020a,planck2020b}. However, this estimation exhibits a discrepancy of approximately $3.08\sigma$ when compared to the value reported by the $\rm{KiDS+ SDSS}$ cosmic shear collaboration, which found $\sigma_{8}=0.766^{+0.020}_{-0.014}$  \cite{kids} \footnote{As mentioned in the introduction, the KiDS constraints have been updated and now demonstrate a significant concordance with the Planck results \cite{newkids}.}.  Our numerical simulations reveal that for the newly introduced branch I, the parameter $\sigma_{8}(z=0)$ falls within the range of $[0.771, 0.800]$ as we vary $\Sigma$ from $10^{-3}$ to $10^{-1}$. Remarkably, this indicates that the $S_8$ tension is alleviated provided the interacting scenario yields lower values compared to the standard model. This intriguing result concerning the clumpiness of the universe can be  tested through galaxy surveys or by analyzing the KiDs+SDSS data on cosmic shear \cite{kids}. Furthermore, branch III produces analogous results, while branches II and IV exacerbate the tensions.  We also observe that our values for $\sigma_{8}(z=0)$ are, in most instances, even smaller than those derived from dark matter-massive neutrino interactions \cite{hivon, mateo1}. Indeed, the exploration of interacting dark energy and dark matter, characterized by an interaction rate $Q \propto \rho_{x}$ within the background and perturbation equations, does not mitigate the $\sigma_8$ tension; a combined analysis using  Planck-2018+DESI+SN+CC datasets yields $\sigma_{8}=0.971^{+0.060}_{-0.086}$ \cite{letter}.

\subsubsection{Dark energy pressure perturbation and Curvature perturbation} \label{subsec:pzeta}

In this section, we investigate the effects of the interaction in the dark-energy pressure perturbation.
We begin by noting that pressure perturbation for the dark energy fluids has two different contributions:
\begin{eqnarray}
\frac{\delta p_{x}}{\bar{\rho_{x}}}=c^2_{sx}\delta_{x}+3\mathcal{H}(c^2_{sx}-c^2_{ax})(1+w_x)\frac{\theta_{x}}{k^2}. \label{nondpx}
\end{eqnarray}

\begin{figure*}[!htbp]
    \centering
    \begin{minipage}[b]{0.48\textwidth}
        \centering
        \includegraphics[width=\textwidth]{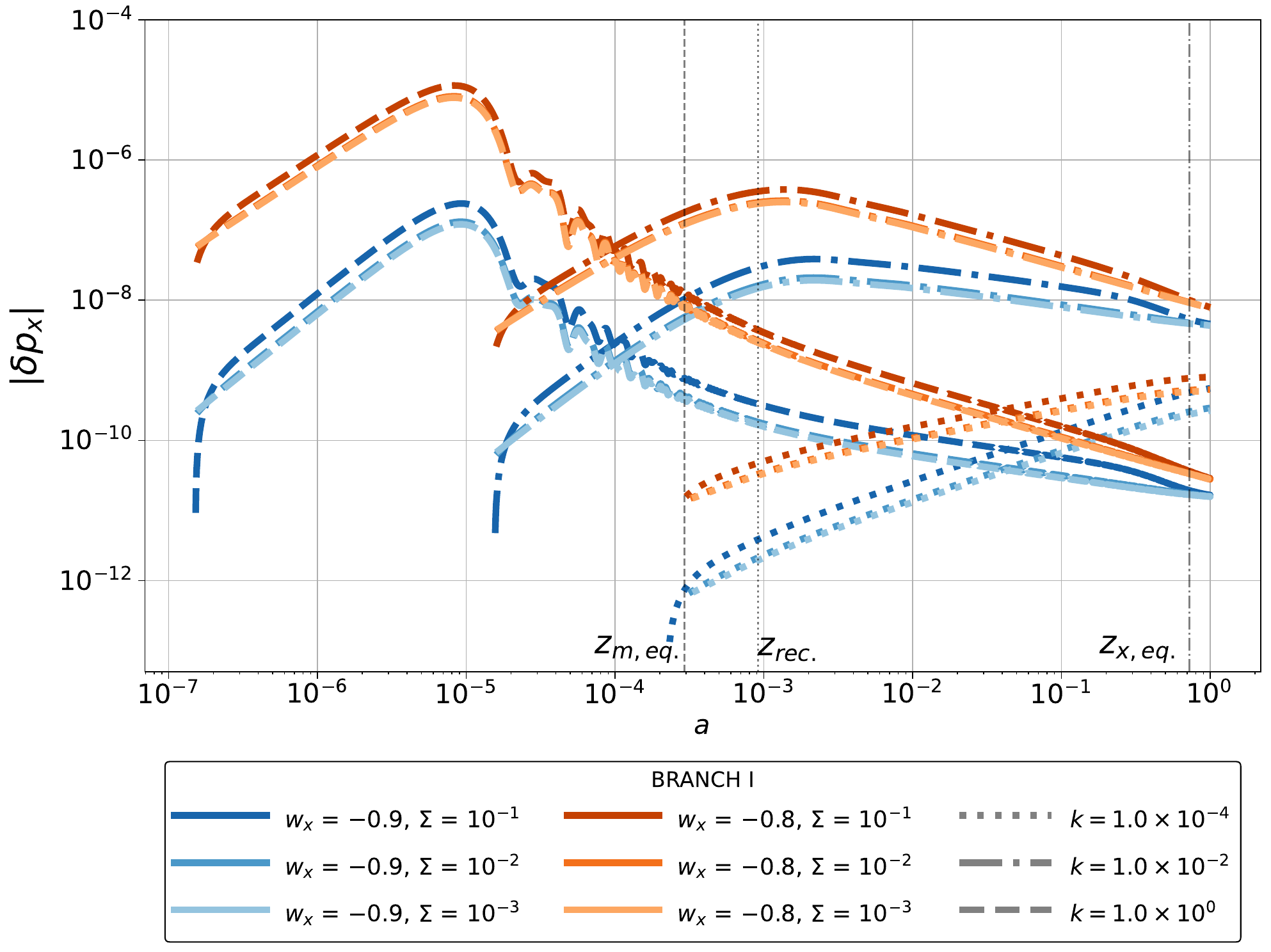}
    \end{minipage}
    \hfill
    \begin{minipage}[b]{0.48\textwidth}
        \centering
        \includegraphics[width=\textwidth]{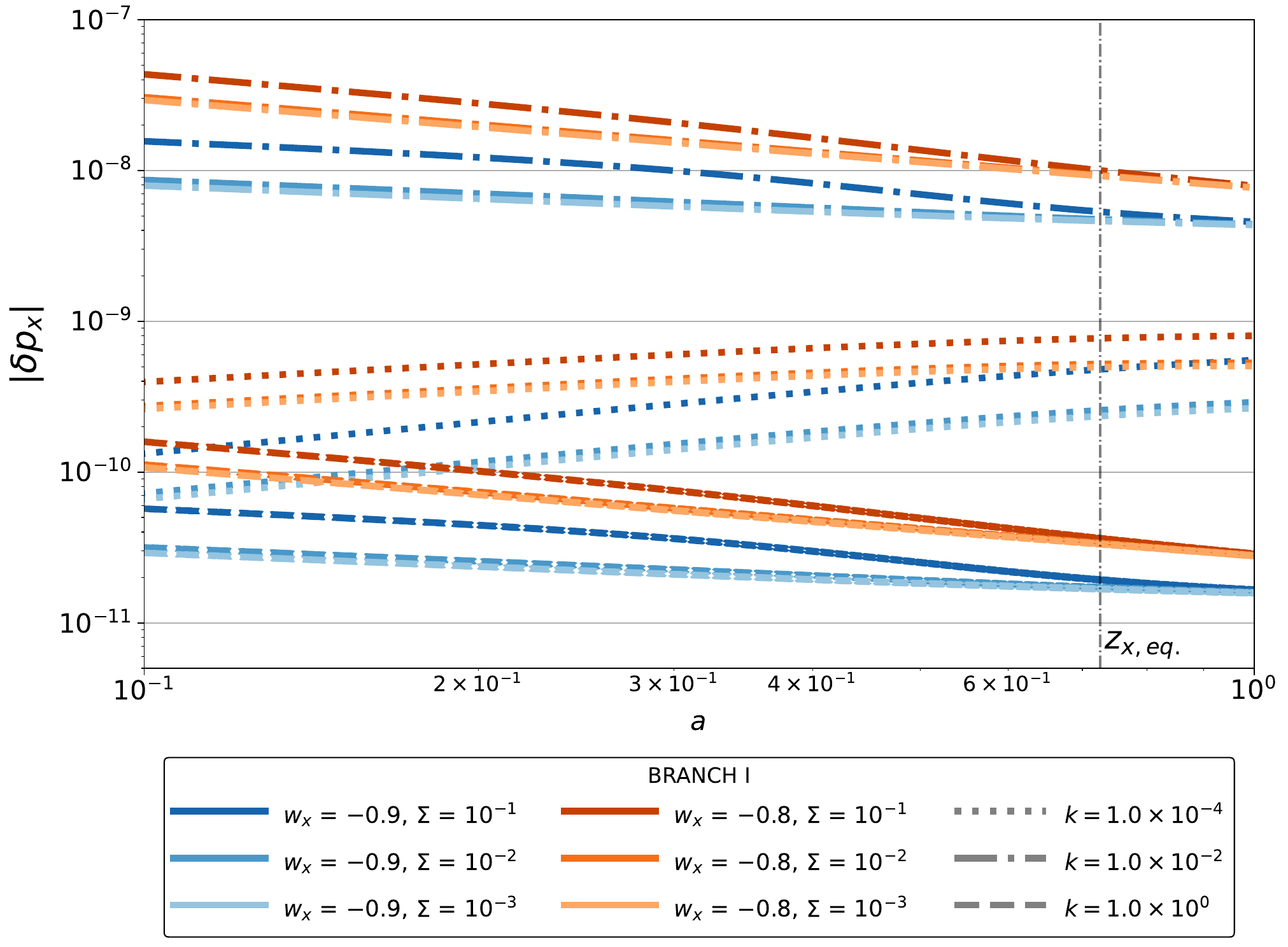}
    \end{minipage}

    \vspace{0.5cm}
    
    \begin{minipage}[b]{0.48\textwidth}
        \centering
        \includegraphics[width=\textwidth]{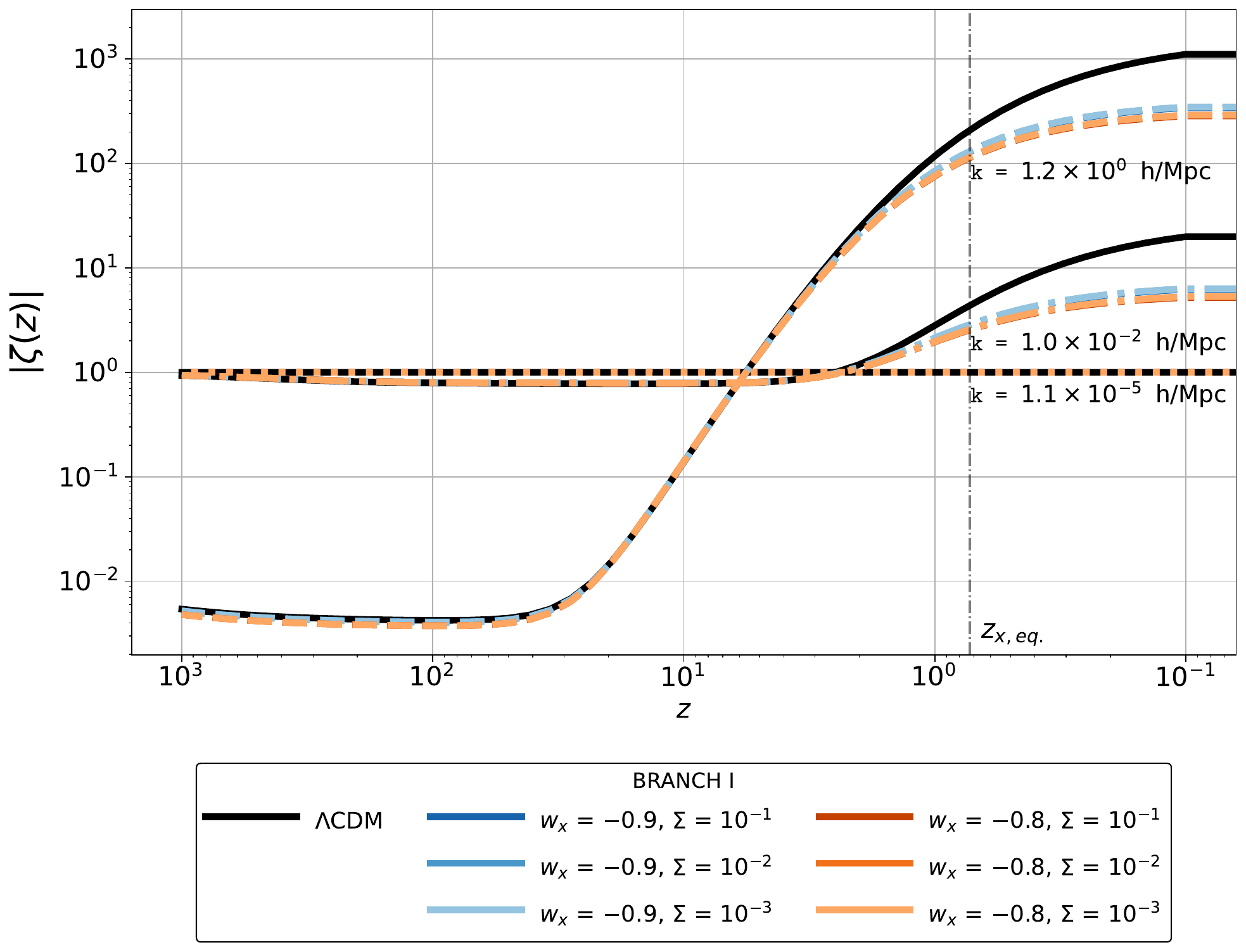}
    \end{minipage}
    \hfill
    \begin{minipage}[b]{0.48\textwidth}
        \centering
        \includegraphics[width=\textwidth]{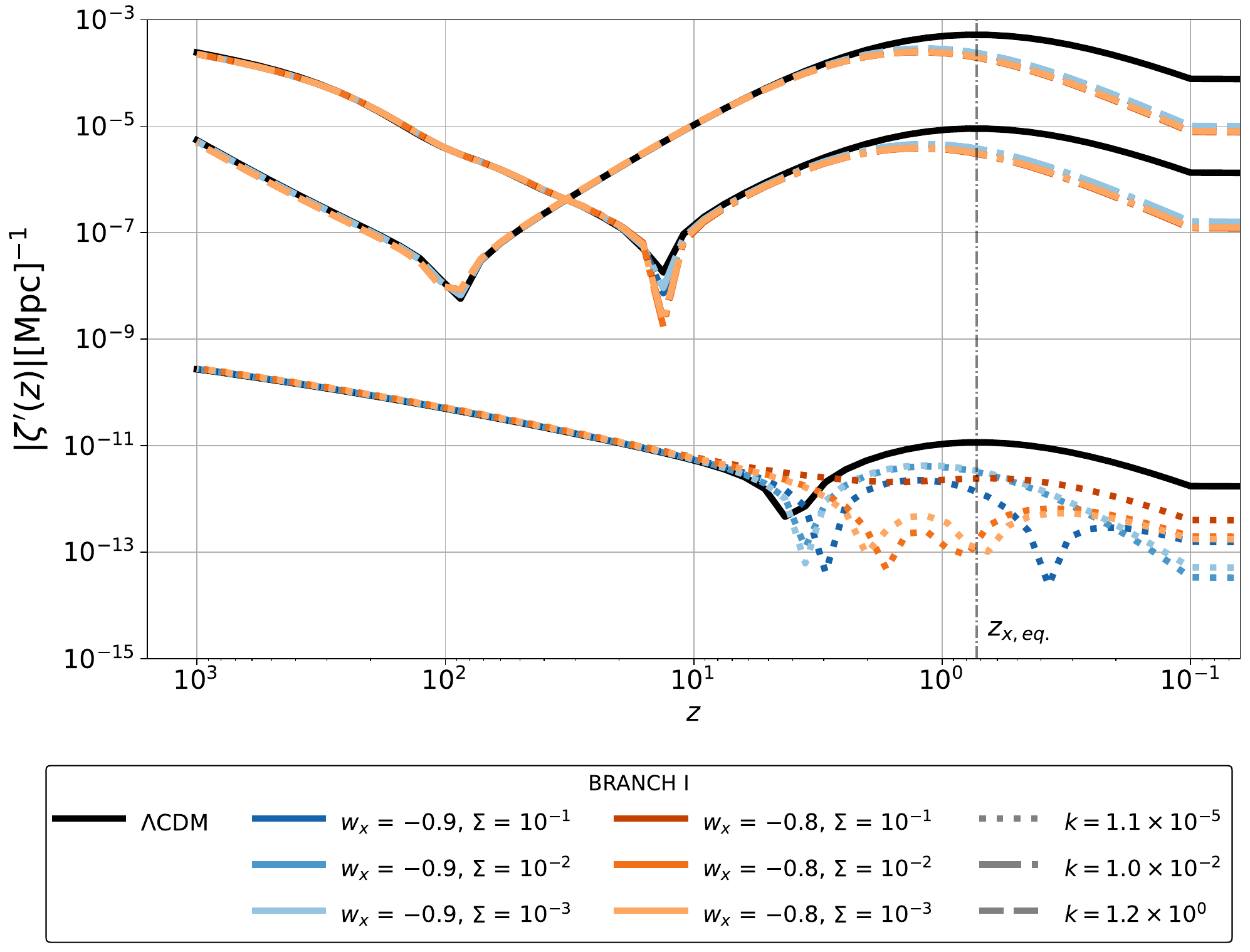}
    \end{minipage}
    
    \caption{Cosmological evolution for models in the \texttt{Frozen-branch-I} of the gravitational potential $\zeta$ (\ref{eq_zeta}) and its derivative with respect to conformal time $\zeta'$, both as a function of redshift $z$, and the dark energy pressure perturbation $\delta p_x$ (\ref{dpx}) as a function of the scale factor $a$. \textbf{Top panel}: On the left, the evolution of the dark energy pressure perturbation, and on the right a zoom-in on late times to highlight the difference in the behavior of different modes. The pressure contrast consists of contributions coming from the density contrast and the velocities, (\ref{dpx}). In what concerns the dark sector interaction, small-scale perturbations exchange energy earlier, as previously introduced, as one can see for the strong coupling case for $w_x = -0.9$. \textbf{Bottom panel}: On the left, the gauge-invariant curvature perturbation $\zeta$ since recombination, and on the right a its derivative $\zeta'$. The curvature perturbation is well-behaved, and, as expected from its definition, it has a larger amplitude for $\mathrm{\Lambda CDM}$ in comparison to \texttt{Frozen-branch-I} models. The small-scale perturbations shown re-enter the horizon earlier in time, and, therefore, grow during matter domination, while intermediate-scale perturbations present delayed growth. Large-scale perturbations are super-horizon and do not grow, for any model. During dark energy domination, enhancement is suppressed for all scales. 
    }
    \label{fig:curv}
\end{figure*}

Eq. (\ref{nondpx}) indicates that the effects of the interaction, i.e. the energy exchange between the dark sector, is imprinted in the pressure perturbation in two ways; in the density contrast and the peculiar velocity.   As we have shown in Fig. \ref{fig:int}, the interaction is relevant for small scales since the early stages for matter domination ($z > 1000$), while for intermediate scales it is only dominant with respect to Hubble expansion during late dark matter ($z > 100$) domination. That is the exact behavior we observe in the top panel of Fig. \ref{fig:curv}~\footnote{The pressure perturbation is most sensitive to the equation of state parameter $w_x$, as seen in the top row of Fig. \ref{fig:curv}. However, this effect does not translate linearly to cosmological observables.}. The amplitude of the pressure perturbation increases with a larger value of the interaction coupling. For small scales, in dotted blue lines, the enhancement is larger and occurs earlier than that of intermediate scales, in dash-dotted blue lines, as the interaction kicks-in earlier for smaller scales -- see the bottom-right plot for the zoomed-in region at late times. Large-scales are unaffected by the interaction.

We note that the pressure perturbation has contributions from both density contrast and velocity perturbation. The contribution of velocities has an explicit $k^{-2}$ factor that suppresses it for smaller scales. However, for large scales, there is no suppression, and the pressure perturbation grows steadily. Therefore, the large amplification seen for small and intermediate scales is sourced by the dark energy density contrast. Nevertheless, it is important to emphasize that the amplitude of dark energy perturbations remains bounded at all times ($|\delta p_{x}|\leq 10^{-4}$); this in turn would indicate the lack of large-scale instabilities caused by $|\delta p_{x}|$. In a way, the interaction in the dark sector smoothed out any nonadiabatic pressure instability, leading $|\delta p_{x}|\leq 10^{-6}$ to redshift $z\geq z_{x, eq}$.

In branch III the effect is reversed: the dark energy pressure perturbation is suppressed as the interaction coupling increases.

Now, we turn our attention to the behavior of the gauge-invariant quantity known as the curvature perturbation, denoted $\zeta$, for various values of the interaction coupling. The total curvature perturbation, assessed on constant density surfaces, serves as a crucial metric for understanding the dynamics of perturbations in our model. As established in the literature \cite{Valiviita:2008iv}, this gauge-invariant formulation allows for a clearer interpretation of the perturbative effects arising from the interaction between dark energy and dark matter:
\begin{equation}
    \zeta = - \eta_{\rm{sync}} - \mathcal{H}\frac{\delta\rho_{\mathrm{tot.}}}{\rho'_{\mathrm{tot.}}} \label{eq_zeta}
\end{equation}
Fig. \ref{fig:curv} shows the behavior of the curvature perturbation in terms of the redshift.   The curvature perturbation, $\zeta(z,k)$,  follows the behavior of the vanilla model until $z_{x,eq}$ for both branches (I-III). \footnote{Notice that for branch I $\zeta$ slightly decreases with $|\Sigma|$, while there is a mild increase with $|\Sigma|$ for branch III.}

However, it is important to note that the different modes consistently exhibit lower amplitudes compared to the concordance model. This trend suggests that, while the overall dynamics remain similar to those of the standard vanilla model, the presence of interactions introduces a suppression effect on the amplitude of curvature perturbations. In order to understand the latter result, we must look at (\ref{eq_zeta}) in a deeper way.  For the $\mathrm{\Lambda CDM}$ scenario, the dark energy has an equation of state $w_{x} = -1$, which neglects its contribution to the pressure perturbation in the denominator of the total contrast (\ref{eq_zeta}),  hence the larger amplitude for late times in comparison to CPL and frozen models, for both phantom and non-phantom cases.  However, the dark energy density for these models is present provided $w_{\mathrm{x}}$ is no longer $-1$. This leads to the suppression of the curvature perturbation for CPL and frozen models when the dark energy density starts to be relevant in comparison to the other components. During dark energy domination this effect is, naturally, more pronounced. Large-scale modes are super-horizon and, therefore, do not present any difference between $\mathrm{\Lambda CDM}$, CPL and frozen models. Notice that the derivative of the curvatuve perturbation is given by  \cite{Valiviita:2008iv}:
\begin{equation}
    \zeta' =-\frac{\mathcal{H}}{(\rho+p)_{\mathrm{tot.}}}\delta p_{\rm{nad}}.\label{eq_zetap}
\end{equation}
We present numerical evidence showing that $\zeta'(z,k)$ remains bounded at all times for various values of $w_{x}$ and $|\Sigma|$  [cf. Fig. \ref{fig:curv}]. Furthermore, our results indicate that the curvature perturbation is conserved on superhorizon scales within the $\delta$-ET  (\texttt{frozen model}) framework under adiabatic conditions; in other words, on large scales, we find that $\zeta' \simeq 0 $. This behavior is consistent with Weinberg's theorem \cite{Weinberg:2008zzc}, which asserts the conservation of curvature perturbations on superhorizon scales. Although we observe nonadiabatic pressure perturbations, they do not grow sufficiently fast to disrupt the conservation of the curvature perturbation. In fact, our findings suggest that the interactions considered in this framework may actually contribute to a reduction in the amplitude of perturbations in comparison to the vanilla model, thereby reinforcing the stability of curvature conservation across the scales we examined.

We conclude this section by highlighting another intriguing property of the $\delta$-ET proposal. It is important to emphasize that the interactions introduced in this framework do not suffer from the so-called dark-sector instabilities. Typically, an additional factor, called the doom factor $d_{x}=\bar{Q}_{x}/[3\mathcal{H}\bar{\rho}_{x}(1+w_{x})]$, which is encoded in the dark energy pressure perturbation, could lead to non-adiabatic instabilities in dark energy perturbations \cite{gavela} \footnote{ It is crucial to notice that it could be found several types of instabilities in the literature. First, strong coupling in a coupled quintessence model leads to an adiabatic, unbounded growth of dark-matter density contrast on small scales \cite{tomi}. Second, a nonadiabatic large-scale instability arises when there is a nonvanishing exchange of energy ($\bar{Q}=\Gamma \bar{\rho}_{c}$) at the background level, particularly for $(w_{x}+1)$  close to zero \cite{Valiviita:2008iv}. Remarkably, this instability occurs independently of the value of the interaction coupling and results in the amplification of curvature perturbations. Lastly, during early cosmic epochs, a significant instability in the dark energy density contrast emerges when $\bar{Q}=\Sigma H \bar{\rho}_{x} $ and $(1+w_{x})\Sigma>0$ \cite{gavela}}. In our analysis, we specifically focus on the scenario with zero energy transfer at the background level. This implies that the doom factor, which is utilized to assess the potential existence of dark energy instabilities, is effectively zero. For a more comprehensive understanding of the origins of such instabilities, the reader is encouraged to consult the references \cite{Valiviita:2008iv, gavela}. This consideration further solidifies the stability of the interactions proposed in the $\delta$-ET proposal.

\subsection{General features for branches II/IV} \label{subsec:ii_iv}
In this section, we  carry on by exploring the other branches of the $\delta$-ET, where the equation of state has fixed values, $w_x = \{-1.06, -1.1\}$ \footnote{Given this choice of $w_x$, setting $w_x = -1.06$ makes it impossible to use $\Sigma = 0.1$ due to model instability. This instability restricts $w_x$ to values less than 0.084 for this specific branch.}. These branches exhibit a phantom-like equation of state,  $w_x < -1$. Such models are characterized by a universe that is dominated at late times by a dark energy component and a violation of the weak energy condition \cite{bigrip}.  Typically, the expansion rate of the universe increases so rapidly that it may result in a future cosmic singularity known as the ``Big Rip''. Despite this singular behavior, these models have not been completely ruled out by recent observations \cite{olga, mateo2, adaf, Asghari:2019qld}.

\begin{figure*}[!htbp]
    \centering
    \begin{minipage}[b]{0.475\textwidth}
        \centering
        \includegraphics[width=\textwidth]{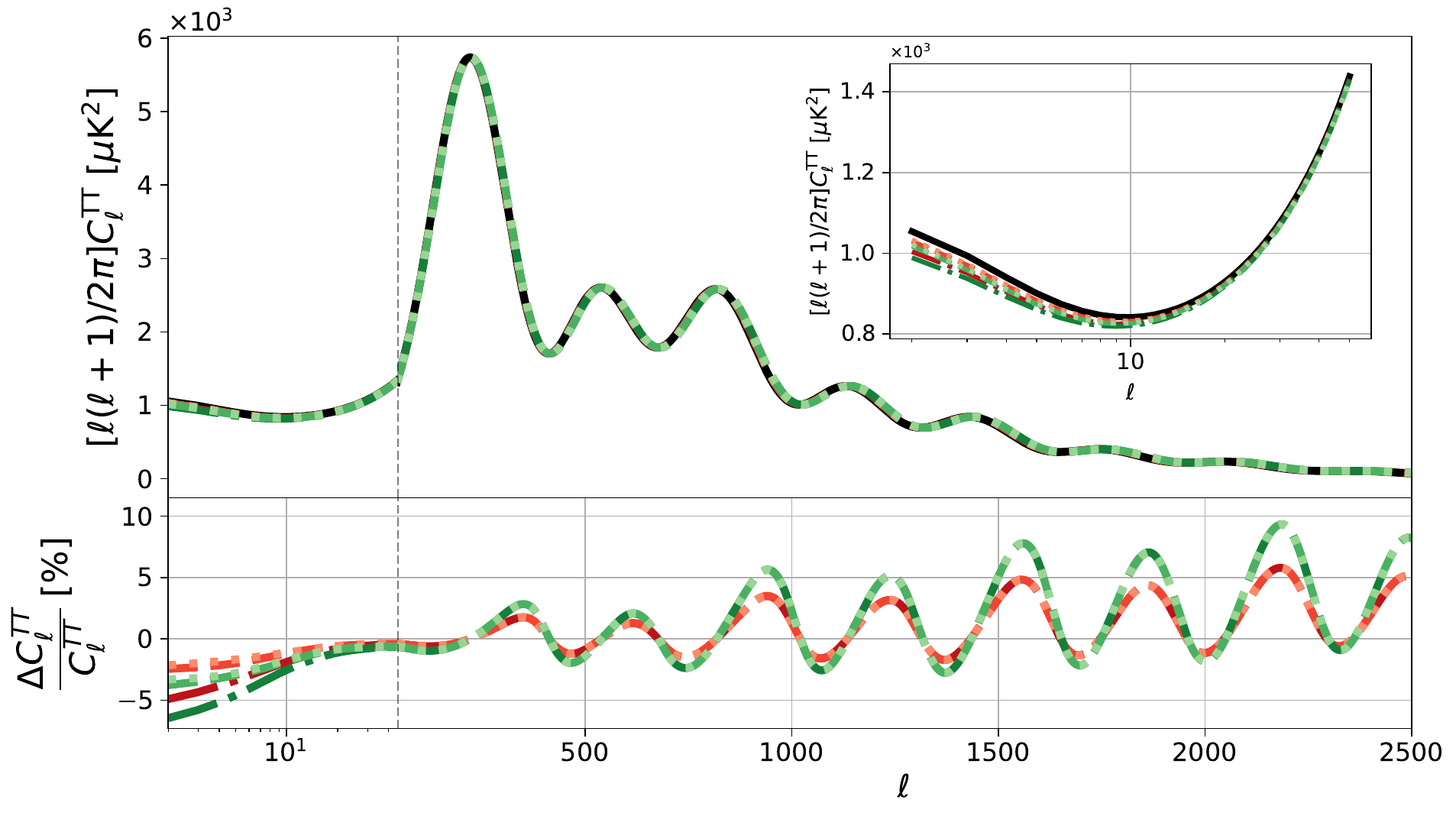}
    \end{minipage}
    \hfill
    \begin{minipage}[b]{0.475\textwidth}
        \centering
        \includegraphics[width=\textwidth]{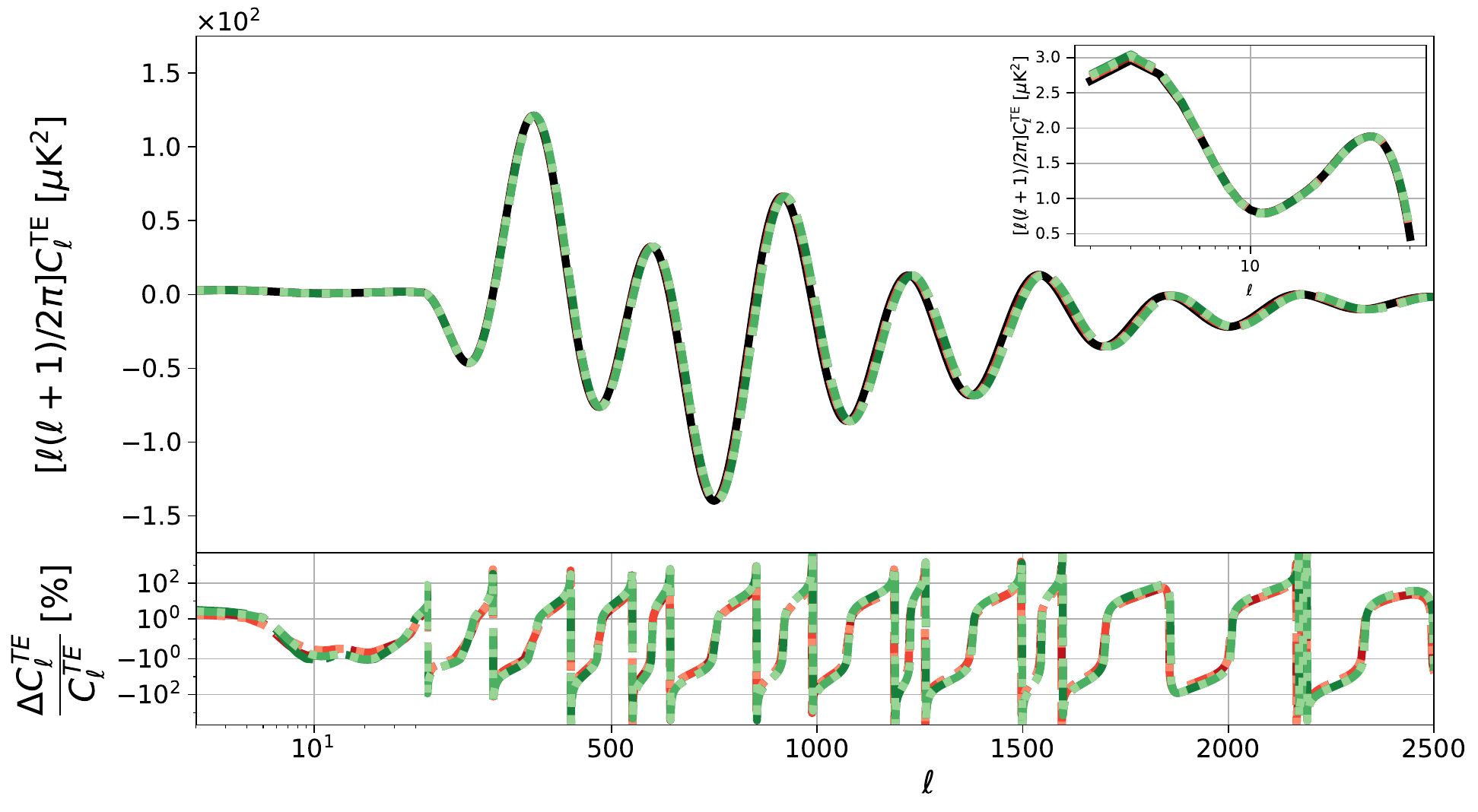}
    \end{minipage}

    \vspace{0.5cm}

    \begin{minipage}[b]{0.475\textwidth}
        \centering
        \includegraphics[width=\textwidth]{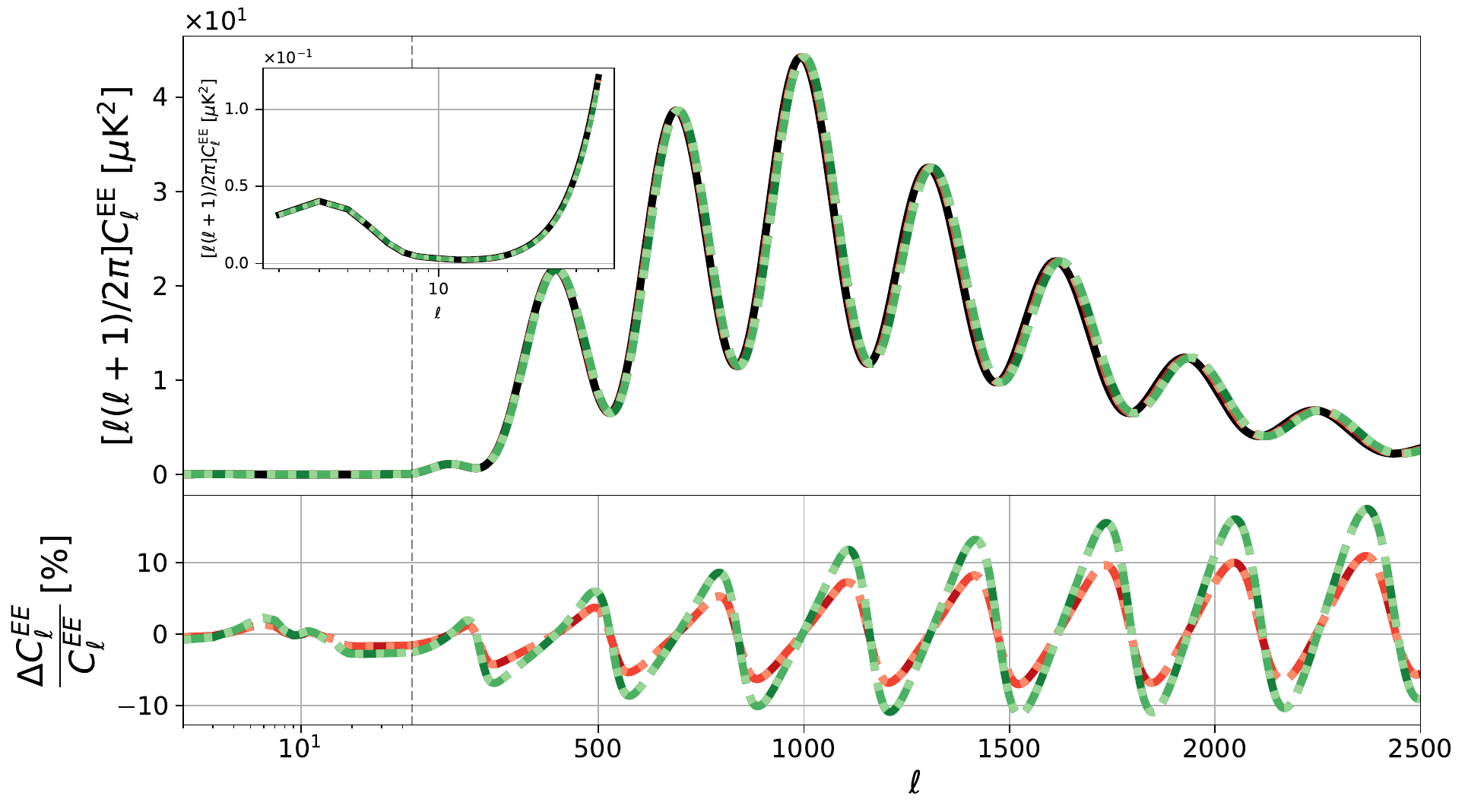}
    \end{minipage}
    \hfill
    \begin{minipage}[b]{0.475\textwidth}
        \centering
        \includegraphics[width=\textwidth]{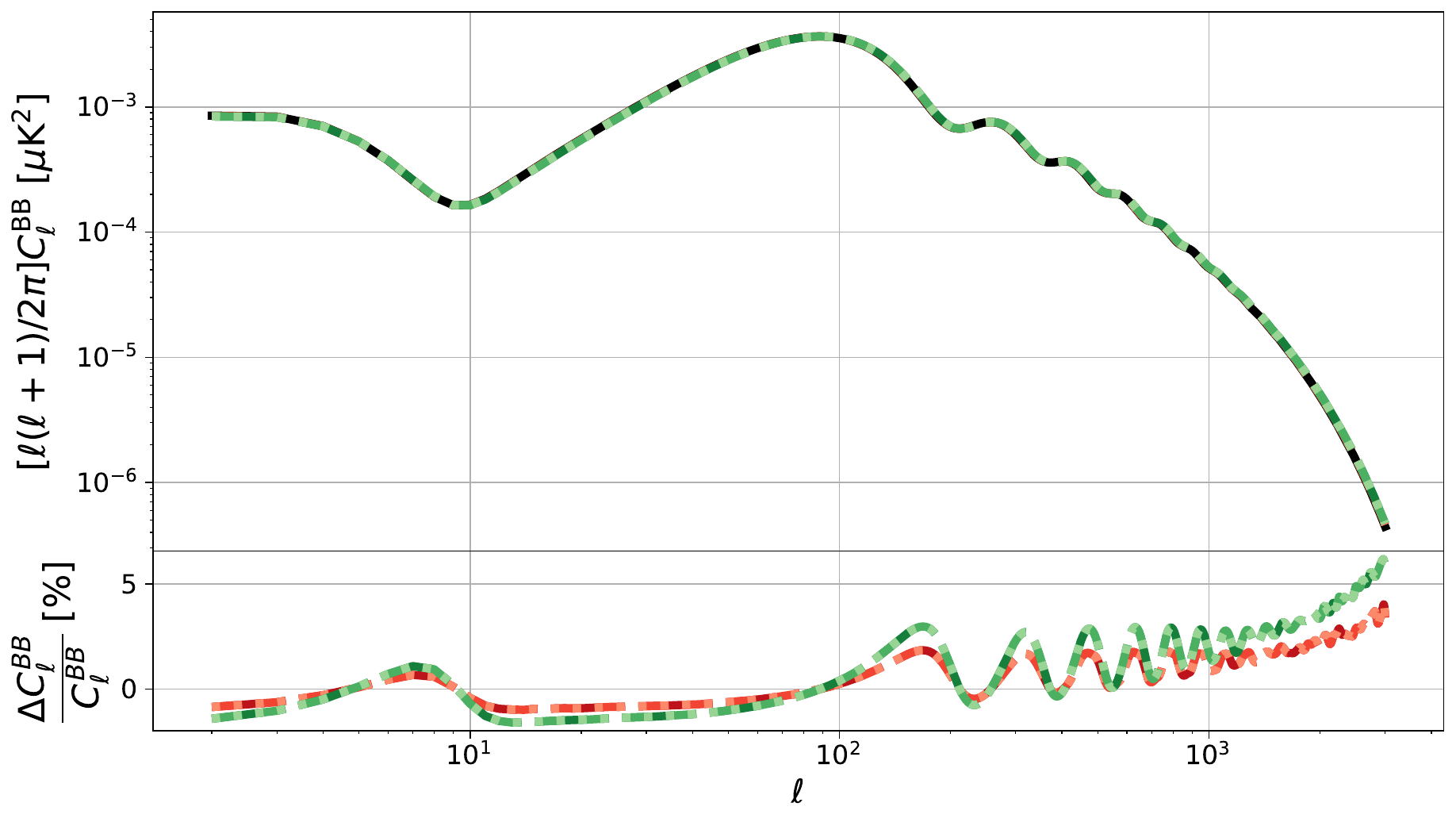}
    \end{minipage}

    \vspace{0.25cm}

    \begin{minipage}[b]{0.5\textwidth}
        \centering
        \includegraphics[width=\textwidth]{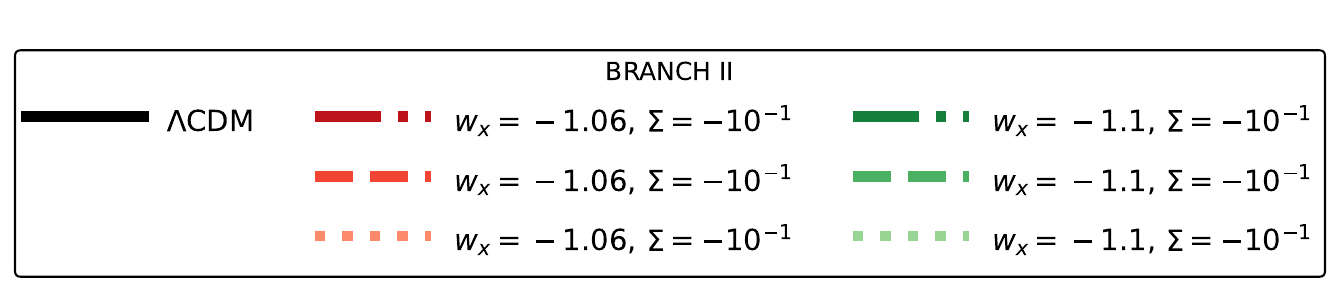}
    \end{minipage}
    
    \caption{
    The CMB angular power spectrum for the $TT, TE, EE$, and $BB$ two-point correlation functions for $\mathrm{\Lambda CDM}$ and the \texttt{Frozen-branch-II} models for a particular range of $\Sigma$ and $w_x$, as indicated. The difference from $\mathrm{\Lambda CDM}$ in each spectrum is illustrated in the lower panel of each plot. This percentage difference emphasizes the role played by interactions in the dark sector, and how such mechanism translates into new features. In the top-left corner plot, we highlight that the $C_{\ell}^{TT}$ presents a larger deviation from $\mathrm{\Lambda CDM}$ at large scales for models with negative coupling constant $\Sigma$. The main contribution of the coupling $\Sigma$ happens at low-$\ell$, as best shown in the $TT$ power spectrum, via the ISW effect, which is weakened for negative $\Sigma$ in branch II.}
    
    \label{fig:cl_class_II}
\end{figure*}

We present the results for the CMB power spectrum and its correlation functions ($TT$, $EE$, $TE$, and $BB$), which are illustrated in Fig. \ref{fig:cl_class_II}. 
Interestingly, the positions of the peaks in all spectra remain unaffected by the coupling $\Sigma$  for branches II and IV, mirroring the behavior observed in branches I and III. As anticipated, a greater deviation of $w_x$ from $\mathrm{\Lambda CDM}$ leads to a more pronounced shift in the peak positions. The damping observed on small scales (high-$\ell$ multipoles) is similarly independent of $\Sigma$. These findings collectively emphasize that the CMB, in general, may not serve as the most effective probe for constraining \texttt{frozen} models, with the notable exception of effects associated with a time-varying Newtonian potential in the late universe. 
Indeed, to resolve the degeneracy in $\Sigma$, it is essential to take into account additional observables and multipole ranges, such as the late integrated Sachs-Wolfe effect. As discussed previously, the variation in the CMB temperature is contingent upon the derivative of the Newtonian potential. Consequently, the auto-correlation function is affected by the matter power spectrum. This behavior is precisely what we observe for low-$\ell$  in the TT power spectrum.
\begin{figure}[h]
    \centering
    \includegraphics[width=0.95\linewidth]{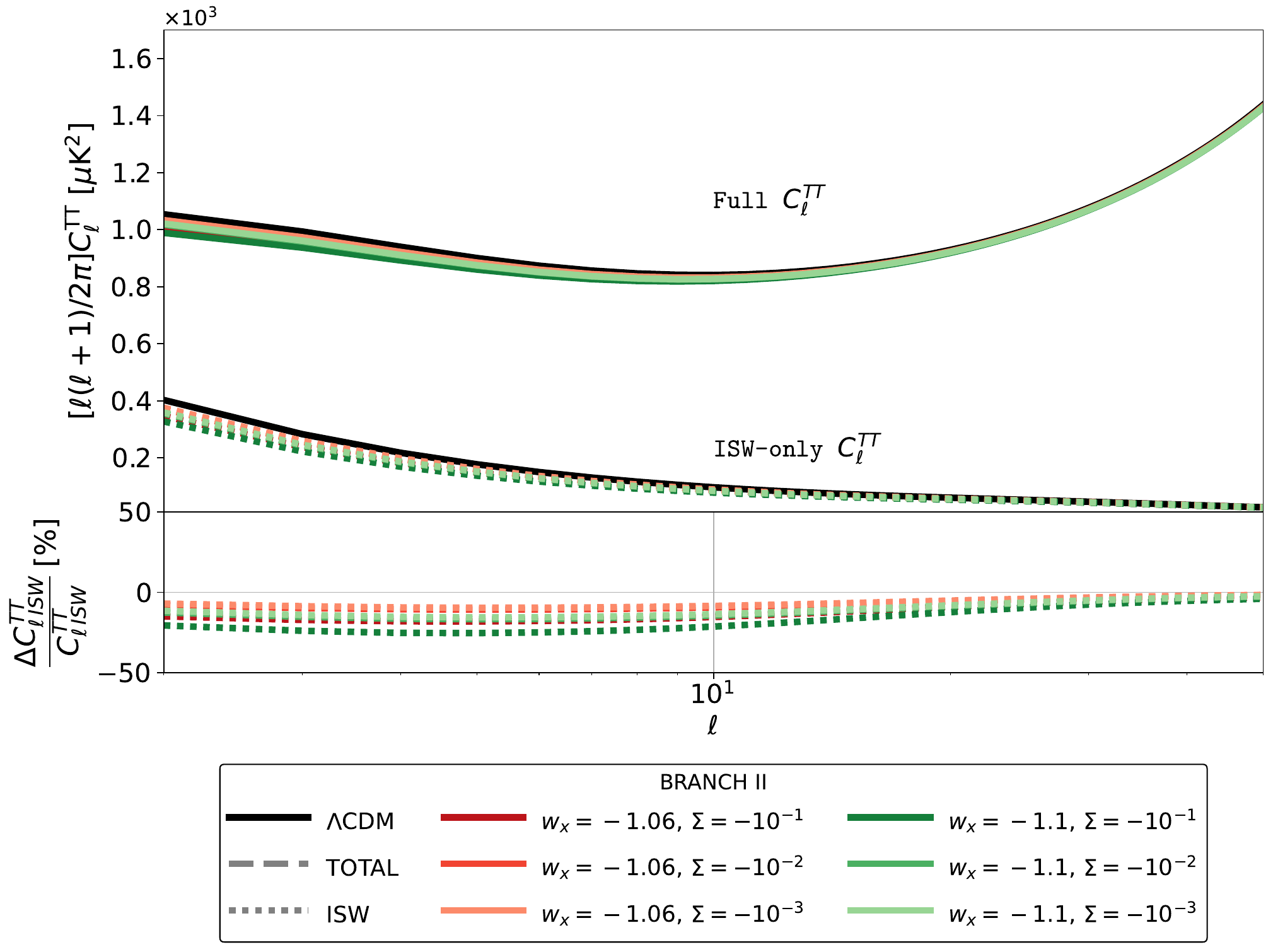}\\
    [1cm]
    \includegraphics[width=0.95\linewidth]{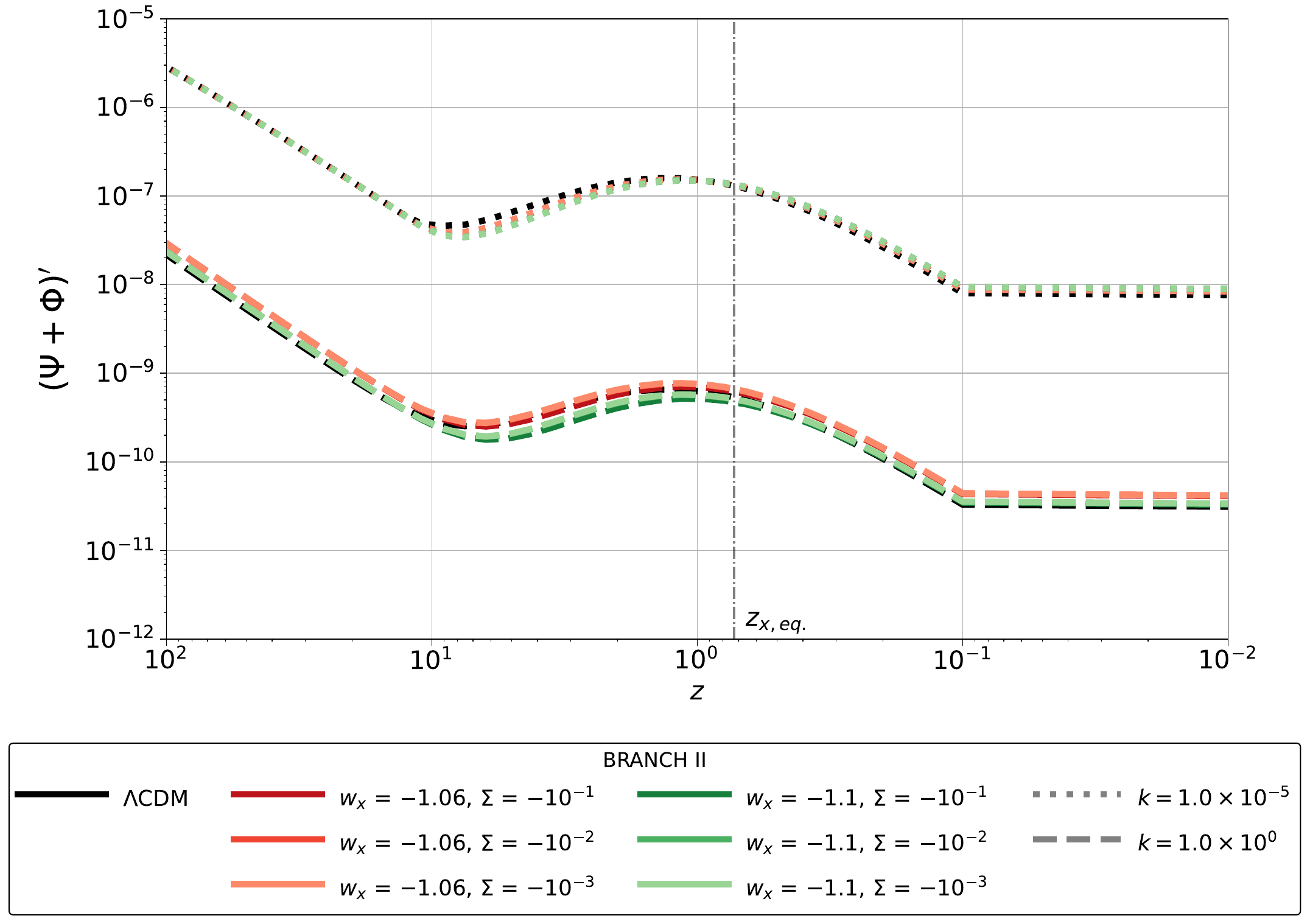}
    \caption{\textbf{Top panel}: In the top, the CMB TT power spectrum and the contribution from the late-time integrated Sachs-Wolfe effect. The lower panel of the plot details the percentage difference between the ISW effect between $\mathrm{\Lambda CDM}$ and \texttt{Frozen-branch-II}. \textbf{Bottom panel}: The derivative of the gravitational potential $(\Psi + \Phi)$. For branch II, the stronger the coupling $\Sigma$, the smaller the effect.}
    \label{fig:LISW_class_II}
\end{figure}

Transitioning to phantom models does not necessarily change all the results we previously obtained. The integrated Sachs-Wolfe (ISW) effect is amplified when $\delta Q_{x}<0$. Consequently, for branch II, the ISW contribution is suppressed, as we have $\Sigma < 0$ and $\delta Q_{x}>0$~\footnote{Note that, once again, the contribution from $h^{'}$ is the dominant one and $h^{'} > 0$.}. The ISW effect remains sensitive to the coupling $\Sigma$ and the equation of state $w_x$, with deviations from $\mathrm{\Lambda CDM}$ becoming more pronounced as $|\Sigma|$ increases. Therefore, branches II and IV exhibit opposite behavior: the ISW effect is suppressed for branch II and amplified for branch IV. Figure \ref{fig:LISW_class_II} illustrates these results for branch II. Moreover, the ISW effect is inherently weaker in phantom models than in $\mathrm{\Lambda CDM}$.

As discussed in the analysis of branches I and III, the ISW effect alone is insufficient, but the use of cross-correlation with galaxies and quasars appears promising.

\begin{figure*}[!htbp]
    \begin{minipage}[b]{0.475\textwidth}
        \centering
        \includegraphics[width=\textwidth]{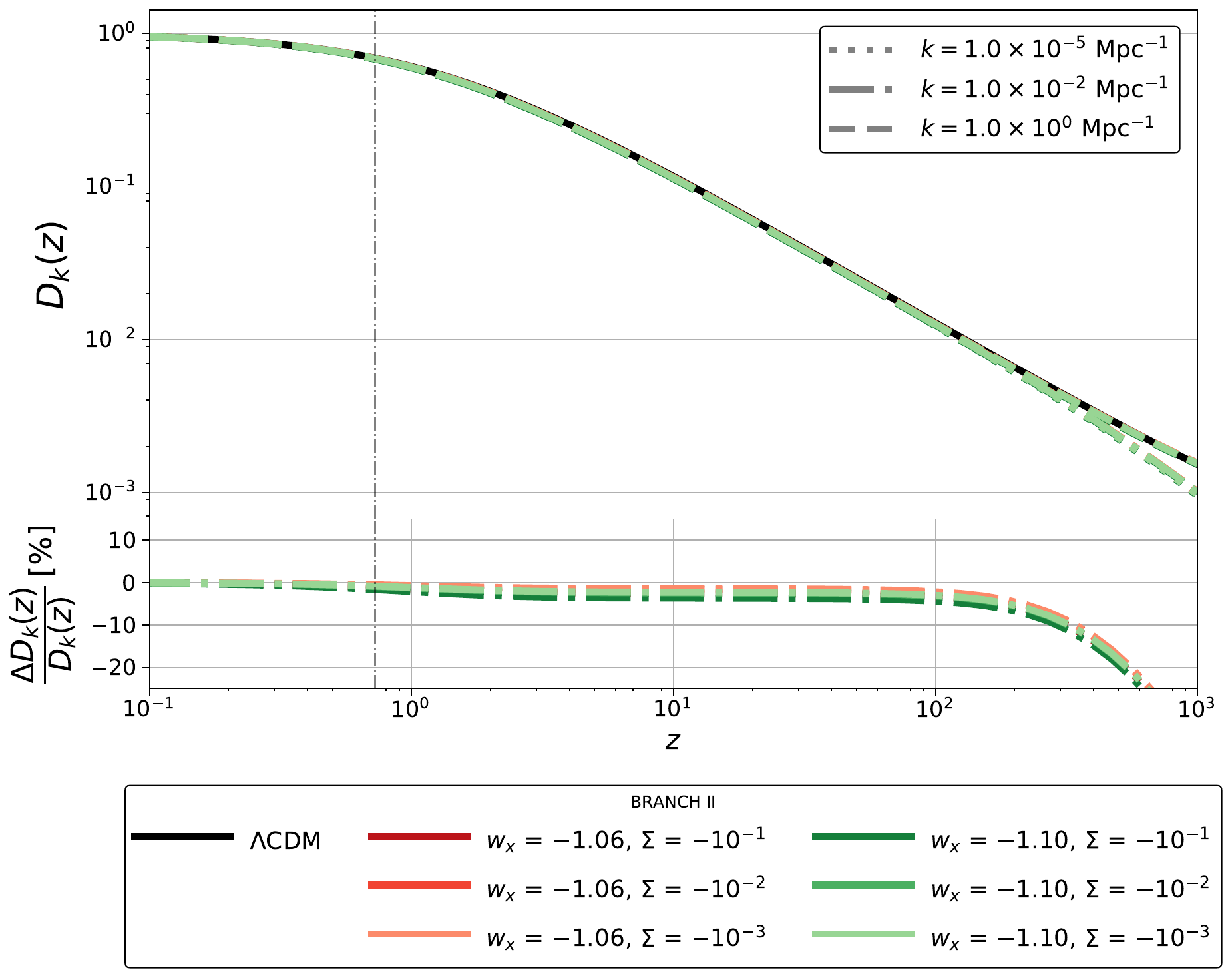}
    \end{minipage}
    \hfill
    \begin{minipage}[b]{0.475\textwidth}
        \centering
        \includegraphics[width=\textwidth]{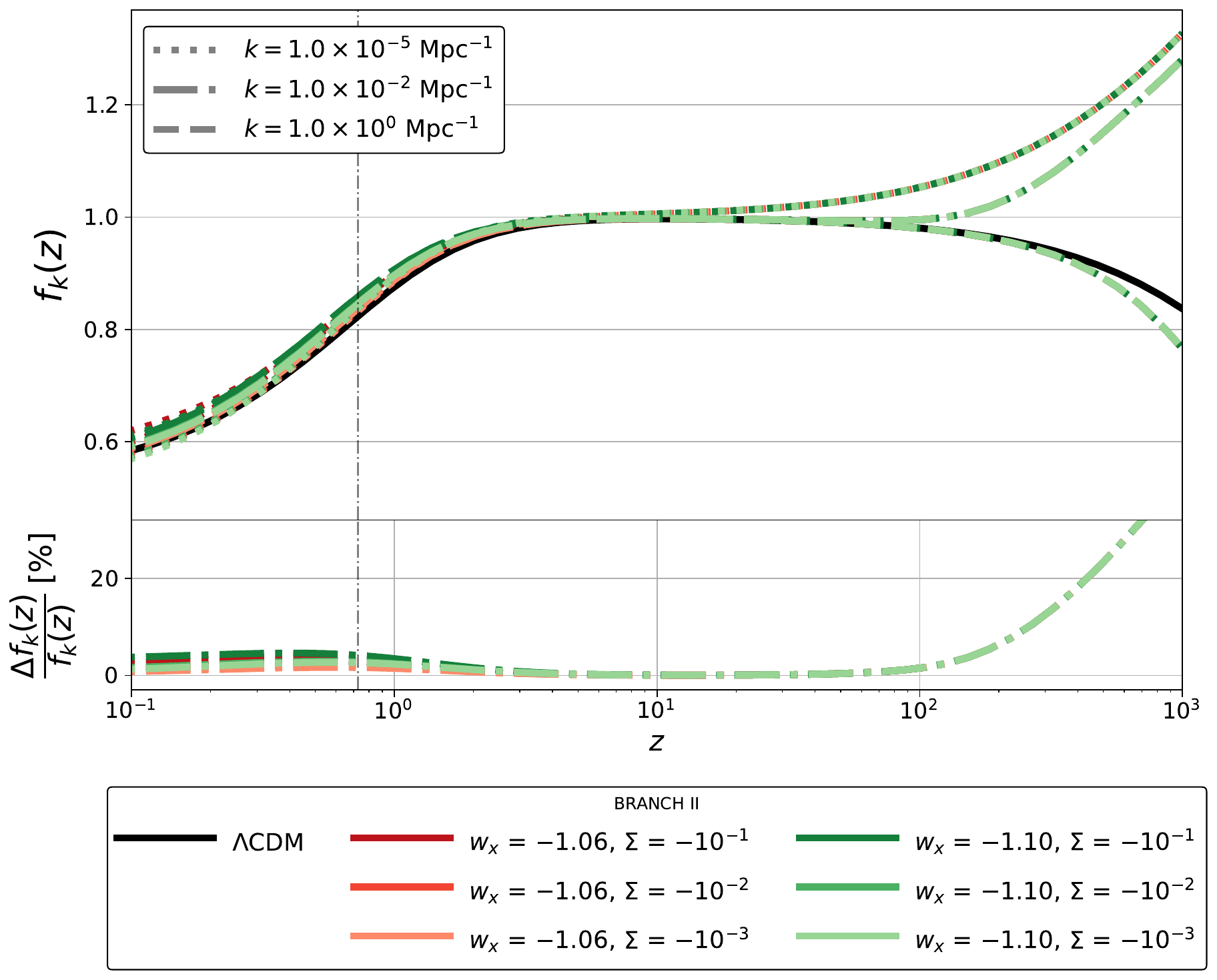}
    \end{minipage}

    \vspace{0.5cm}

    \begin{minipage}[b]{0.475\textwidth}
        \centering
        \includegraphics[width=\textwidth]{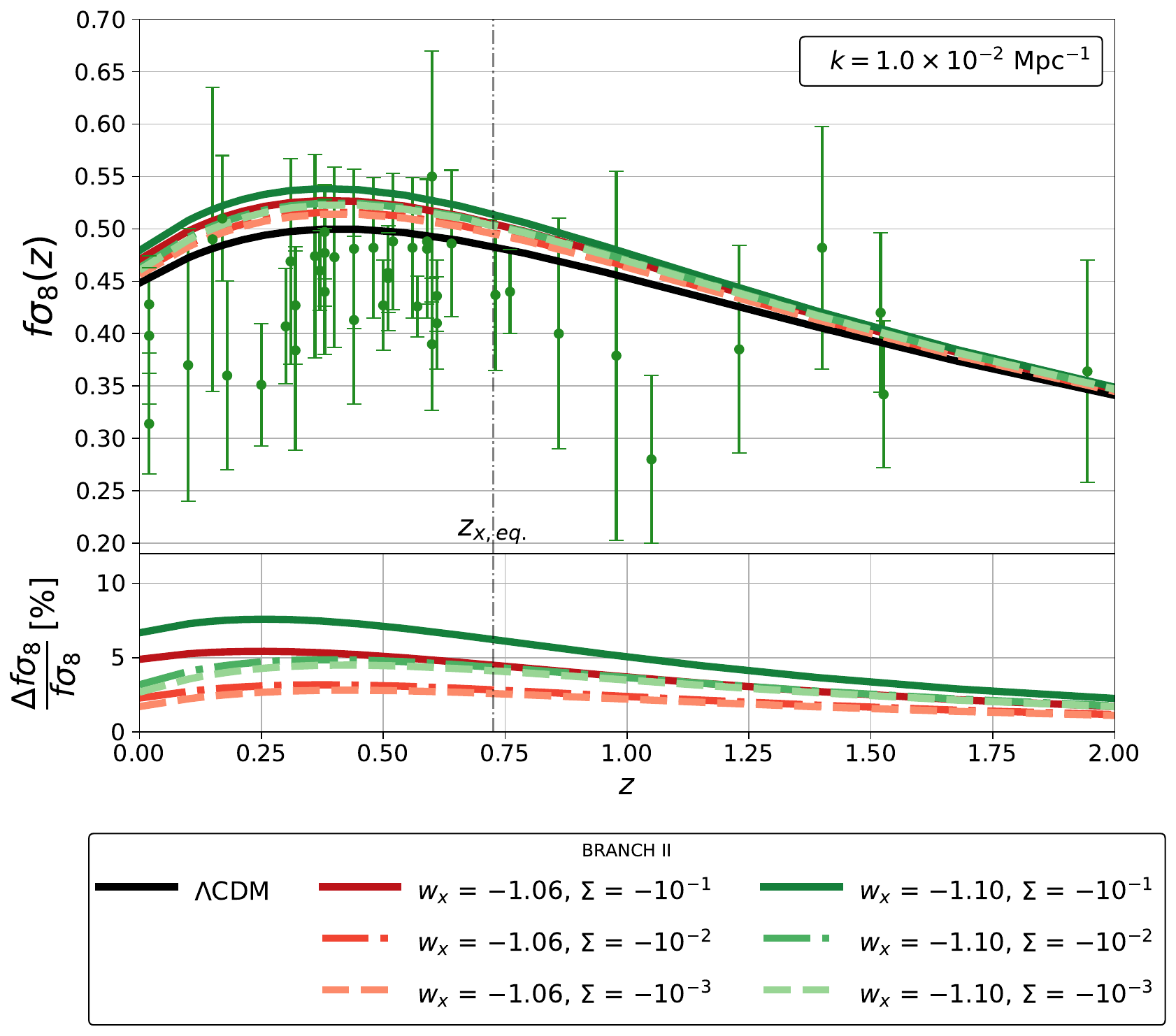}
    \end{minipage}
    \hfill
    \begin{minipage}[b]{0.475\textwidth}
        \centering
        \includegraphics[width=\textwidth]{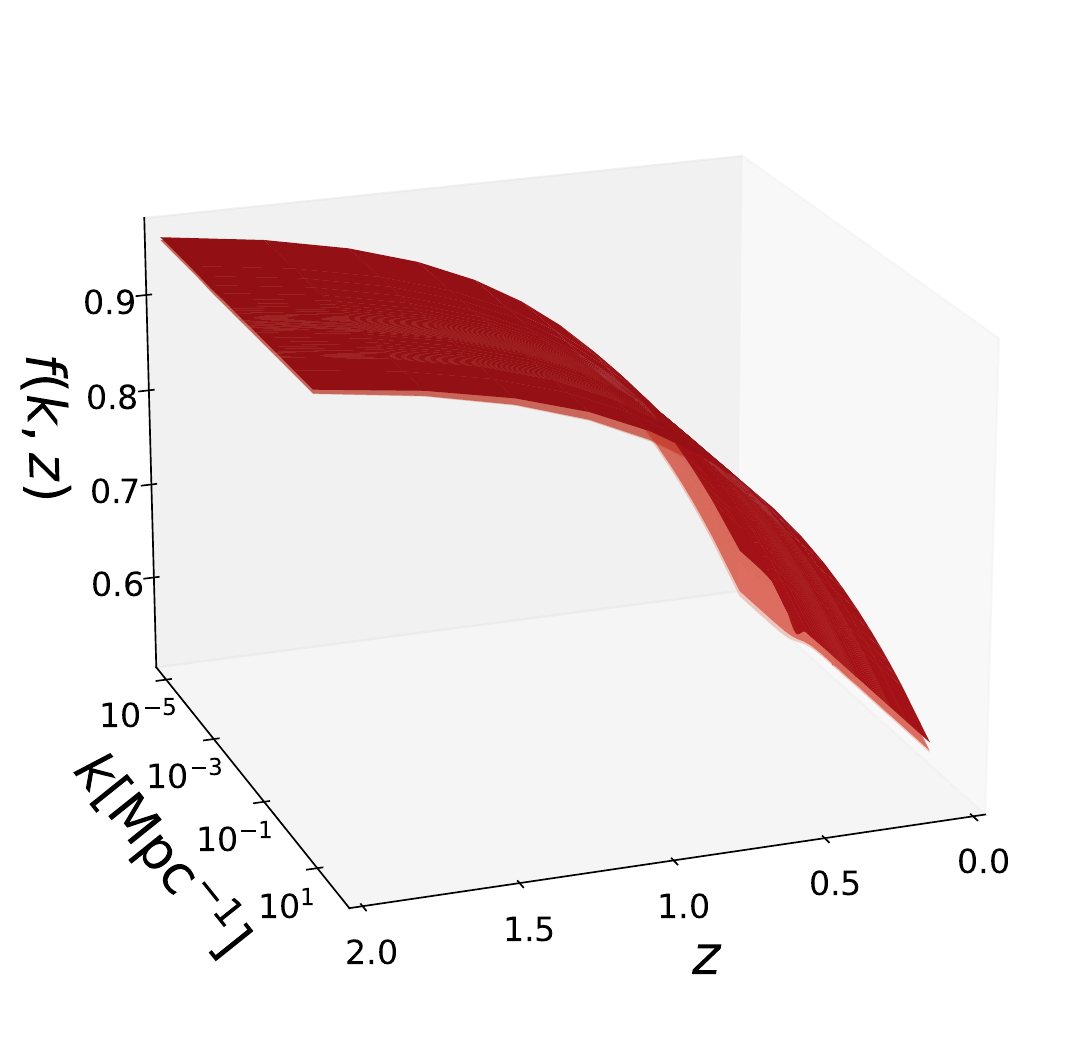}
    \end{minipage}

    \caption{\textbf{Top panel}: \textit{Left}: The growth factor $D_{k}(z)$ for models in branch II, for three different $k$ values, following the same prescription as Fig. \ref{fig:fs8}. Similarly, we present the percentage difference to $\mathrm{\Lambda CDM}$ for intermediate scales, which is only relevant for very large $z$. We note an almost scale independence of $D_{k}(z)$, and therefore the difference to the reference model is consistent across scales. \textit{Right}: The growth factor $f_{k}(z)$ for models in branch II, for the chosen large, intermediate, and small scales. There is a large scale dependence for $f_{k}(z)$ for large $z$. However, for $z < 10$ -- which is the relevant range for observations -- the difference is below $10\%$ across the models. We also note an almost scale independence of $f_{k}(z)$ for this low redshift range. \textbf{Bottom panel}: \textit{Left}: The cosmological evolution of the $f \sigma_8$ observable in terms of redshift for $k = 1.0 \times 10^{-2} \mathrm{MPc}^{-1}$ for models in \texttt{branch II}. We highlight the percentage difference to $\mathrm{\Lambda CDM}$, which \textbf{can grow} larger than 5 $\%$ at late times for models with strong coupling $|\Sigma|$. Here, the redshift $z_{x,eq}$ signifies the point at which the density of dark matter is equal to the density of dark energy. In our model, $f \sigma_8$ varies slightly with wavenumber in the chosen region of $k$ and $z$, as explained next. \textit{Right}: The 3D plot for $f_{k}(z)$, highlighting the fact that $f_{k}(z)$ does not vary much with $k$ for the relevant redshift region, $z < 2$. Therefore, there is an almost scale-independence of $f_{k}(z)$, and, consequently, of $f \sigma_8(z)$ as well.}
    \label{fig:fs82}
\end{figure*}

If the CMB lacks sufficient constraining power for the model, as it is largely not dependent on $\Sigma$, the same cannot be said for the matter perturbation sector.
Compared to the $\mathrm{\Lambda CDM}$ model, phantom branches lead to a larger amplitude of the matter power spectrum. As described in Sec. \ref{subsec:pdek}, branch II shows the power spectrum amplification for negative coupling ($\Sigma < 0$), while branch IV yields a suppression for positive coupling ($\Sigma > 0$). The latter results seem to be consistent with the fact that the energy exchange occurs from dark energy to dark matter when $\delta Q_x < 0$ ($\Sigma > 0$).
\begin{figure}[h]
    \centering
    \includegraphics[width=0.95\linewidth]{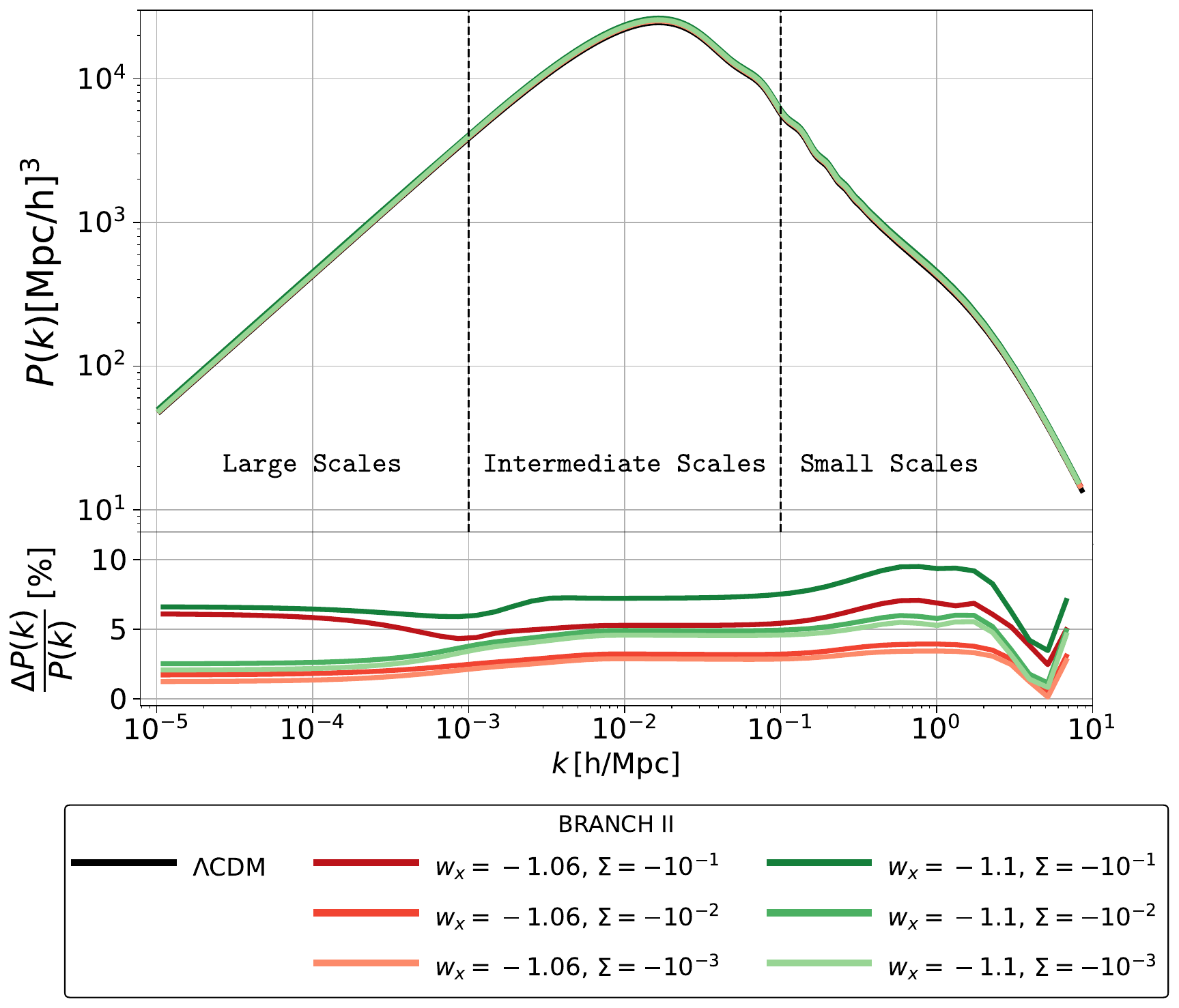}\\
    [1cm]
    \includegraphics[width=0.95\linewidth]{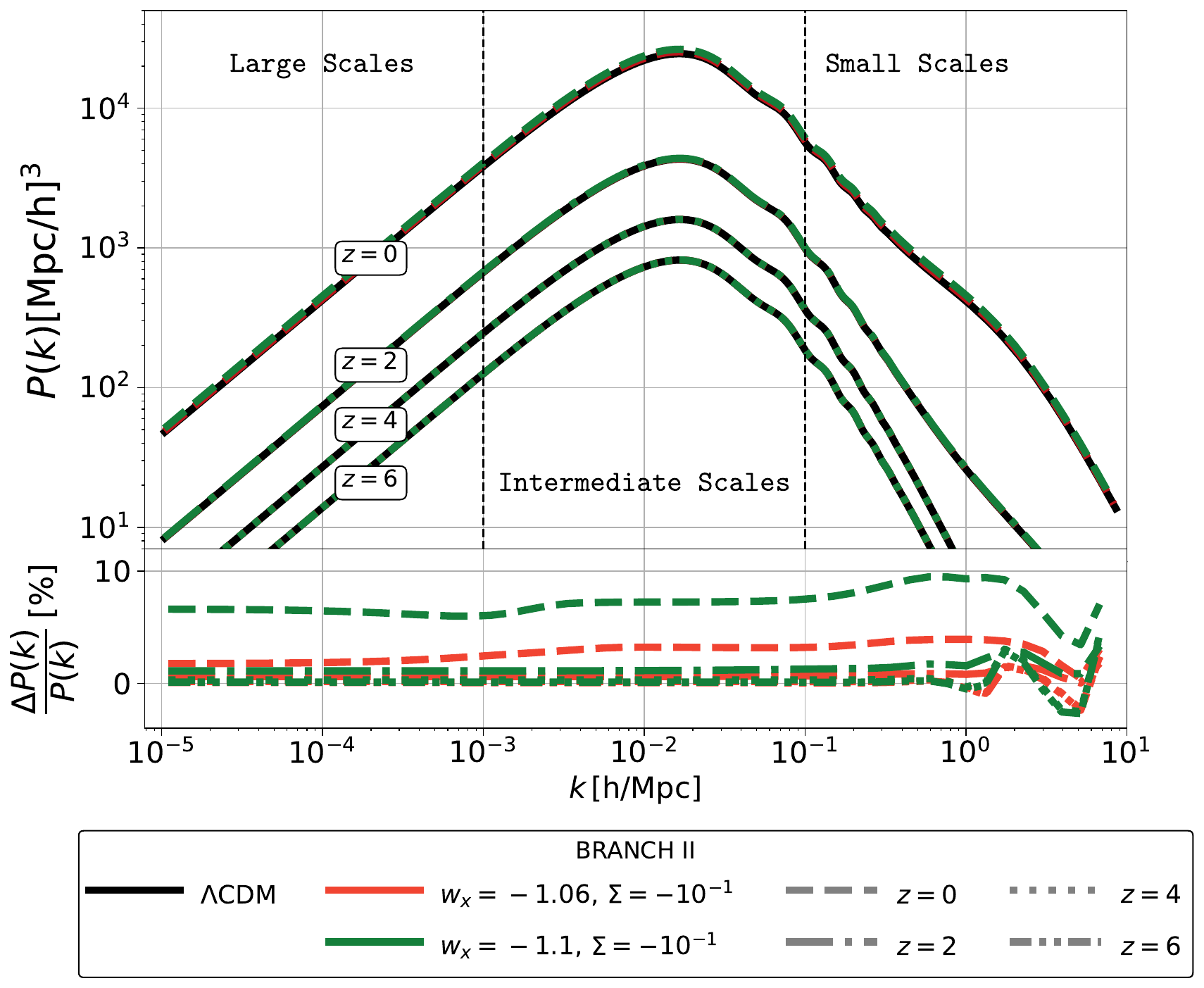}
    \caption{
    \textbf{Top panel}: The non-linear matter power spectrum at $z = 0$ for $\mathrm{\Lambda CDM}$ and the \texttt{Frozen-branch-II} models as a function of wavenumber $k$. In the lower panel, we present the $\%$ difference to $\mathrm{\Lambda CDM}$ in the non-linear power spectrum. Stronger coupling leads to stronger deviation from $\mathrm{\Lambda CDM}$. \textbf{Bottom panel}: Evolution of the non-linear power spectrum for four different redshift values (in different line styles), for the pairs $\Sigma = -0.1$, $w_x = -1.06$ and $\Sigma = -0.1$, $w_x = -1.1$. The difference to $\mathrm{\Lambda CDM}$ grows in time, reaching the maximum value at $z = 0$ for both $w_x$. That corroborates the result that the interaction is relevant only at late-times.
    }
    \label{fig:Pk_class_II}
\end{figure}
Similarly to the previous section, the deviation from $\mathrm{\Lambda CDM}$ model becomes significant only at late times, which aligns with the time evolution of the interaction: it occurs during matter domination for small scales and after $z = 100$ for intermediate scales (the late stages of matter domination). Furthermore, the clustering of matter in phantom models is stronger than in the typical $\mathrm{\Lambda CDM}$ model, taking into account the enhancement of the matter power spectrum in this scenario. When the interaction within the\texttt{frozen model} is efficient, the energy exchange between dark matter and dark energy enhances the power spectrum of matter for $\Sigma < 0$ and suppresses it for $\Sigma > 0$. Consequently, it amplifies $f \sigma_8$ for $\Sigma < 0$ and suppresses it for $\Sigma > 0$. These results are illustrated in Figs. \ref{fig:Pk_class_II} and \ref{fig:fs82}.

As seen in Fig. \ref{fig:fs82}, the deviation from $\mathrm{\Lambda CDM}$ becomes significant only for stronger coupling, exceeding the level $5\%$ at late times. In contrast, a weaker coupling results in a deviation below $5\%$.

As with the matter power spectrum, for strong coupling, the deviation increases primarily after dark energy domination. This aligns with the period when the interaction becomes relevant compared to the Hubble expansion. 
Furthermore, we can develop a diagnostic method based on these combined estimators to analyze how a model deviates from the concordance model or the CPL model with constant $w_{x}$ ($w\mathrm{CDM}$), where the fixed point represents the optimal value of the concordance/CPL model.

A key finding from the analysis of branches II and IV, complementing the previous section, is that when $\Sigma$ and $(1 + w_x)$ have the same signs, as in branches I and II, the dark energy pressure perturbation $\delta p_x$ increases with the magnitude of the coupling $|\Sigma|$, while in models III and IV it is suppressed for stronger coupling.  This new behavior deviates from the previous analysis, which linked the sign of the interaction term $\delta Q_x$ (and the coupling strength $\Sigma$) directly to the enhancement or suppression of an observable [cf. Fig. \ref{fig:curv}]. The discrepancy arises due to $\theta_x$. Specifically, for branches I and II, $\theta_x$ increases with the magnitude of $\Sigma$, whereas it decreases for branches III and IV, see \ref{appendix:c}. Consequently, the $\theta_x$-dependent term in Eq. (\ref{nondpx}) shows contrasting trends: it increases for branches I and II and decreases for branches III and IV. Additionally,  regarding $\zeta$ and $\zeta'$, branches II and IV align with the behavior described in the previous section: $\zeta$ and $\zeta'$ increase (decrease) with positive (negative) $\Sigma$. However, this effect lessens for large-scale, super-horizon perturbations [cf. Fig. \ref{fig:curv}].
\begin{figure*}[!htbp]
    \centering
    \begin{minipage}[b]{0.48\textwidth}
        \centering
        \includegraphics[width=\textwidth]{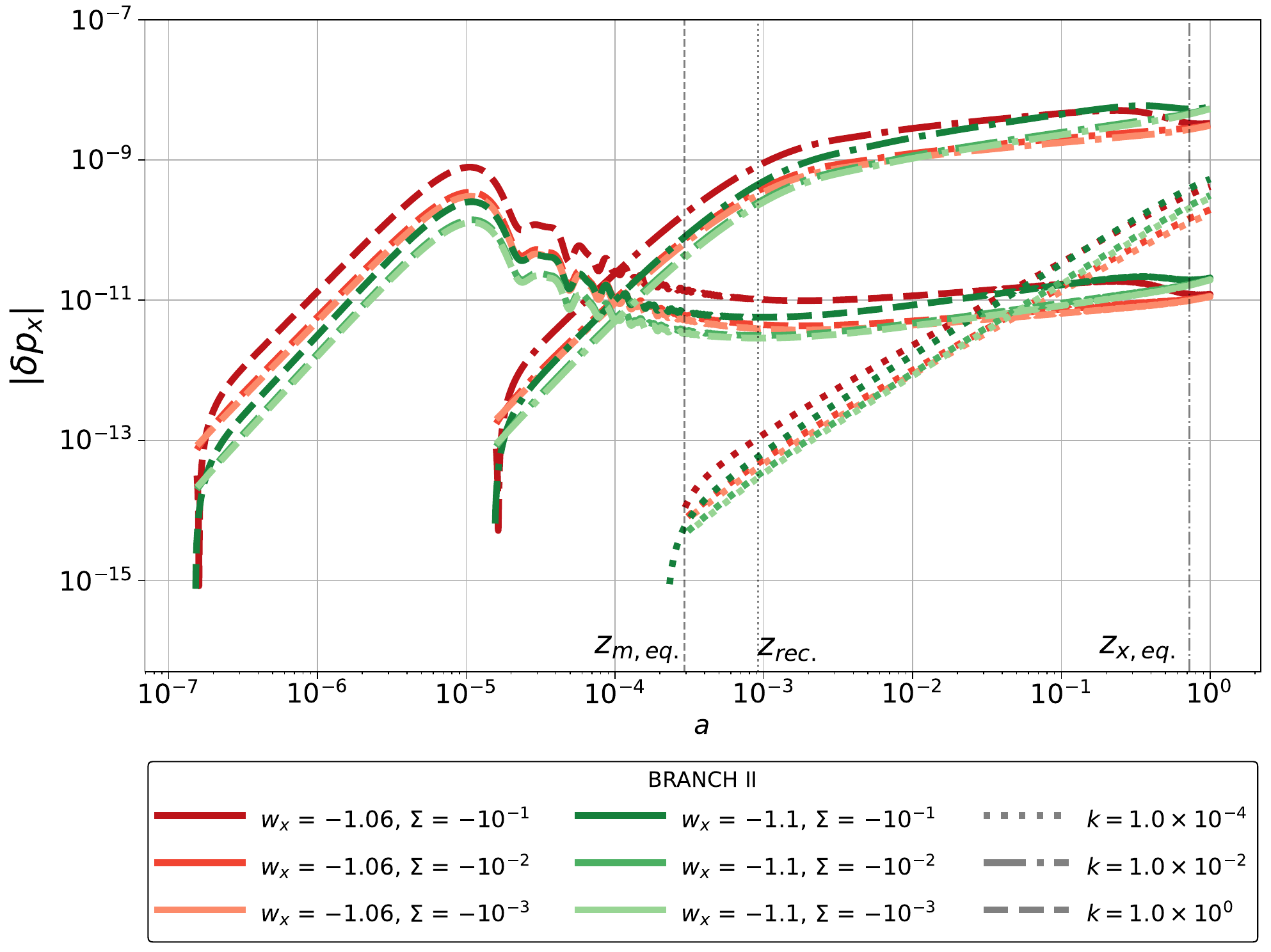}
    \end{minipage}
    \hfill
    \begin{minipage}[b]{0.48\textwidth}
        \centering
        \includegraphics[width=\textwidth]{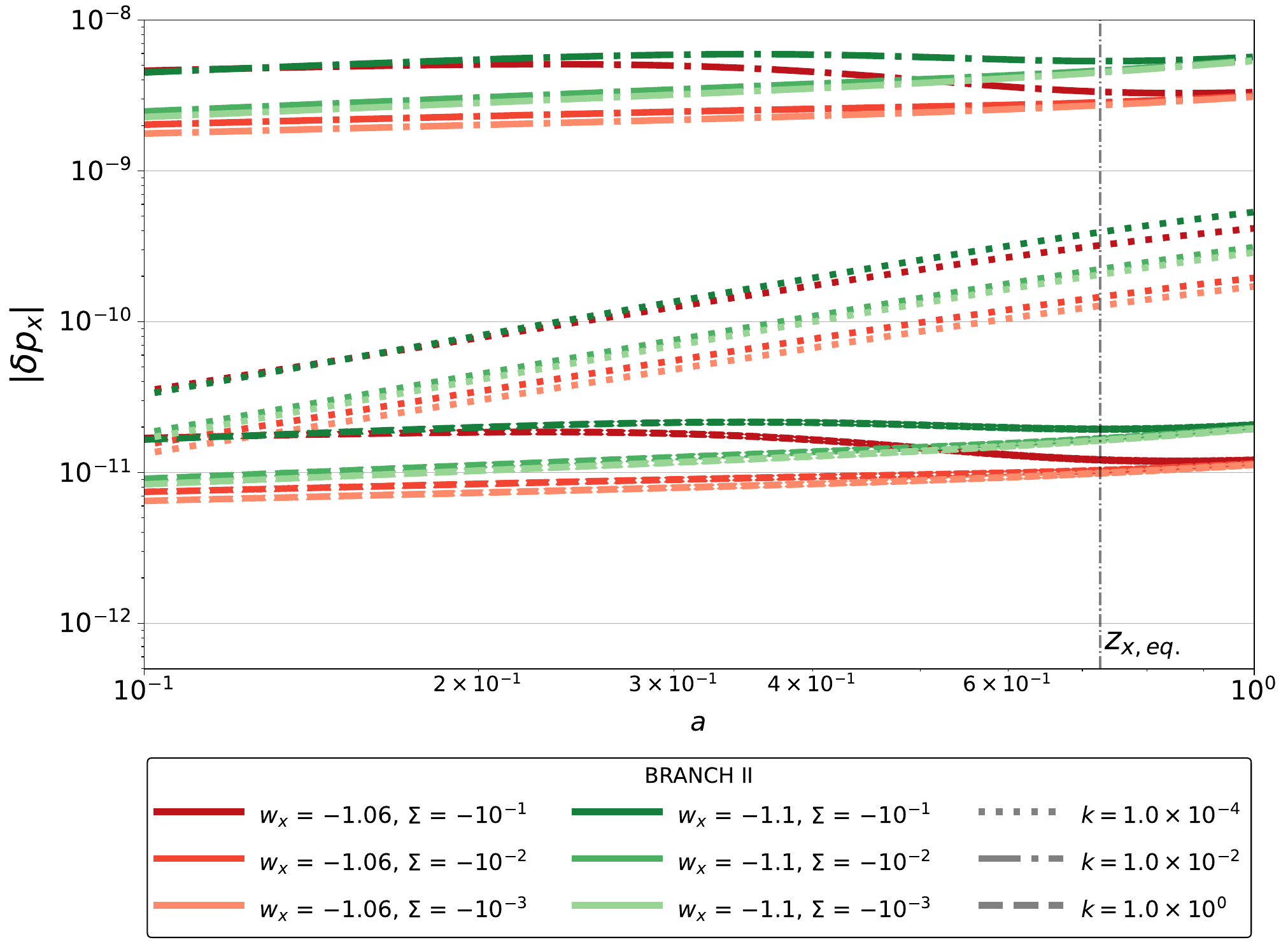}
    \end{minipage}

    \vspace{0.5cm}
    
    \begin{minipage}[b]{0.48\textwidth}
        \centering
        \includegraphics[width=\textwidth]{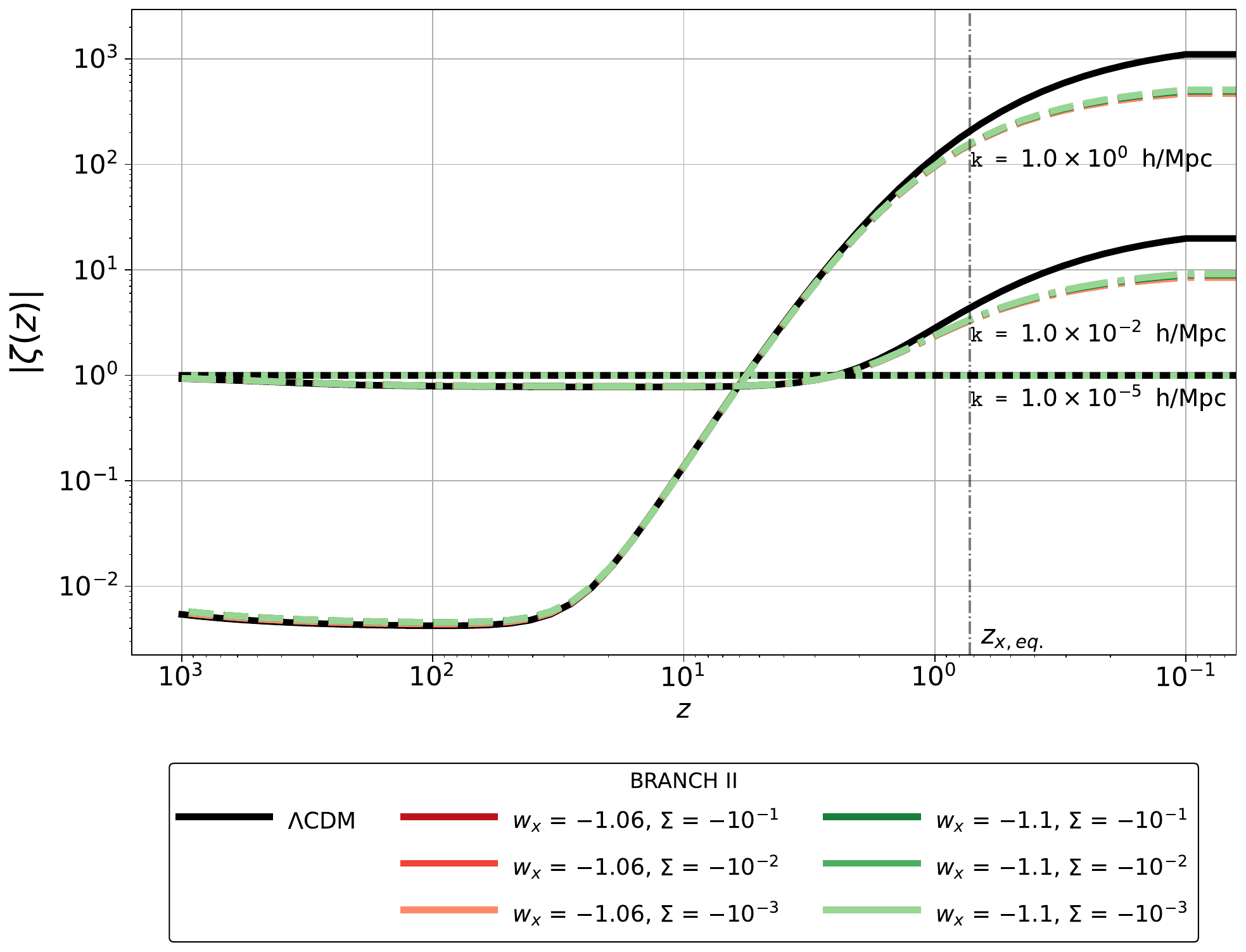}
    \end{minipage}
    \hfill
    \begin{minipage}[b]{0.48\textwidth}
        \centering
        \includegraphics[width=\textwidth]{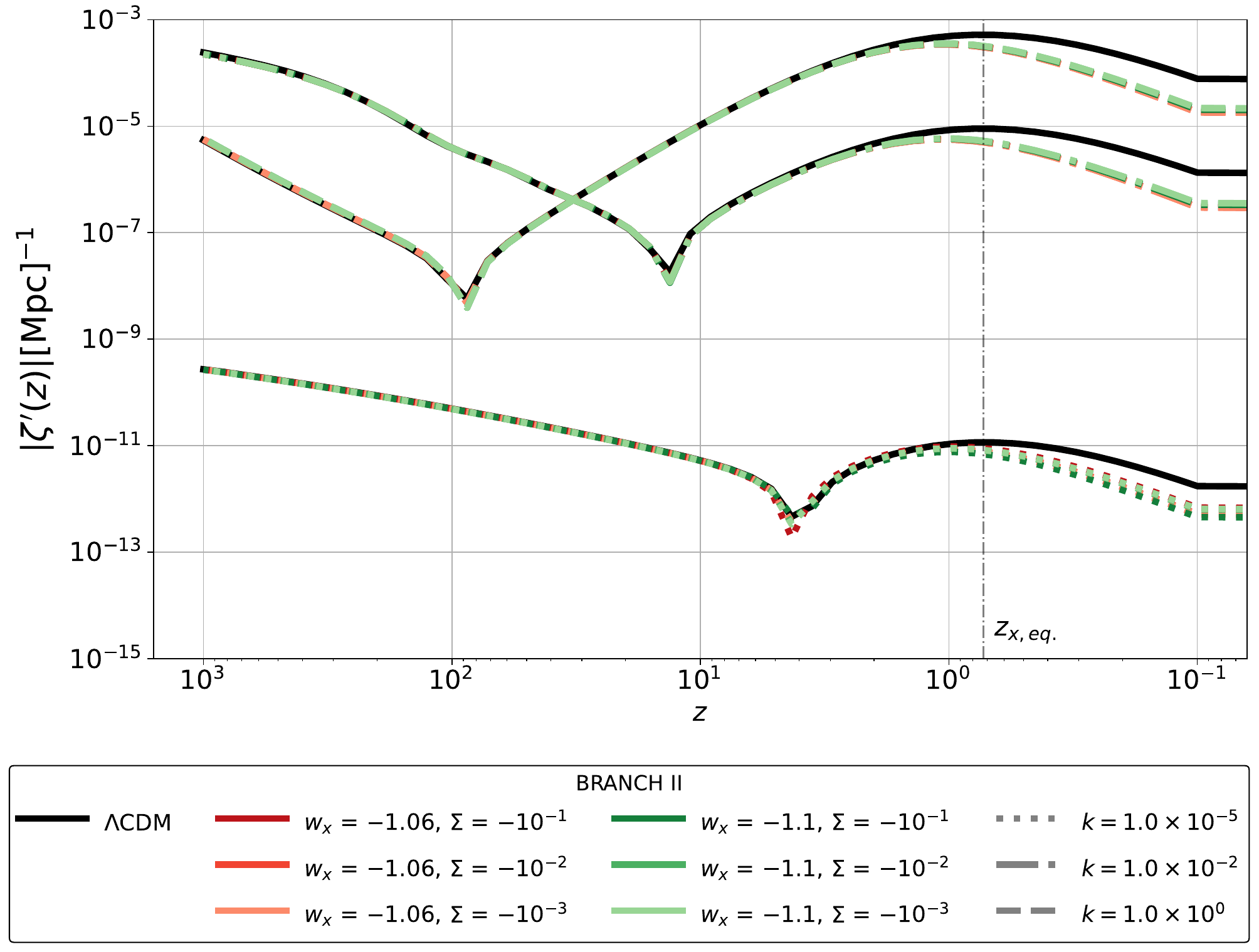}
    \end{minipage}
    
    \caption{Cosmological evolution for models in the \texttt{Frozen-branch-II} of the gravitational potential $\zeta$ (\ref{eq_zeta}) and its derivative with respect to conformal time $\zeta'$, both as a function of redshift $z$, and the dark energy pressure perturbation $\delta p_x$ (\ref{dpx}) as a function of the scale factor $a$. \textbf{Top panel}: On the left, the evolution of the dark energy pressure perturbation, and on the right a zoom-in on late times to highlight the difference in the behavior of different modes. Small-scale perturbations exchange energy earlier, mirroring the behavior of branch I, as present in the strong coupling case for $w_x = -1.1$. \textbf{Bottom panel}: On the left, the gauge-invariant curvature perturbation $\zeta$ since recombination, and on the right a its derivative $\zeta'$. It shows a well-behaved nature, similar to branch I. Large-scale perturbations are super-horizon and do not grow, and the enhancement is suppressed during dark energy domination.}
    \label{fig:curv2}
\end{figure*}

\subsection{ Parameter Ranges and Cosmological Tensions} \label{sec_e}

In this subsection, we present a comprehensive summary of our principal findings, delineating the parameter ranges for each model, discussing their respective observational signatures, and outlining the methodologies for observational validation.

For Branch I, we examine the parameter space defined by  $w_{x} \in [-0.9, -0.8]$ (indicative of a non-phantom equation of state) and $\Sigma \in [10^{-3}, 10^{-1}]$ (representing small coupling). Our analysis reveals that within this range, larger coupling values correspond to a reduction in $S_{8}$, thereby alleviating the $S_{8}$ tension relative to other proposed models. Additionally, we identify a potential amplification of the ISW amplitude, which may be detectable through cross-correlation studies between CMB data and large-scale structure (LSS) observations, or by investigating the signal in the vicinity of significant cosmic structures, such as the Eridanus supervoid \cite{planck2016isw, cspot}. In Branch III, characterized by  $\Sigma \in [-10^{-1},-10^{-3}]$, we find that while the tension associated with $S_{8}$ is mitigated, the amplitude of the ISW effect is diminished when compared to the standard model. For the phantom-like models ( $w_{x} \in [-1.1, -1.06]$) encompassing branches II and IV, we observe analogous trends; namely, an increase in $\Sigma$ leads to a decrease in $S_{8}$ while simultaneously enhancing the ISW effect. In summary, our work highlights two distinct physical signatures that can be rigorously tested through cross-correlation techniques and examinations of superstructures, providing critical insights into the interactions between dark energy and dark matter at late times.

\section{Summary} \label{sec:sum}

This work presents a novel interaction mechanism within the dark sector that remains inactive at the background level but emerges in the perturbative regime. Similarly to a phase transition, it results in a modified cosmological model in which energy exchange becomes significant at late cosmic times. The proposed framework retains the early universe’s resemblance to the standard $\mathrm{\Lambda CDM}$ model, but introduces a distinctive perturbative behavior as the universe evolves. This new mechanism has potential observational consequences in cosmic structures, including the CMB power spectrum, matter clustering, and redshift-space distortions. Although we share some similarities with momentum-transfer models, our approach differs significantly, particularly in its effects on the continuity equations and initial conditions during the radiation era. 

We have shown in Sec. \ref{sec:bg} that the background evolutions directly follow the basic Friedmann equations for a universe that consists of photons (and other relativistic degrees of freedom), baryons, dark matter, and a dark energy fluid that has an equation of state parameter different from a cosmological constant, $w_x \neq -1$, following CPL parameterization \cite{Chevallier:2000qy, Linder:2002et}, i.e. the so-called $w\mathrm{CDM}$ scenario. Then, we incorporate the interaction between dark matter and dark energy in Sec. \ref{sec:cp}, restricting to the perturbation sector. The modified Euler and continuity equations are the foundation for understanding the new interaction introduced in Sec. \ref{sec:nm}. Note that the interaction described by the \texttt{frozen model} is scale dependent, as described in Equations \eqref{IFV} and \eqref{eq:kappa}.

As the background is unchanged from the \texttt{vanilla} scenario, the new complete case we analyze in this paper was named the \texttt{frozen vanilla model}.

The new initial conditions for a system in which there is no interaction between its dark components at the background level, but has it for the perturbative sector, are different from those previously considered in the literature \cite{Ma:1994dv, Valiviita:2008iv, gavela}, as shown in Section \ref{sec:ic}. This allowed us to numerically evolve the Boltzmann equations for selected \texttt{frozen vanilla} models using \texttt{CLASS} \cite{ju1, ju2, ju3}. The selection and characterization of these models, which we have separated into four different branches according to their values of $(1 + w_x)$ and $\Sigma$ -- are detailed in Sec. \ref{subsec:ps}. To summarize, non-phantom models have been designated as branches I and III, while phantom models are in branches II and IV. The other key parameter of the models, $(1 + w_x)$, was chosen on the basis of the latest literature on the interaction of dark energy with dark matter and the most recent data releases, such as DESI.

The main takeaways from the exploratory analysis are:

\begin{itemize}
    \item The CMB power spectra at high and intermediate $\ell$ are sensitive to the parameters of the CPL model ($w_x$), while the effects of the interaction are subdominant  (see Fig. \ref{fig:cl_class_I}).

    \item The interaction influences the ISW effect at low-$\ell$ by altering the evolution of matter and dark energy perturbations. Consequently, the sign of the interaction $\delta Q_x$, which is directly related to $\Sigma$, plays a crucial role in determining how the ISW effect deviates from the standard $\mathrm{\Lambda CDM}$ model, c.f. Fig. \ref{fig:LISW_class_I}.
    In branch I, where $\Sigma > 0$ (negative interaction, with energy flowing from dark energy to dark matter), the ISW effect is already around $15\%$ and $35\%$ for $w_x = -0.9$ and $w_x = -0.8$, respectively. This indicates an enhanced ISW effect (for base-level CPL) relative to $\mathrm{\Lambda CDM}$, Fig. \ref{fig:LISW_class_I}. The percentage difference from $\mathrm{\Lambda CDM}$ is around $50\%$ for strong coupling, $|\Sigma| \sim 10^{-1}$.
    When the interaction is reversed (branch III), the ISW effect weakens with increasing coupling.  A similar trend is observed for phantom models. For negative $\Sigma$, where energy flows from dark matter, the ISW effect is suppressed, exceeding $25\%$ for strong coupling in branch II, Fig. \ref{fig:LISW_class_II}. Conversely, in branch IV, similar to branch III, the trend reverses, shifting from a negative to a positive percentage difference relative to $\mathrm{\Lambda CDM}$. These results are in line with the behavior of $(\Phi + \Psi)'$, which governs how the ISW effect works.
    
    \item For $\Sigma < 0$, the matter power spectrum increases, as this condition implies $\delta Q_x > 0$. Thus, the matter power spectrum is suppressed for larger coupling in branches I and IV (for I, see. Fig. \ref{fig:Pk_class_I}), whereas it is enhanced in branches II and III (for II, see Fig. \ref{fig:Pk_class_II}).
    
    Due to their respective values of $w_x$, the matter power spectrum in branch I (non-phantom) remains consistently below that of $\mathrm{\Lambda CDM}$. In contrast, in branch III (non-phantom, reversed interaction), it exceeds $\mathrm{\Lambda CDM}$ for larger coupling. Similarly, in branch II, the matter power spectrum consistently exceeds that of $\mathrm{\Lambda CDM}$. However, in branch IV, it is suppressed below $\mathrm{\Lambda CDM}$ for larger coupling.
    
    \item Consequently, the $f \sigma_8$ observable
    grows with $\Sigma < 0$, bottom-right plot of Fig. \ref{fig:fs82}, contrary to what happens for $\Sigma > 0$, bottom-right plot of Fig. \ref{fig:fs8}. The deviation to $\mathrm{\Lambda CDM}$ also grows in time, consistent with the fact that the interaction is relevant mainly during dark energy domination. We stress that this observable is far more sensitive to the interaction than others (for example CMB).  This behavior is important to the data analysis of the scenario -- which will be present in a following work -- as it breaks the degeneracy to vanilla CPL models. 
    
    As expected from the matter power spectrum results, the deviation to $\mathrm{\Lambda CDM}$ is stronger for branches I and II, where there is no reversion of the percentage difference to $\mathrm{\Lambda CDM}$. Whereas in branches III and IV the interaction leads to clustering more in line with the $\mathrm{\Lambda CDM}$ predictions.

   The interaction is scale-dependent, and, consequently, the growth function and growth rate will be scale-dependent as well, Figs. \ref{fig:fs8} and \ref{fig:fs82}. However, this dependence is small at low redshifts. This allows one to analyze the $f \sigma_8$ observable at the relevant range of redshifts, $ 0 < z < 2$, as detailed in the previous item.
    
    \item The dark energy pressure perturbation, differently from the results for the matter power spectrum, grows for stronger coupling for both signs of $\Sigma$ for the unusual case present in the literature -- where $\Sigma$ and $(1 + w_x)$ have the same signs, as in branches I and II, Figs. \ref{fig:curv} and \ref{fig:curv2}. On the other hand, for branches III and IV, the pressure perturbation decreases with $|\Sigma|$.

    \item The gravitational potential $\zeta$ and its conformal time derivative, $\zeta'$ are well behaved for every coupling and scale and are closely related to the $\mathrm{\Lambda CDM}$ results. They only differ, albeit slightly, during dark energy domination, solely for small and intermediate scales. That leads to a smaller value of both $\zeta$ and $\zeta'$ with respect to $\mathrm{\Lambda CDM}$ at those scales, following the expected result of CPL cosmology. 
    
    At large scales, which are super-horizon, for $\zeta$ the effect is almost unnoticeable. It can only be inspected using $\zeta'$. We conclude that the behavior is the same for all scales.

\end{itemize}

In conclusion, our findings suggest that the frozen vanilla model offers a promising approach to addressing modern cosmological challenges, such as the $S_8$ tension. The model is grounded in a robust $\delta$-ET framework and is strongly motivated by recent observations. In future work, we will perform a Bayesian analysis of this scenario, incorporating state-of-the-art cosmological data -- such as DESI -- to determine whether the frozen model is favored over $\mathrm{\Lambda CDM}$ and other dark matter–dark energy interaction models. In this context, we anticipate incorporating a combination of multiple datasets to effectively constrain $w_x$ and $\Sigma$. A comprehensive MCMC analysis will be undertaken that encompasses both high-redshift and low-redshift datasets. 
To achieve this, we will integrate observations from SNe Ia, BAO, CMB and  $f\sigma_{8}$. Given the distinct imprints this model leaves on the evolution of cosmic structure,  $f\sigma_8$ is expected to be the most powerful observable for placing stringent constraints on the model parameters.  Finally, we will compute the Bayesian evidence to quantify which models are statistically favored by the datasets in comparison to the vanilla model.

\begin{acknowledgments}
The exploratory analysis was performed using a modified version of the public repository \texttt{CLASS} code\footnote{\url{https://github.com/lesgourg/class_public}}. The authors thank CNPq, FAPES, and Fundação Araucária for financial support.  S.L. is supported by grant PIP 11220200100729CO CONICET and grant 20020170100129BA UBACYT.

\end{acknowledgments}
\appendix

\section{Appendix A: Microscopic Model details}\label{appendix:a}
In Section \ref{sec:nm}, we have applied the effective theory (ET) method at the level of perturbed field equations and energy-momentum conservation. Nonetheless, the fundamental concepts are equally valid for both the Lagrangian formulation and the perturbed field equations. As a potential microscopic realization of the  $\delta$-ET model, we could consider the dynamics of $N$ interacting fields in conjunction with fluids through a Lagrangian framework. The latter framework can be articulated as the sum of background terms and several perturbative contributions.  We write the Lagrangian as, $\mathcal{L}_{\delta-\rm{EFT}}=\mathcal{L}_{\rm{b}}[\phi_{A}, \partial_{\mu}\phi_{A}]+\mathcal{L}_{\rm{fluids}}[n, \rho, u^{\mu}]+ \mathcal{L}_{\rm{int}}[n,  \rho, u^{\mu},\phi_{A}]$, where $n$ stands for the fluid number density, $\rho$ is the fluid density, and the four-fluid velocity is $u^{\mu}$. Here  $\mu$ denotes a spacetime index and $A$  serves as an index representing an internal color space.  The model involves a non-trivial combination of interacting scalar fields and fluids. The  structure of $\mathcal{L}_{\rm{int}}$ reveals the intricate dependencies and couplings that emerge in the presence of these perturbations, thereby shaping the dynamics of the field interactions and influencing the overall behavior of the cosmological perturbations. In fact, we could potentially add a smooth Heaviside function  to  make sure that the interacting terms, $\mathcal{L}_{\rm{int}}[n,  \rho, u^{\mu},\phi_{A}]$, are only relevant at late times. In any case, the specific form of $\mathcal{L}_{\rm{int}}$ will account for the $\delta Q_{x}$ terms in the continuity equations; thus,  mixed terms involving  $u^{\mu}$ and $\nabla_{\mu}\delta\phi_{A}$ have to be included.  While our procedure differs from that outlined in \cite{Aoki:2025bmj}, we do share certain similarities. Looking ahead, we will introduce a model grounded entirely in a Lagrangian formulation, which will be explored further in the context of the standard effective field theory approach as discussed in \cite{skordis}, \cite{Aoki:2025bmj}.

\section{Appendix B: The relation between the synchronous gauge and the cold dark matter frame}
\label{appendix:b}
In this section, we begin by mentioning the basic features of a scalar gauge transformation, along with the transformation rules for the metric components and the energy-momentum tensor. An infinitesimal gauge transformation can be written as $\tilde{x}^{\mu}=x^{\mu} + \epsilon^{\mu}$, with $| \epsilon^{\mu}| \ll1$. In particular, for a scalar gauge transformation, the latter vector takes the following form: $\epsilon^{\mu}(x, \tau)= (\epsilon^{0}, -\partial^{i}\epsilon^{s})$, $\nabla \times \nabla \epsilon^{s}=0$.  

Under a gauge transformation, the metric changes as follows  \cite{Kodama:1984ziu}:
\begin{eqnarray}
\delta \tilde{g}_{\mu\nu}(\tilde{x})=\delta g_{\mu\nu}(x)+ a^{2}[-\epsilon^{\rho}_{,\mu} \eta_{\rho\nu} -\epsilon^{\sigma}_{,\nu} \eta_{\mu\sigma}-2{\cal{H}}\eta_{\mu\nu} \epsilon^{0} ],~~~~~~~
\end{eqnarray}
where $\eta_{\rho\nu} $ denotes the Minkowski metric.  After performing a Fourier transformation on all the perturbed variables and the gauge scalar transformation as follows: $\epsilon^{0}(x, \tau)=\epsilon^{0}_{k}(\tau)e^{+ikx}$,  $\epsilon^{s}(x, \tau)=\epsilon^{s}_{k}(\tau)e^{+ikx}/k$, the gauge transformation rules for the individual components of the 4-scalar are:
\begin{eqnarray}
\tilde{\phi}_{k}&=& \phi_{k}-\epsilon'^{0}_{k} - {\cal{H}}\epsilon^{0}_{k},\\
\tilde{B}_{k}&=&B_{k}-\epsilon'^{s}_{k} +k\epsilon^{0}_{k},\\
\tilde{E}_{k}&=&E_{k}-k\epsilon^{s}_{k} ,\\
\tilde{\psi}_{k}&=&\psi_{k}+{\cal{H}}\epsilon^{0}_{k}. 
\end{eqnarray}
The synchronous gauge is obtained by fixing $\tilde{\phi}_{\text{sync},k}=0$ and $\tilde{B}_{\text{sync},k}=0$ in the above equations, while $\tilde{E}_{\text{sync},k}=-(\frac{h}{2}+3\eta)/k^2$ and $\tilde{\psi}_{\text{sync},k}=\eta$ \cite{Ma:1994dv}.  The former equalities determine how the gauge scalar functions must be selected; in fact, they represent first-order differential equations for  $\epsilon^{0}_{k}$ and $\epsilon^{s}_{k}$ :
\begin{eqnarray}
\epsilon'^{0}_{k} + {\cal{H}}\epsilon^{0}_{k}&=&\phi_{k},\label{gsys1}\\
\epsilon'^{s}_{k} +k\epsilon^{0}_{k}&=&-B_{k}\label{gsys2}
\end{eqnarray}
The general solutions associated with the the system of equations (\ref{gsys1}-\ref{gsys2}) is given by: 
\begin{eqnarray}
\epsilon^{0}_{k} &=&\frac{C_{1}}{a}+ \frac{1}{a} \int{d\tau a \phi_{k}(\tau)} , \label{c1}\\
\epsilon^{s}_{k} &=& C_{2}-\int{d\tau [B_{k}(\tau)+ k\epsilon^{0}_{k}(\tau)] } \label{c2}.
\end{eqnarray}
One way to understand the solutions displayed in Eqs.  (\ref{c1}-\ref{c2}) is done by noting that the synchronous gauge is not completely fixed; in general, there are two arbitrary constants that must be chosen to remove this residual freedom. This fact can also be verified by integrating the Euler equation for the cold dark matter component. The lack of momentum transfer between dark matter and dark energy leads to: $\theta'_{\text{cdm}} + \mathcal{H}\theta_{\text{cdm}}=0$, whose solution is given by $\theta_{\text{cdm}}=\theta_{0}(k, \tau_{i})/a$ .  Choosing $\theta_{0}(k, \tau_{i}) = 0$ is equivalent to performing a scalar gauge transformation in which cold dark matter is at rest, corresponding to $C_{1} = 0$ \cite{Ma:1994dv}. This point can be double-checked by performing two consecutive scalar gauge transformations and using the transformation rule for the potential velocity ($\tilde{v}_{k}=v_{k} +\epsilon'^{s}_{k}$). The initial curvature perturbation (amplitude) allows us to fix $C_{2}$. Nevertheless, the possibility of choosing a cold dark matter comoving frame becomes not possible once the dark components exchange momentum. For this reason, an option to fix this residual freedom is to choose the center of mass of the dark sector ($v_{T} = 0$) \cite{mateo2}.

\section{Appendix C: Density contrast and Peculiar Velocities} \label{appendix:c}

In this appendix, we present the results for the density contrasts of dark energy, dark matter, radiation, and photons, as well as the velocity perturbations -- the relative velocity to dark matter -- for dark energy, baryons, and photons.

\begin{figure*}[!htbp]
    \centering
    \begin{minipage}[b]{0.48\textwidth}
        \centering
        \includegraphics[width=\textwidth]{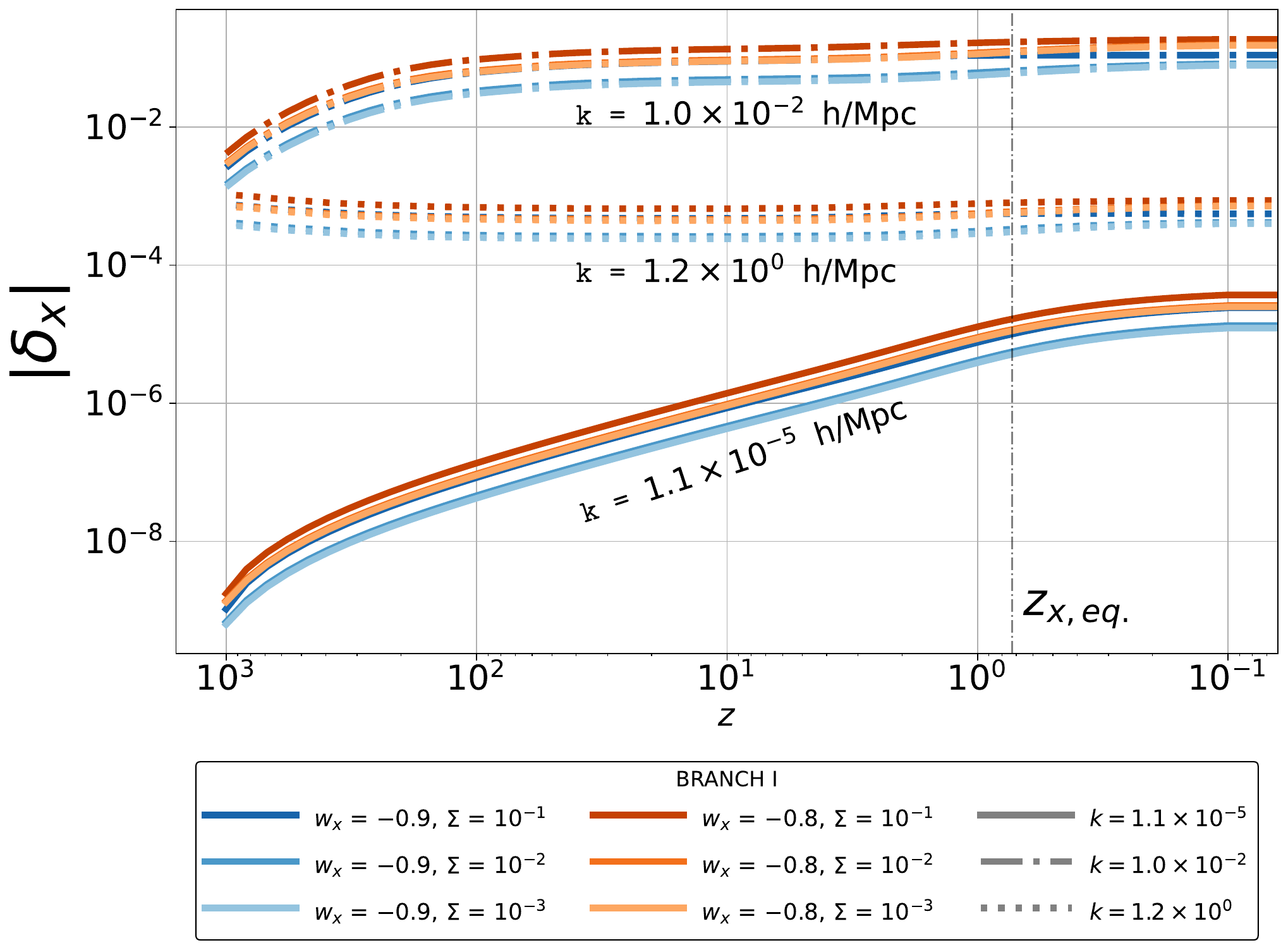}
    \end{minipage}
    \hfill
    \begin{minipage}[b]{0.48\textwidth}
        \centering
        \includegraphics[width=\textwidth]{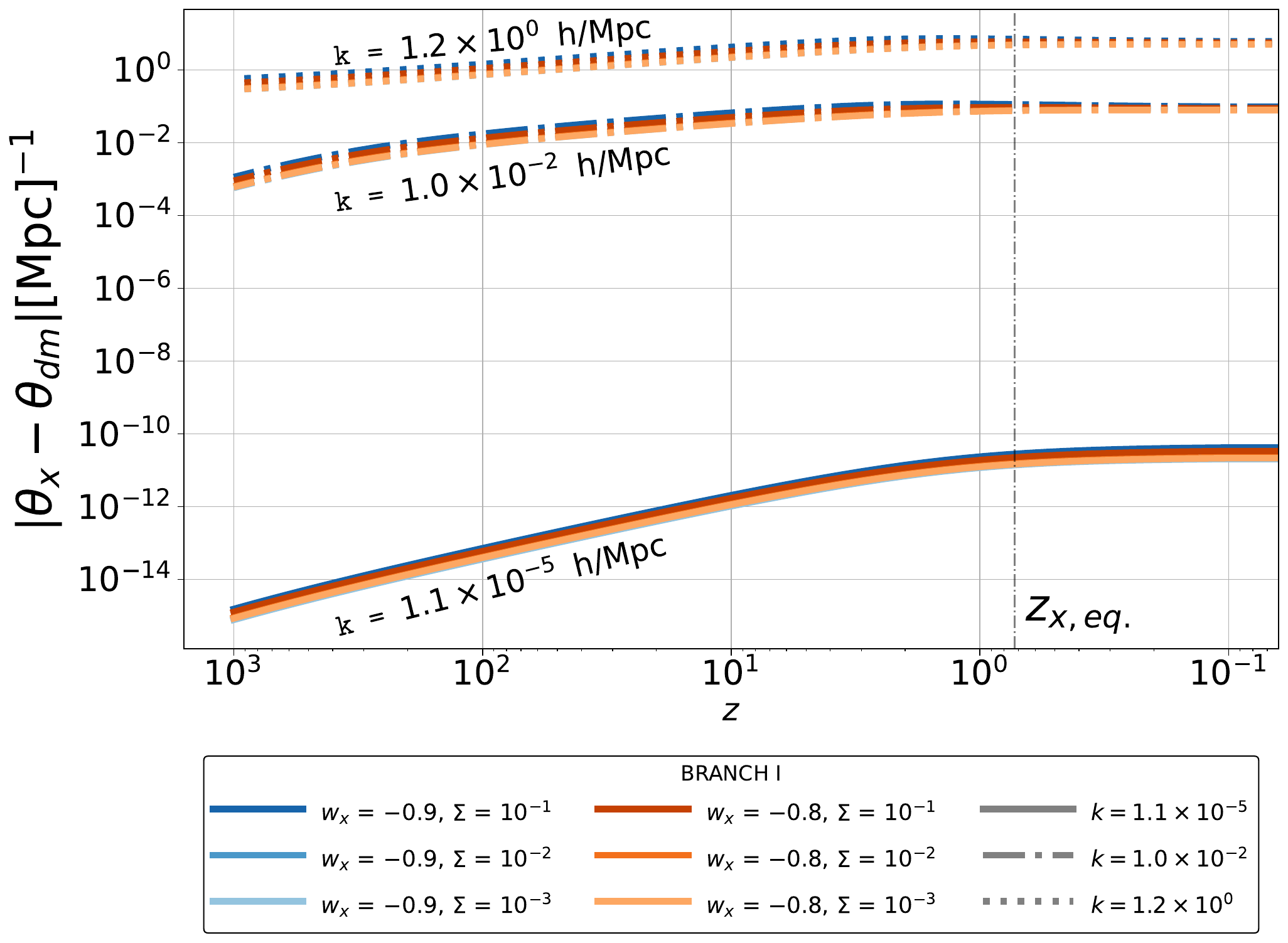}
    \end{minipage}

    \vspace{0.5cm}
    
    \begin{minipage}[b]{0.48\textwidth}
        \centering
        \includegraphics[width=\textwidth]{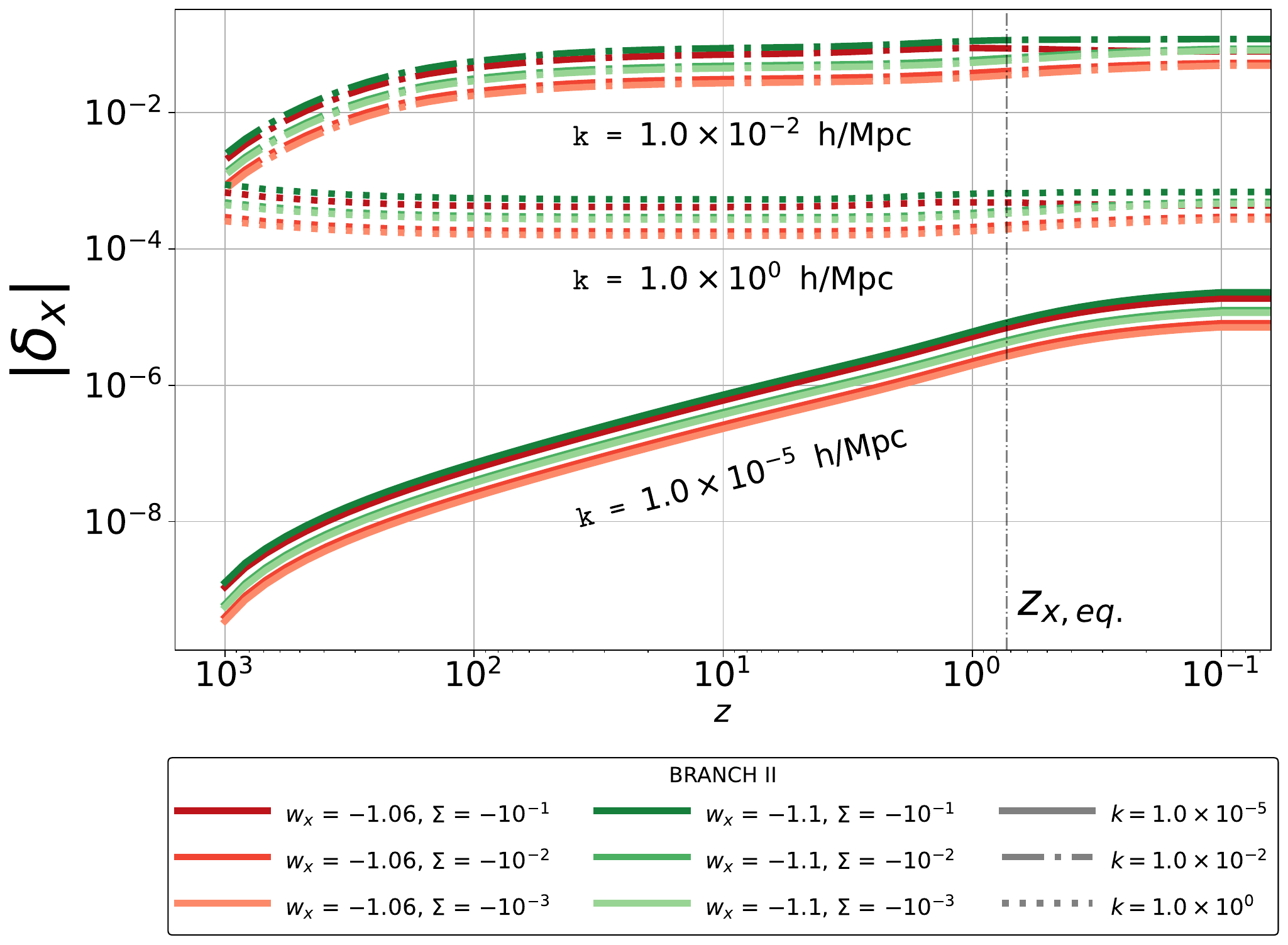}
    \end{minipage}
    \hfill
    \begin{minipage}[b]{0.48\textwidth}
        \centering
        \includegraphics[width=\textwidth]{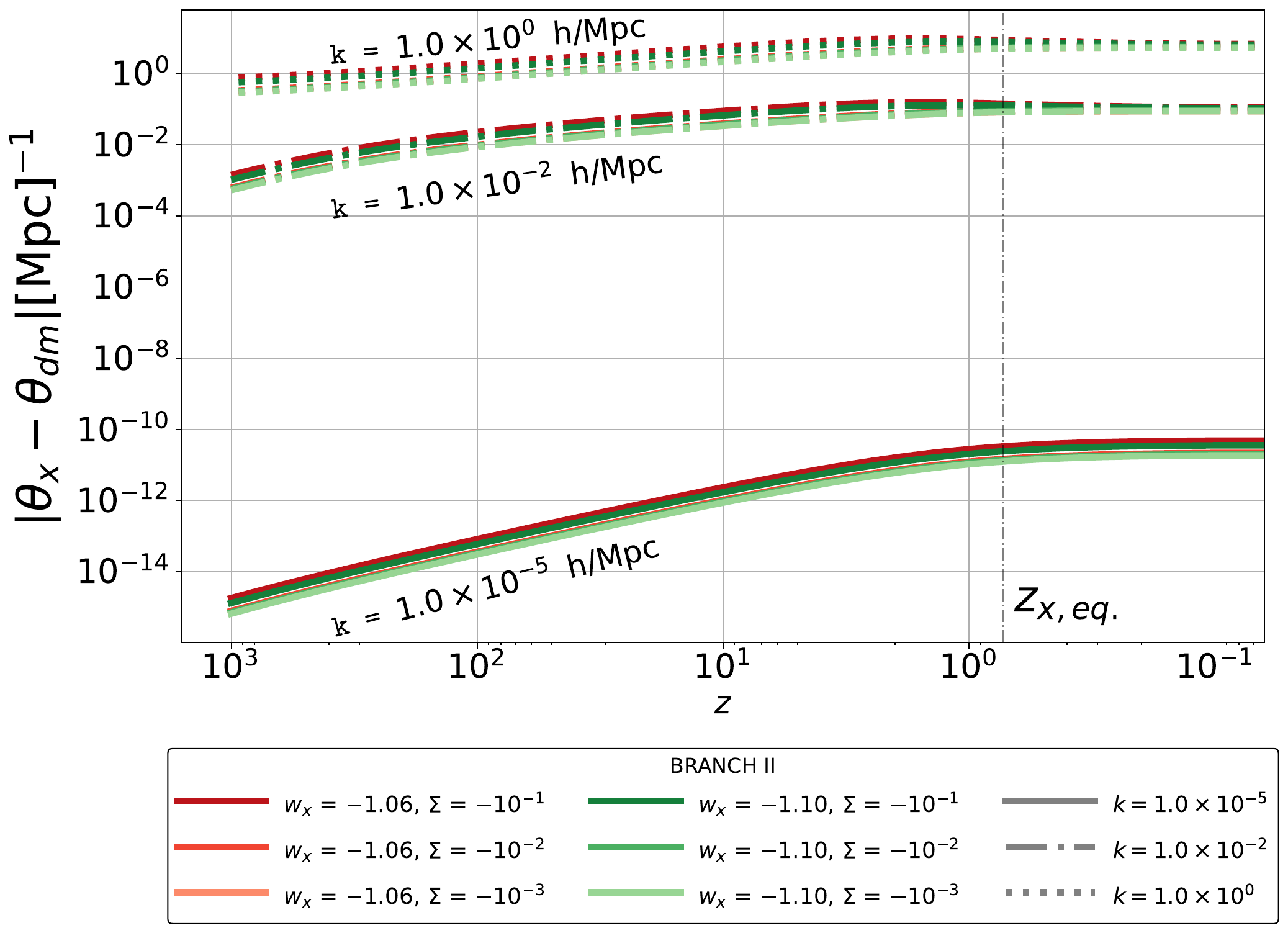}
    \end{minipage}
    
    \caption{\textbf{Top panel}: \textit{Left}: The cosmological evolution of the density contrast for dark energy for \texttt{Frozen-branch-I} models. The density contrast with the magnitude of coupling $|\Sigma|$. \textit{Right}: The relative velocity between dark energy and dark matter for \texttt{Frozen-branch-I}. It also grows with $|\Sigma|$. \textbf{Bottom panel}: \textit{Right}: Density contrast of dark energy for branch II. Note that for late times, after dark energy domination, the sign of $\delta_x$ changes. \textit{Right}: Relative velocity between dark energy and dark matter for branch II. Once again, it grows with $\Sigma$. }
    \label{fig:DELTA_THETA_X}
\end{figure*}

First, we analyze the results for the dark energy component in Fig. \ref{fig:DELTA_THETA_X}. For branches I (pictured in the top left panel) and III, $\delta_x$ is positive, while for branches II (pictured in the bottom left panel), it is negative.

In branches I and II, higher values of $\Sigma$ (in module) lead to an increased contrast in dark energy density,$|\delta_{x}|$.

Concerning the velocity perturbation, for both branches, $\theta_x$ is negative and larger $|\Sigma|$ leads to an increased magnitude of relative velocity in the dark sector. This indicates a loosening of the coupling between these components. In addition, the dark energy velocity is an important factor in the intrinsic pressure perturbation. The contribution is inherently scale dependent and is largely suppressed for intermediate- and large-scale modes with $k^2$.   

\begin{figure*}[!htbp]
    \centering
    \begin{minipage}[b]{0.45\textwidth}
        \centering
        \includegraphics[width=\textwidth]{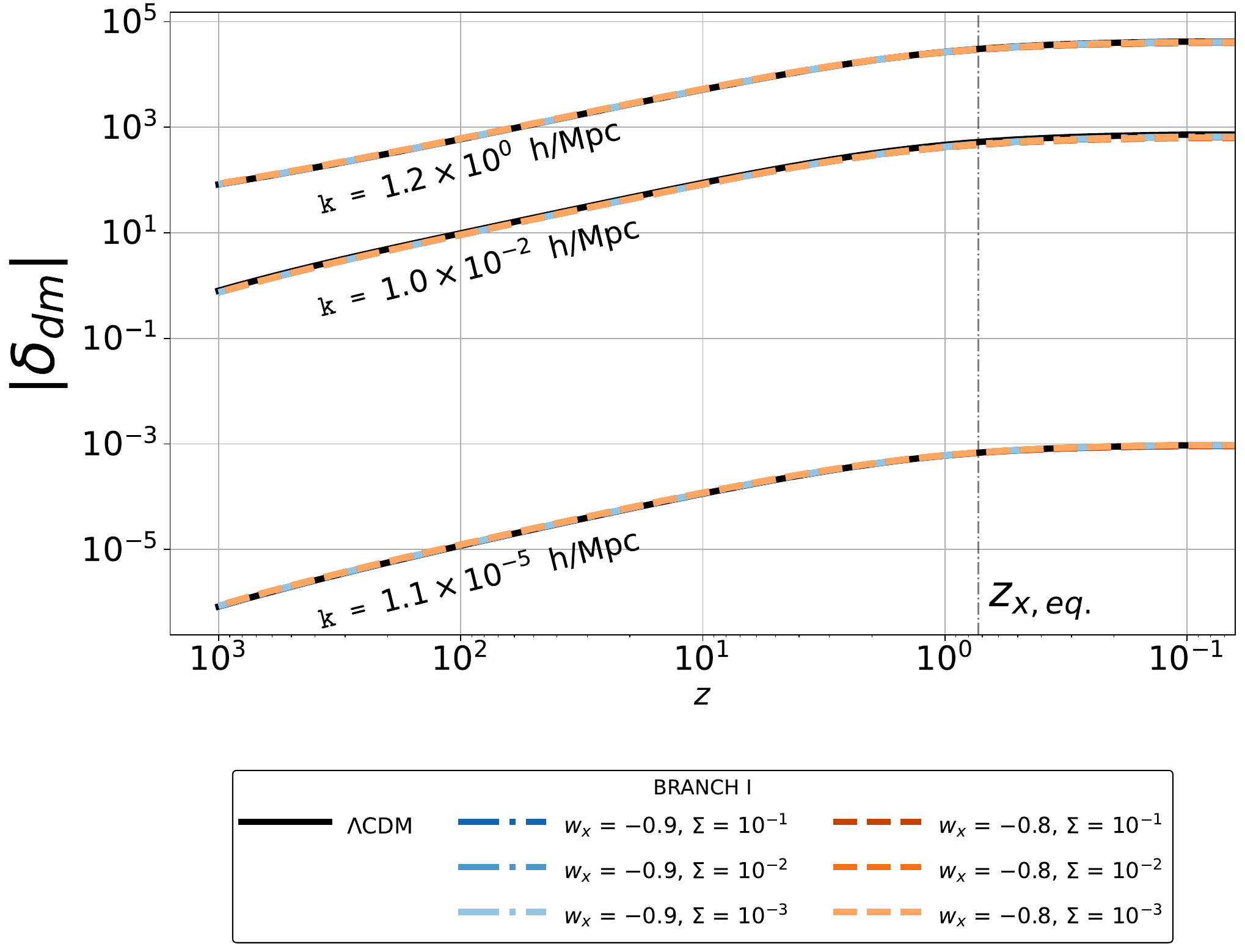}
    \end{minipage}
    \hfill
    \begin{minipage}[b]{0.45\textwidth}
        \centering
        \includegraphics[width=\textwidth]{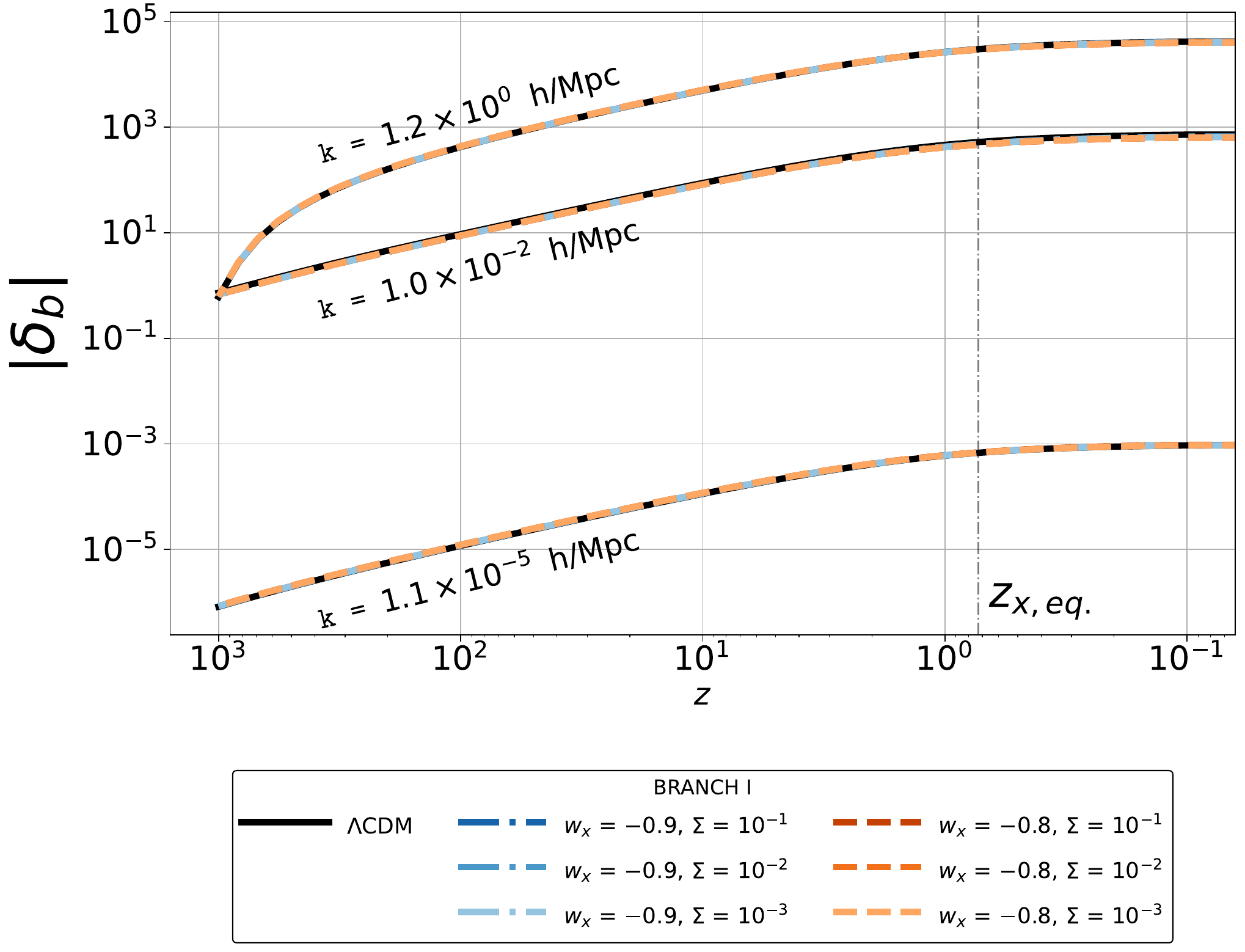}
    \end{minipage}

    \vspace{0.5cm}

    \begin{minipage}[b]{0.45\textwidth}
        \centering
        \includegraphics[width=\textwidth]{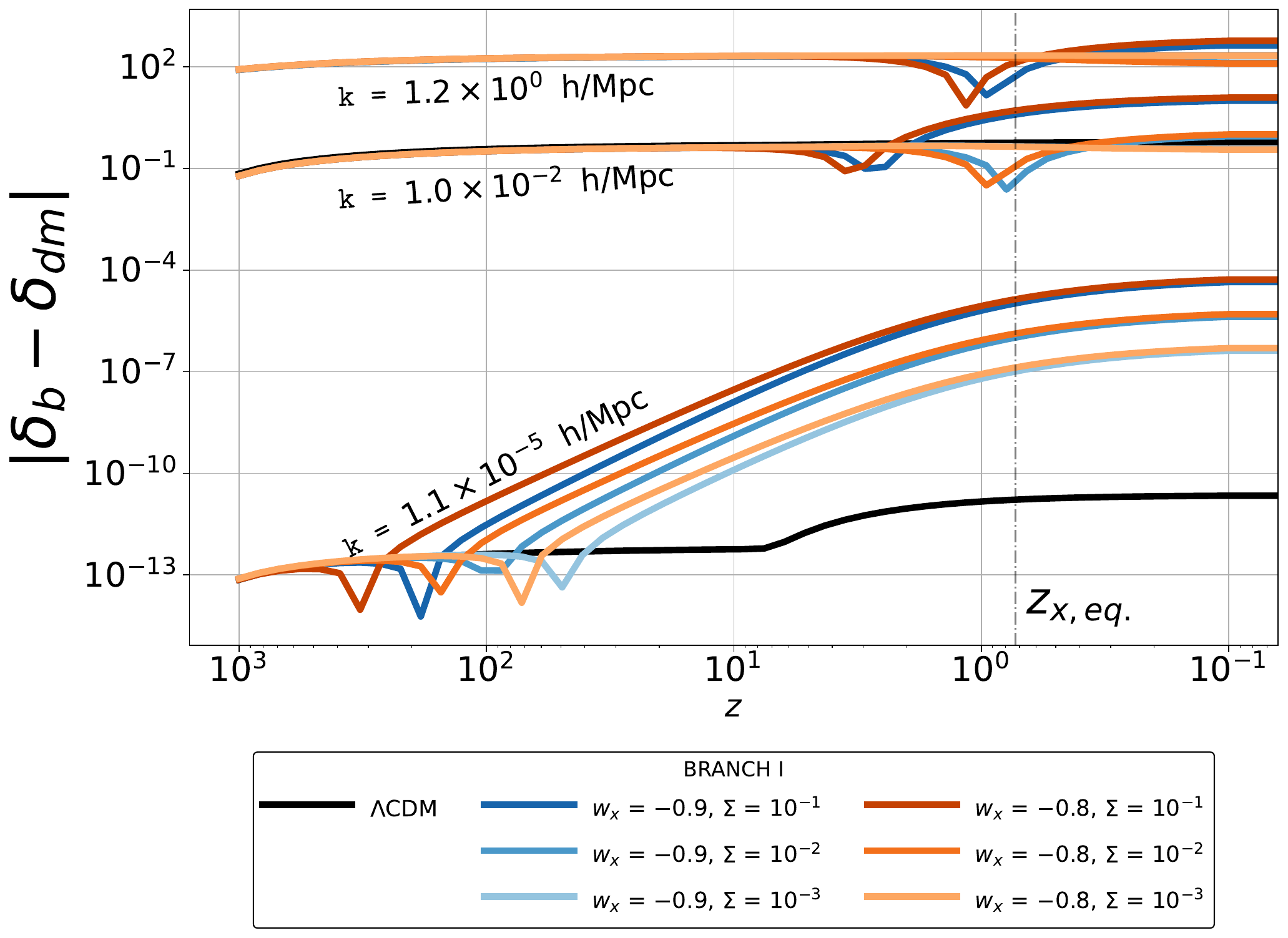}
    \end{minipage}
    \hfill
    \begin{minipage}[b]{0.45\textwidth}
        \centering
        \includegraphics[width=\textwidth]{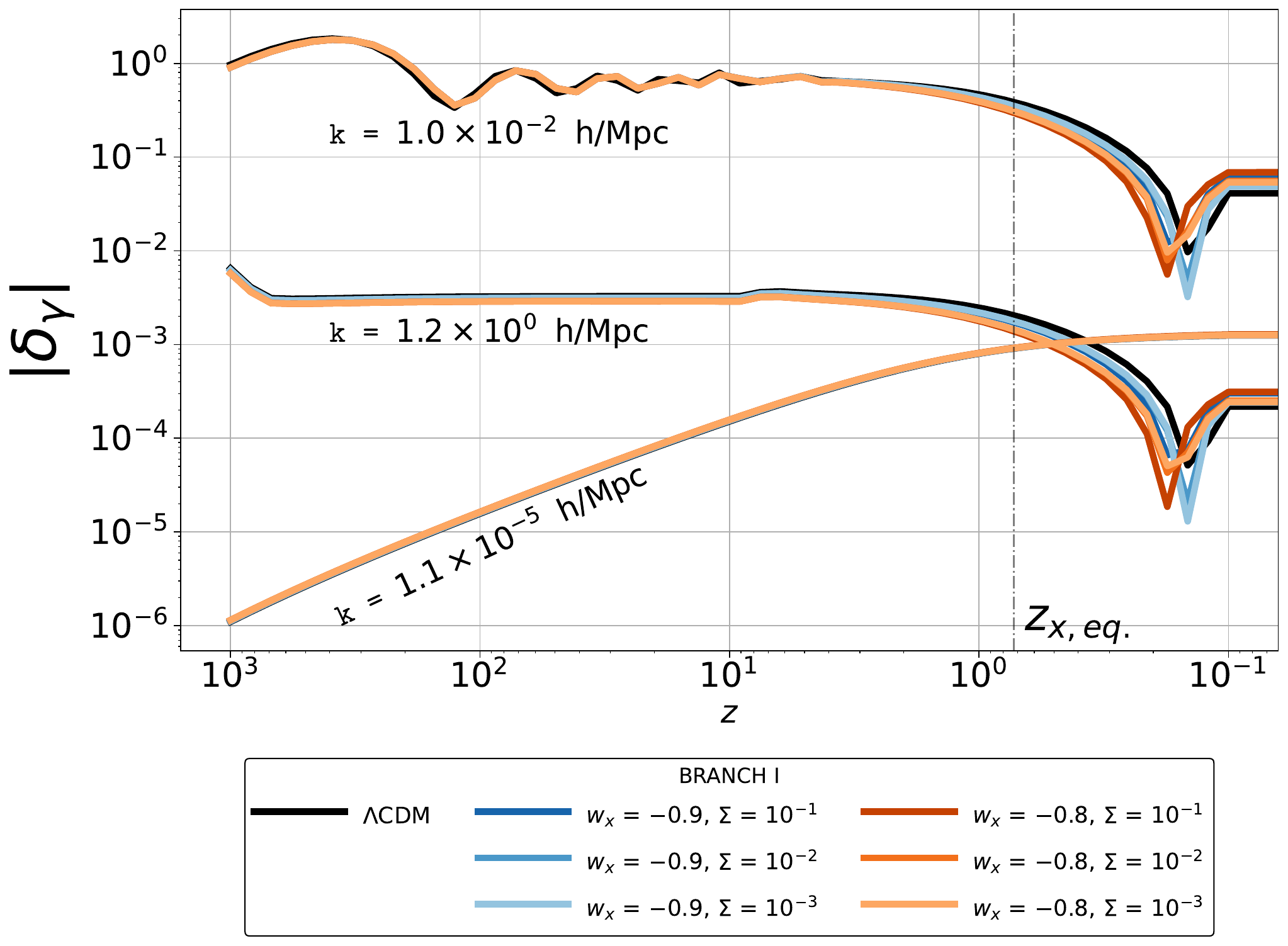}
    \end{minipage}
    \caption{The density contrasts for dark matter, baryons, their relative contrast, and the density contrast for photons for \texttt{Frozen-branch-I} models.}
    \label{fig:DELTAS-BRANCH-I}
\end{figure*}

\begin{figure*}[!htbp]
    \centering
    \begin{minipage}[b]{0.45\textwidth}
        \centering
        \includegraphics[width=\textwidth]{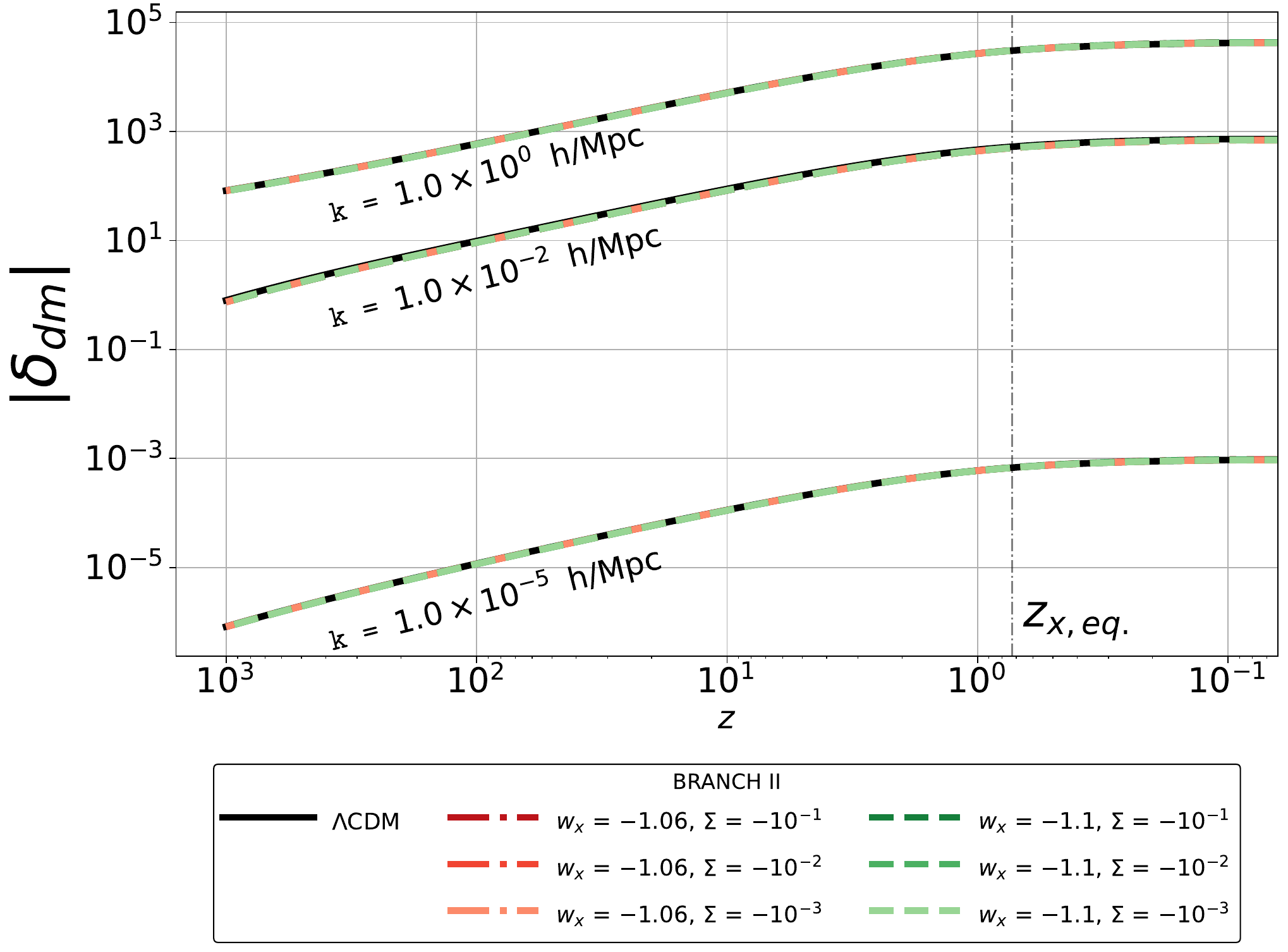}
    \end{minipage}
    \hfill
    \begin{minipage}[b]{0.45\textwidth}
        \centering
        \includegraphics[width=\textwidth]{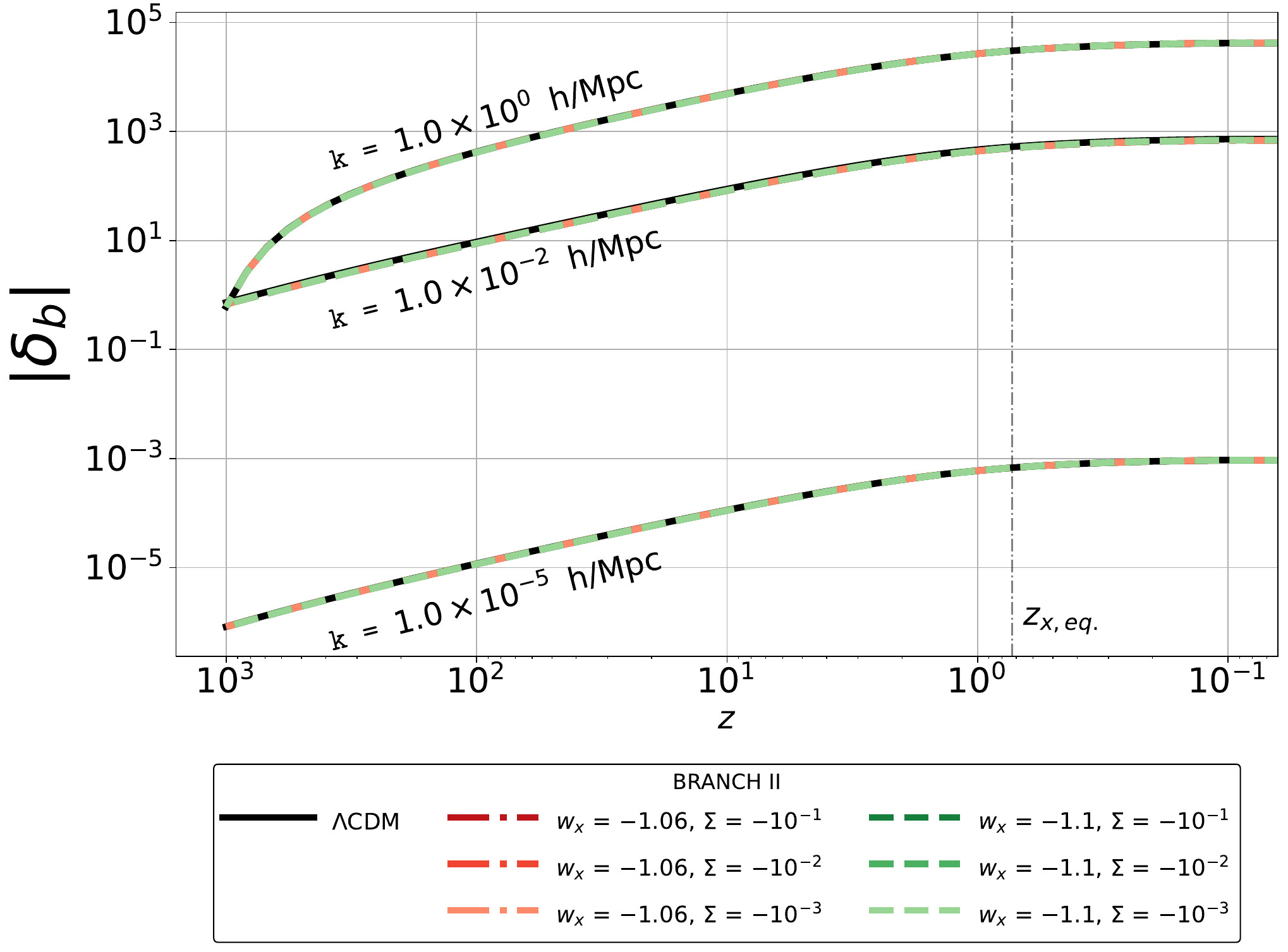}
    \end{minipage}

    \vspace{0.5cm}

    \begin{minipage}[b]{0.45\textwidth}
        \centering
        \includegraphics[width=\textwidth]{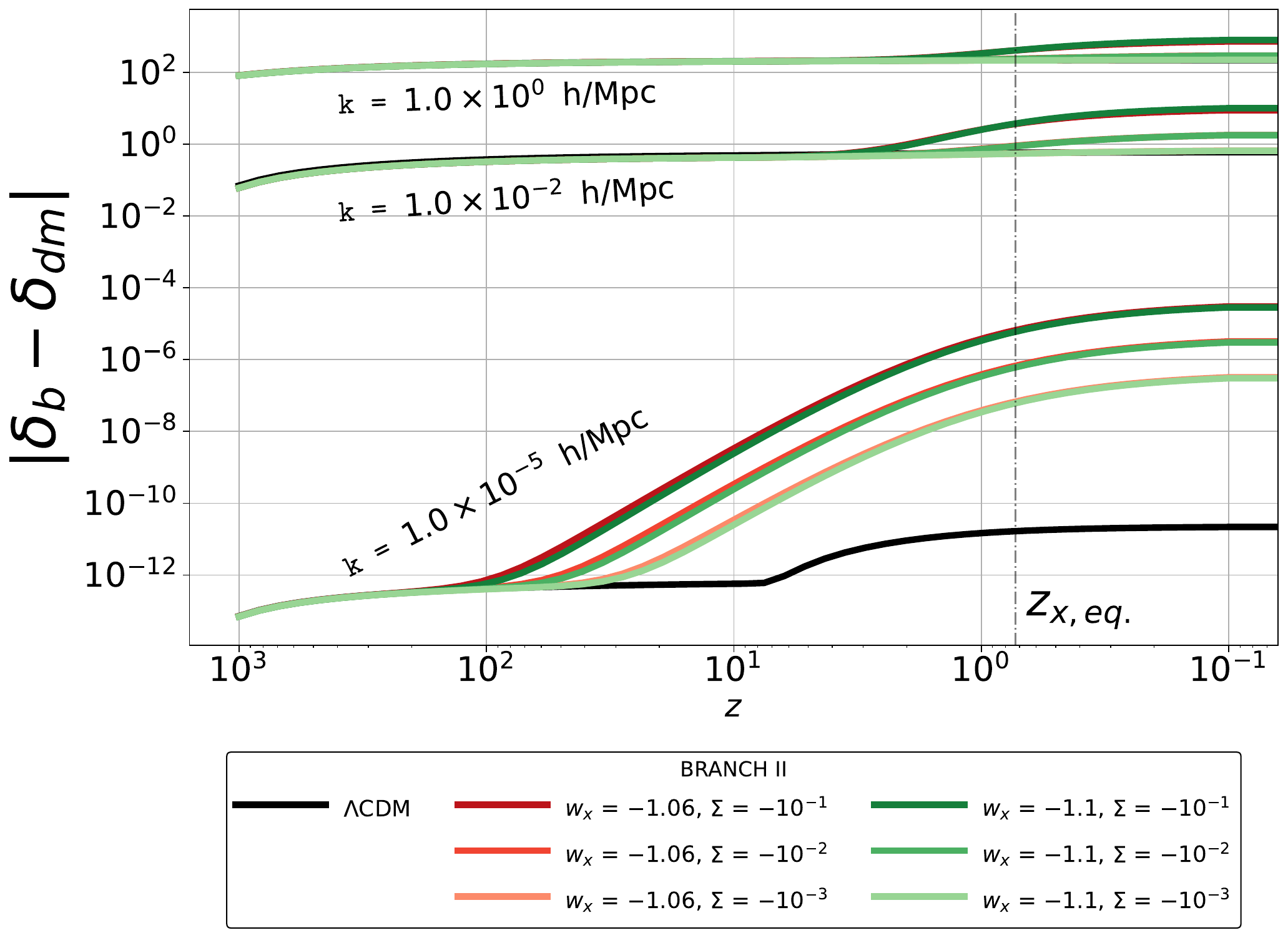}
    \end{minipage}
    \hfill
    \begin{minipage}[b]{0.45\textwidth}
        \centering
        \includegraphics[width=\textwidth]{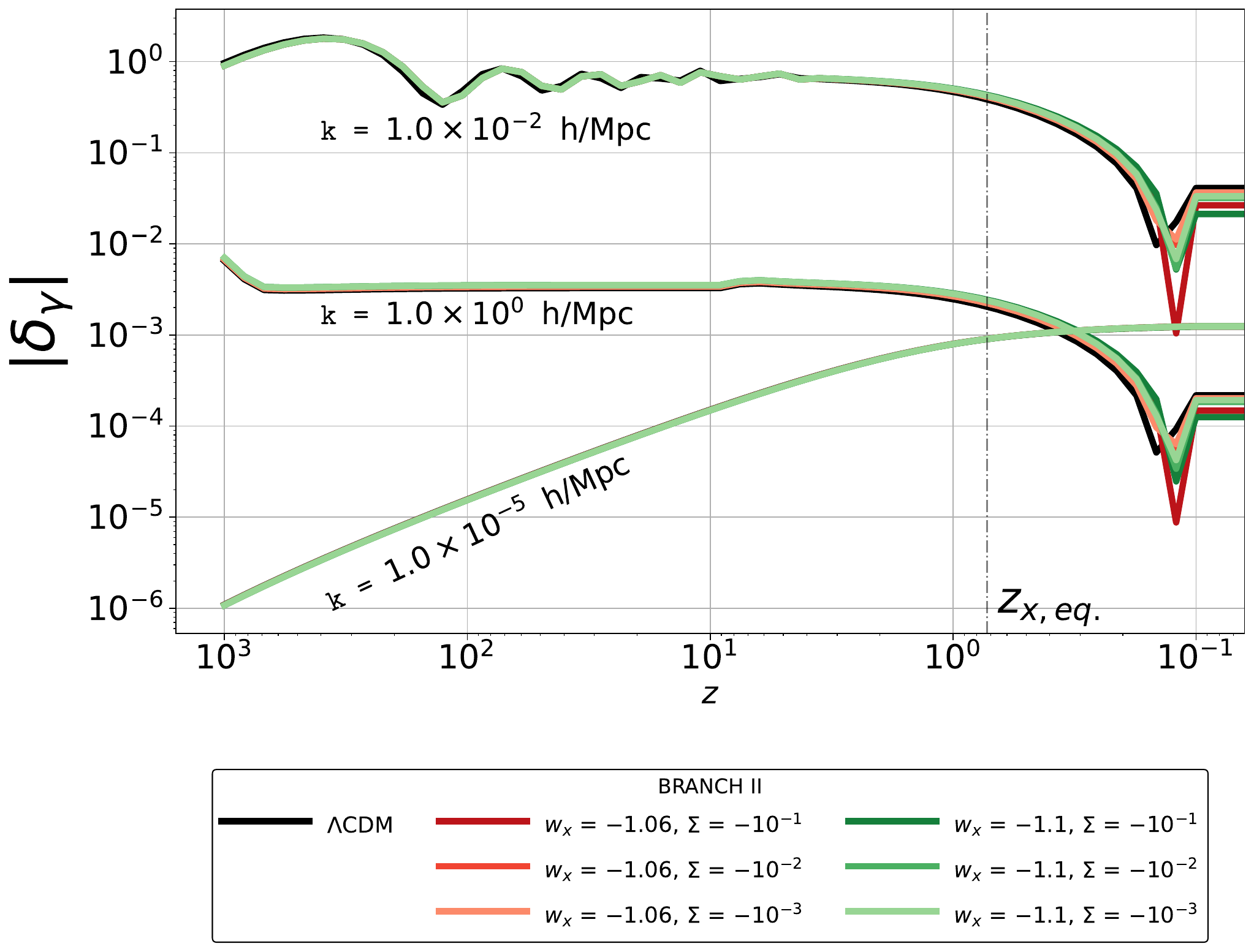}
    \end{minipage}
    \caption{The density contrasts for dark matter, baryons, their relative contrast, and the density contrast for photons for \texttt{Frozen-branch-II}.}
    \label{fig:DELTAS-BRANCH-II}
\end{figure*}

The dark matter and baryon density contrasts are represented for branches I and II in the top row of Figs. \ref{fig:DELTAS-BRANCH-I} and \ref{fig:DELTAS-BRANCH-II}. No notable feature is present: the amplitude of the density contrast grows in time for every model and scale. $\delta_{dm}$ is negative, representing underdensities (voids). The same applies to $\delta_b$. In the bottom rows of Fig. \ref{fig:DELTAS-BRANCH-I} and \ref{fig:DELTAS-BRANCH-II} we present the relative difference between the density contrasts of baryons and dark matter. 

Branch I, Fig. \ref{fig:DELTAS-BRANCH-I}, presents a change in the sign of this relative difference in late times for all scales at strong coupling, for intermediate and large scales for medium coupling, and for large scales at weak coupling. For the photons, there is a change in sign for all couplings at large and intermediate scales, but no change for large scales. For branch II there is no change in sign for the baryons, but the photons present the same behavior as that of branch I.

\begin{figure*}[!htbp]
    \centering
    \begin{minipage}[b]{0.45\textwidth}
        \centering
        \includegraphics[width=\textwidth]{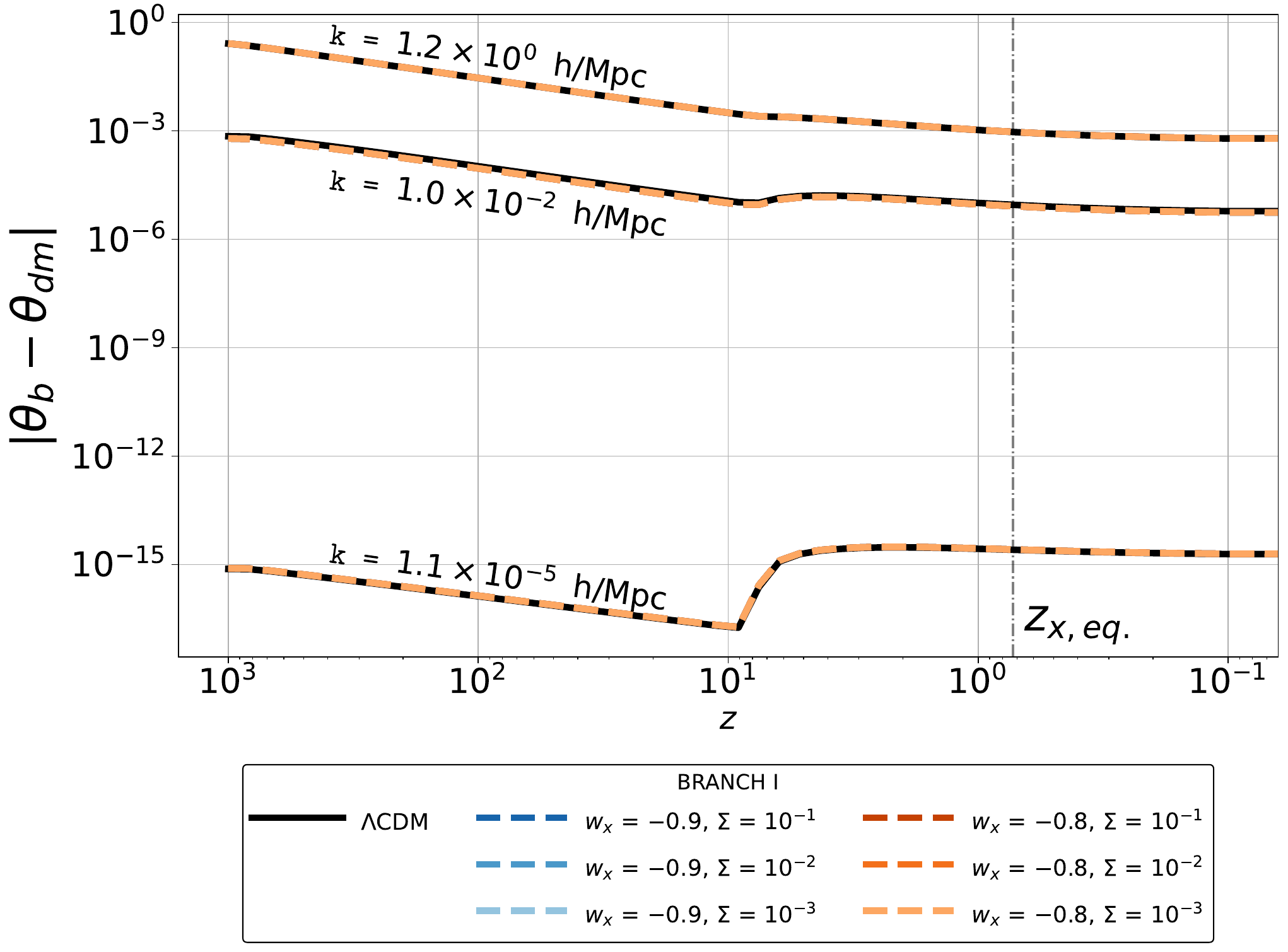}
    \end{minipage}
    \hfill
    \begin{minipage}[b]{0.45\textwidth}
        \centering
      
        \includegraphics[width=\textwidth]{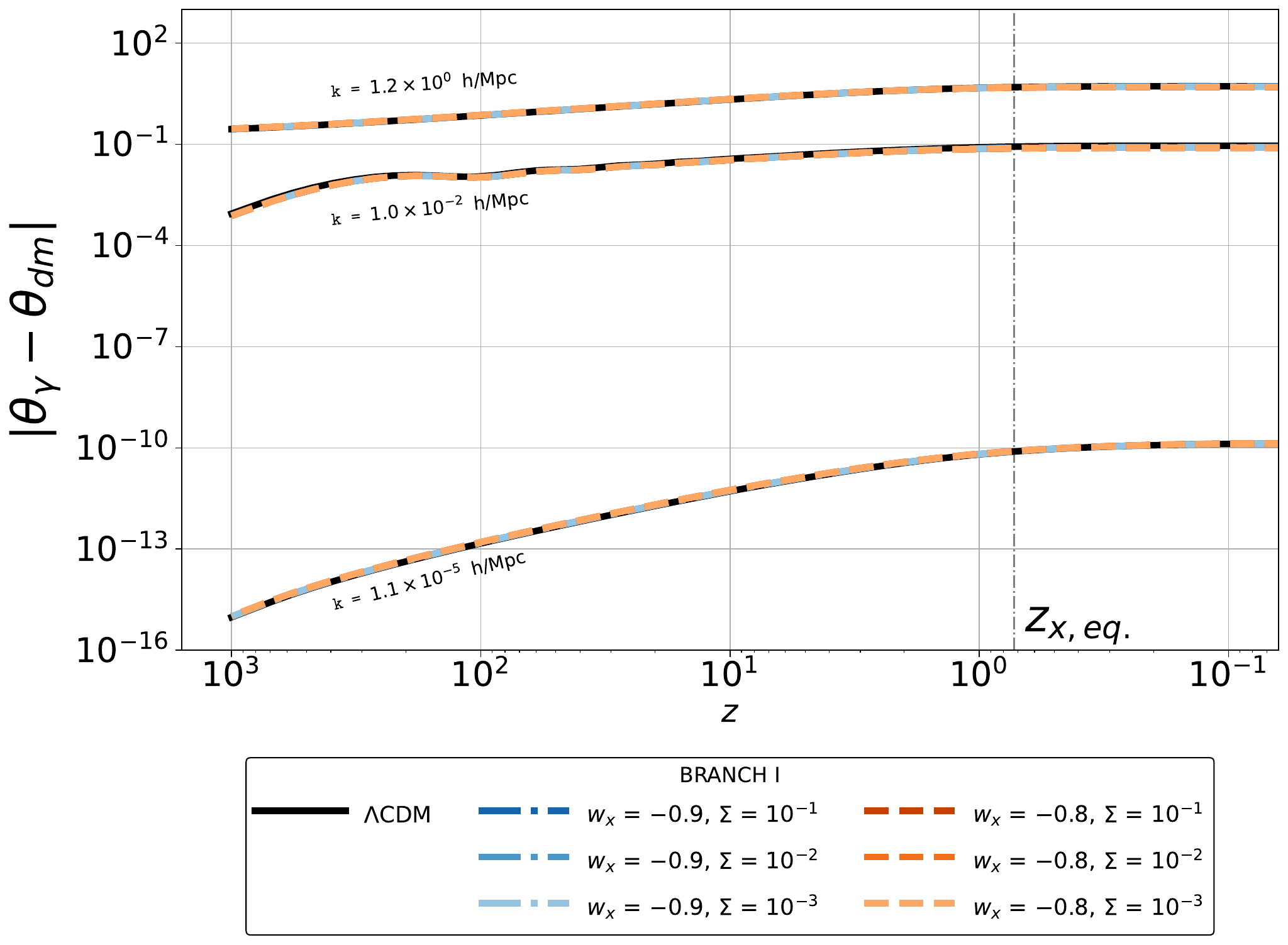}
    \end{minipage}

    \vspace{0.5cm}

    \begin{minipage}[b]{0.45\textwidth}
        \centering
        \includegraphics[width=\textwidth]{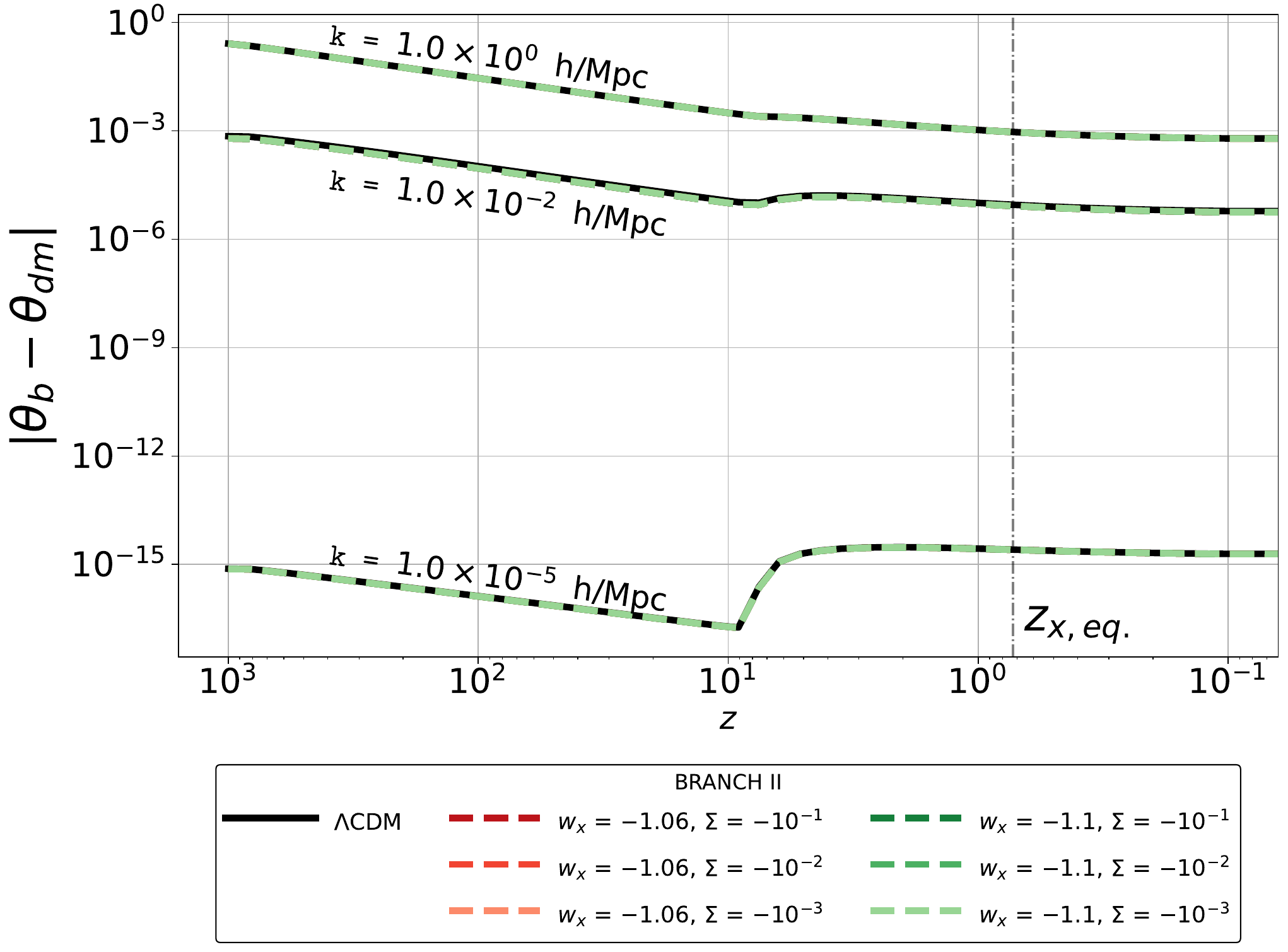}
    \end{minipage}
    \hfill
    \begin{minipage}[b]{0.45\textwidth}
        \centering
        \includegraphics[width=\textwidth]{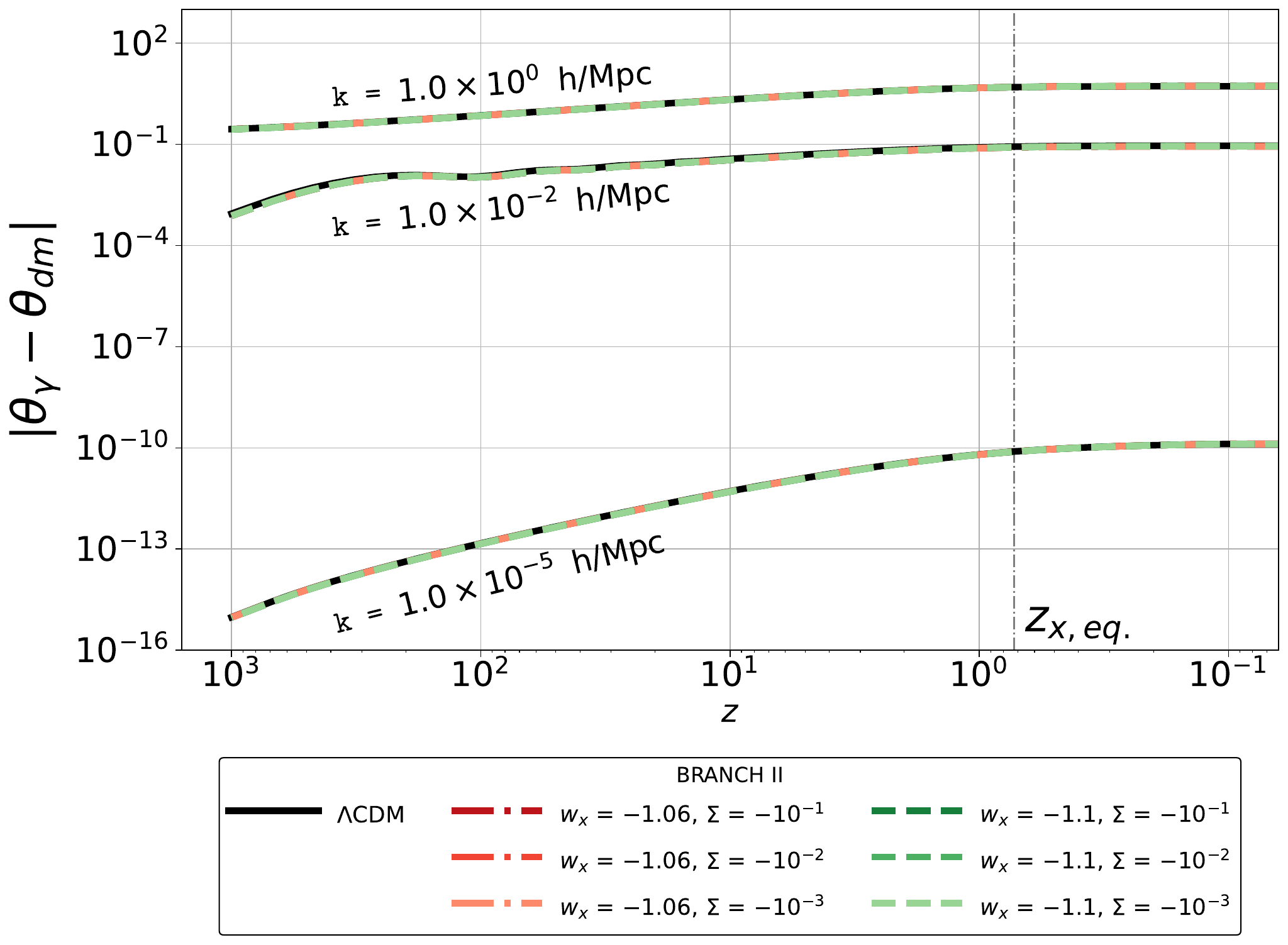}
    \end{minipage}
    \caption{\textbf{Top panel}: The relative velocities between baryons and photons with respect to dark matter for branch I. \textbf{Bottom panel}: Relative velocities between baryons and photons with respect to dark matter for branch II.}
    \label{fig:THETAS-BRANCH-I-II}
\end{figure*}

In Fig. \ref{fig:THETAS-BRANCH-I-II} we show the relative velocities among dark matter, baryons, and photons. At all times, they remain nearly constant across all models, regardless of whether the coupling is positive or negative. This consistency extends to photons as well; since they decouple after recombination, they are unaffected by interactions within the dark sector. As a result, these numerical simulations align with the theoretical predictions that underpin the effective theory known as the \texttt{frozen model}. 

\newpage
\newpage

\section{Appendix D: Clustering observables: $S_8$ and $f \sigma_8$} \label{appendix:d}

\begin{figure*}[!htbp]
    \centering
    \begin{minipage}[b]{0.475\textwidth}
        \centering
        \includegraphics[width=\textwidth]{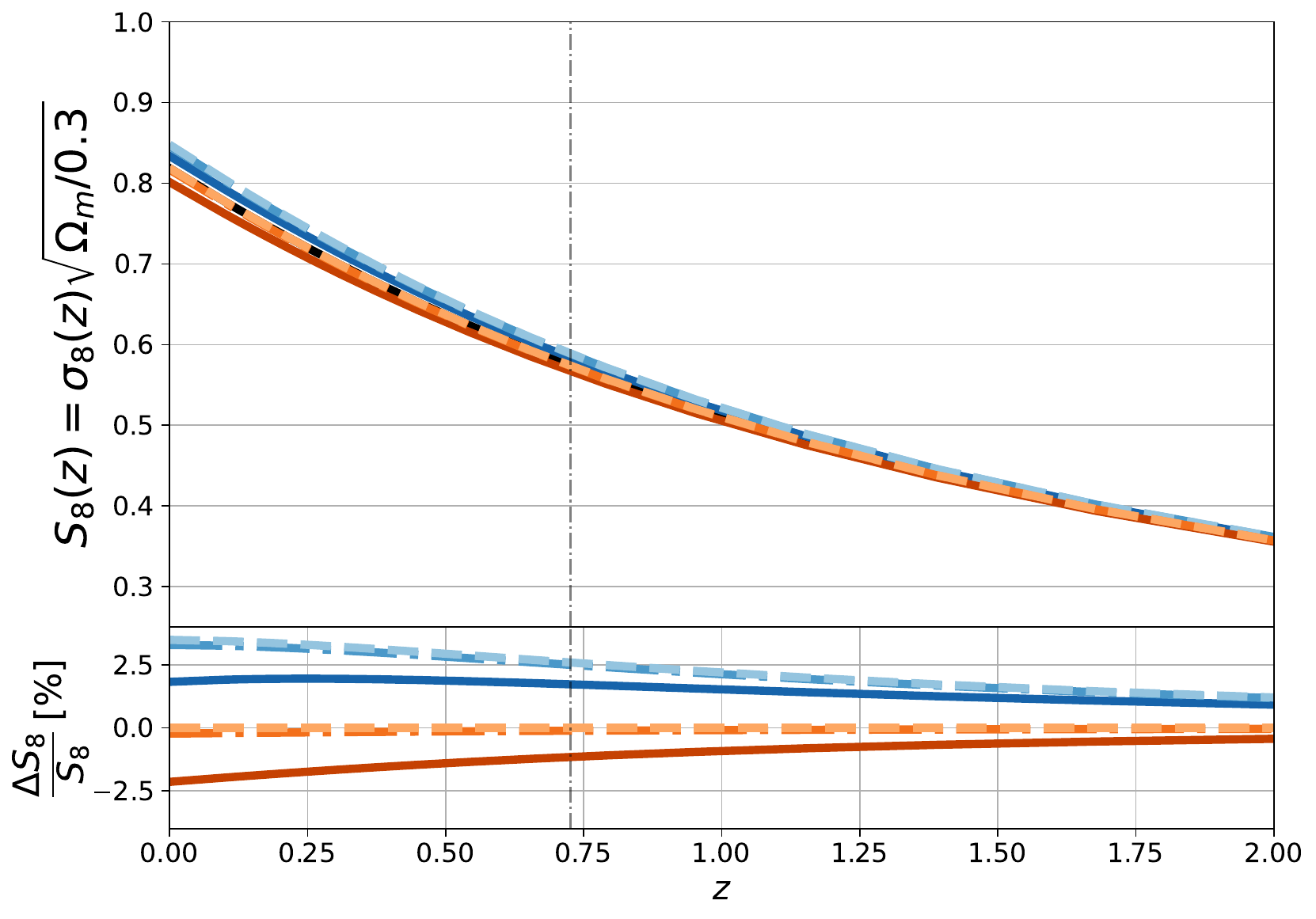}
    \end{minipage}
    \hfill
    \begin{minipage}[b]{0.475\textwidth}
        \centering
        \includegraphics[width=\textwidth]{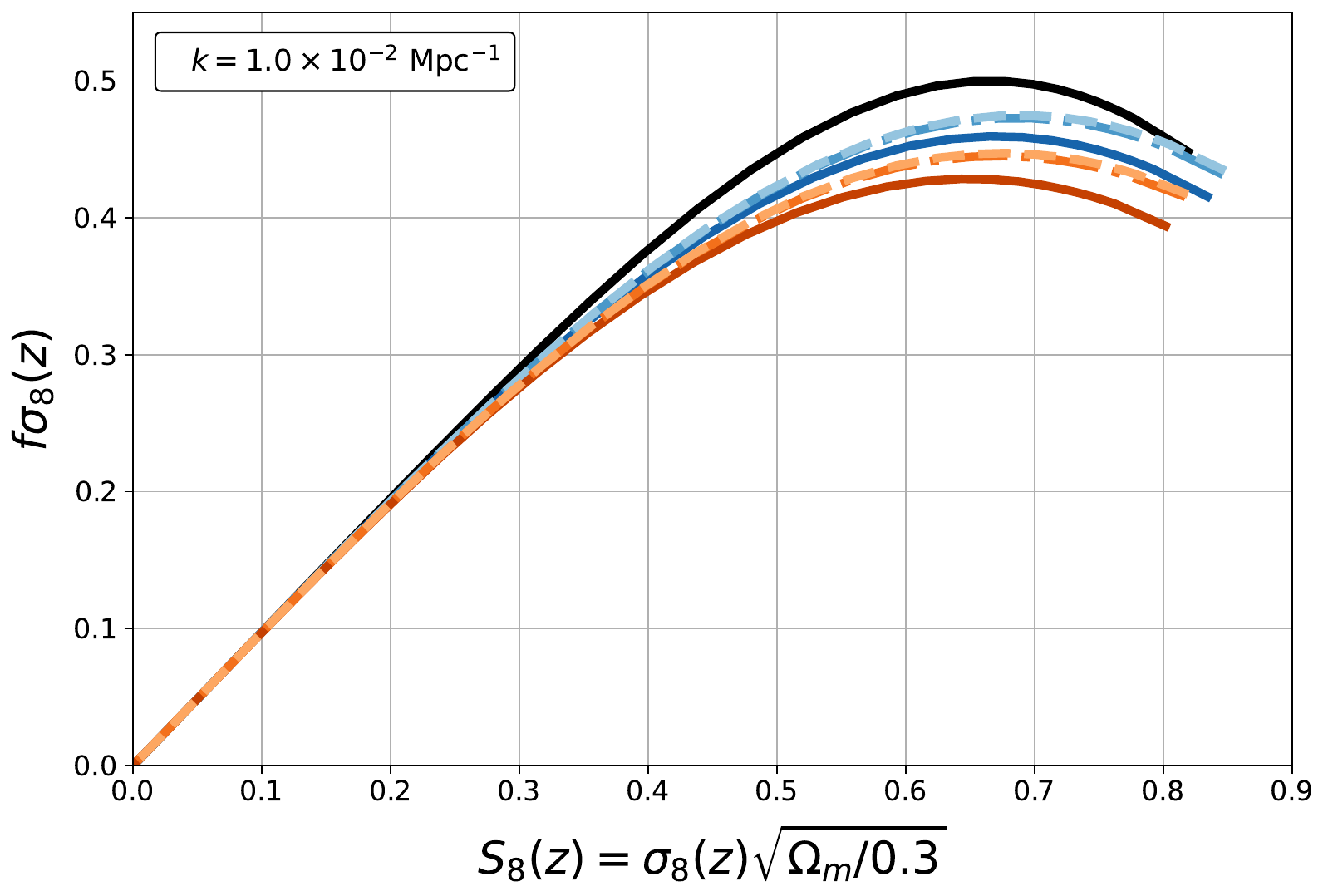}
    \end{minipage}

    \vspace{0.5cm}

    \begin{minipage}[b]{0.475\textwidth}
        \centering
        \includegraphics[width=\textwidth]{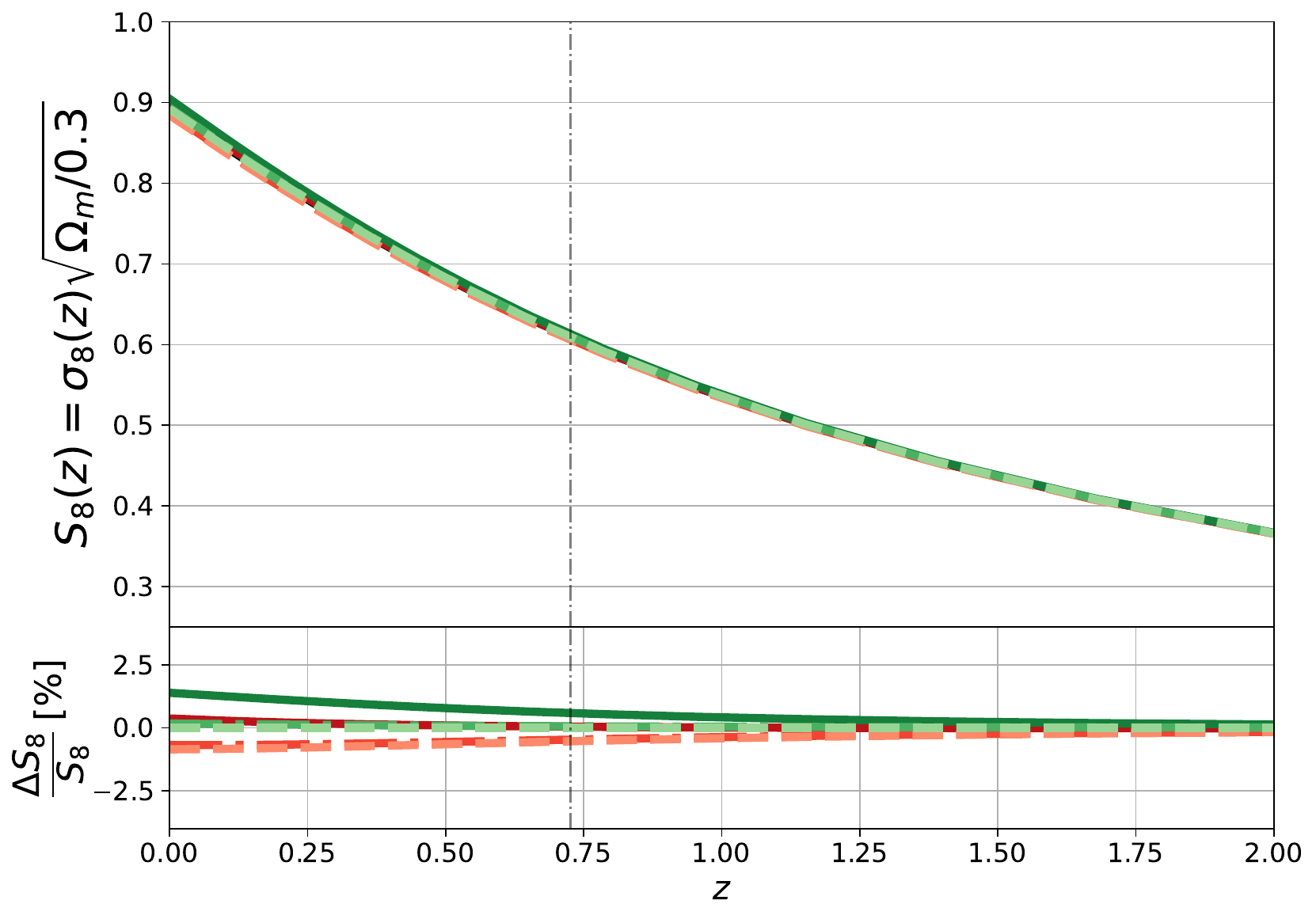}
    \end{minipage}
    \hfill
    \begin{minipage}[b]{0.475\textwidth}
        \centering
        \includegraphics[width=\textwidth]{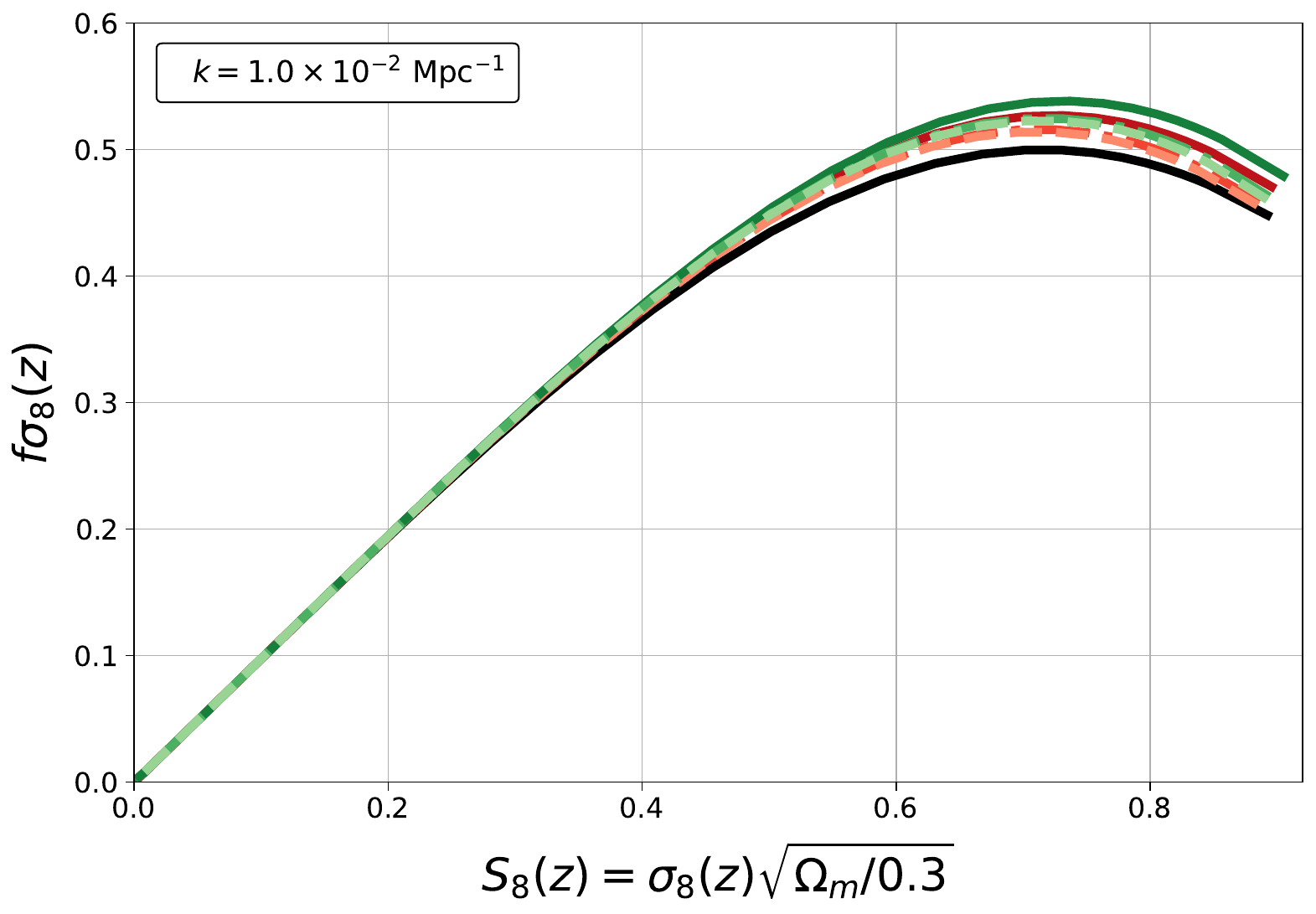}
    \end{minipage}

    \vspace{0.25cm}

    \begin{minipage}[b]{0.475\textwidth}
        \centering
        \includegraphics[width=\textwidth]{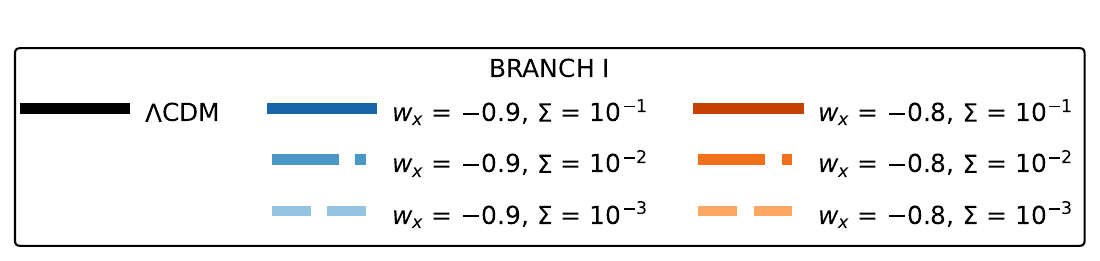}
    \end{minipage}
    \hfill
    \begin{minipage}[b]{0.475\textwidth}
        \centering
        \includegraphics[width=\textwidth]{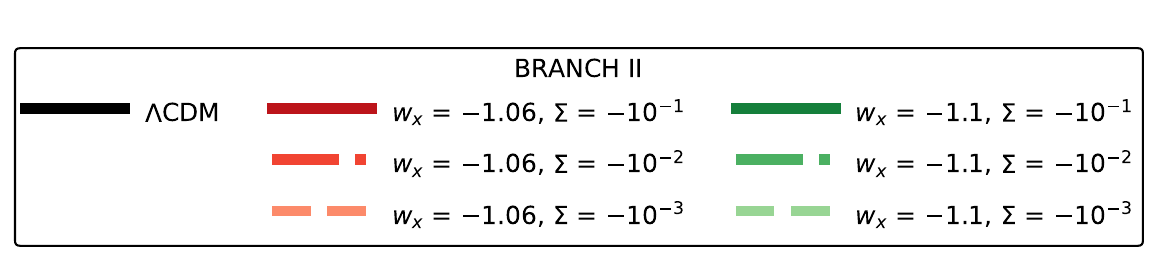}
    \end{minipage}
    
    \caption{\textbf{Top panel}: \textit{Left}: The cosmological evolution of the $S_8$ observable in terms of redshift for models in \texttt{branch I}. 
    \textit{Right}: The parametric relation between $f\sigma_8$ and $S_8$, which enhances the distinction between the $\mathrm{\Lambda CDM}$ model and the \texttt{branch I} models with strong coupling $|\Sigma|$. \textbf{Bottom panel}: \textit{Left}: The cosmological evolution of the $S_8$ function in terms of redshift for models in \texttt{branch II}. 
    \textit{Right}: The parametric relation between $f\sigma_8$ and $S_8$, which enhances the distinction between the $\mathrm{\Lambda CDM}$ model and the \texttt{branch II} models with strong coupling $|\Sigma|$.}
    \label{fig:fs8ap}
\end{figure*}

In this appendix, we emphasize the relationship between $S_8$ and $f \sigma_8$ within branches I and II. Fig. \ref{fig:fs8ap} illustrates that the observable $S_{8}(z)$ exhibits a slight deviation from the vanilla model when varying the interaction coupling $|\Sigma| \in [10^{-3}, 10^{-1}]$ for a fixed value of $w_{x}$. In particular, the relative percentage difference from the vanilla model remains below $3\%$ for $z\leq 0.75$ across branches I and II. However,  by examining the $f\sigma_{8}(z)-S_{8}(z)$ plane, we can leverage this connection as a valuable tool to differentiate among various interacting models. For branch I, the analysis of the functions $f\sigma_{8}(z)$ and $S_{8}(z)$ within the intermediate redshift range ($ 0.4 \leq z \leq0.9$) reveals that the parametric curve presents a pronounced arch with a lower amplitude compared to the vanilla model when $w_{x}$ is set at a higher value for a given $|\Sigma|$.  As the amplitude of this arch increases, the interaction coupling decreases. Conversely, in branch II, the scenario is reversed; here, the parametric curve displays an arch with a higher amplitude relative to the vanilla model.  For a phantom-like equation of state $w_{x}=\{-1.1, -1.08\}$, it is observed that as the interaction coupling $|\Sigma|$ increases, the arch becomes more pronounced, with a higher amplitude at its peak. 

\end{document}